\documentclass[pdflatex,sn-mathphys-num]{sn-jnl}

\usepackage{graphicx}%
\usepackage{multirow}%
\usepackage{amsmath,amssymb,amsfonts}%
\usepackage{amsthm}%
\usepackage{mathrsfs}%
\usepackage[title]{appendix}%
\usepackage{xcolor}%
\usepackage{textcomp}%
\usepackage{manyfoot}%
\usepackage{booktabs}%
\usepackage{algorithm}%
\usepackage{algorithmicx}%
\usepackage{algpseudocode}%
\usepackage{listings}%
\newcommand{\kuramoto}{Kuramoto}
\newcommand{\modelname}{\textit{HoloBrain}}
\newcommand{\gnnmodelname}{\textit{HoloGraph}}
\newcommand{\first}[1]{\textcolor{red}{#1}}   
\newcommand{\second}[1]{\textcolor{blue}{#1}} 
\newcommand{\third}[1]{\textcolor{green}{#1}}       
\newcommand{\fourth}[1]{\textcolor{orange}{#1}}           

\theoremstyle{thmstyleone}%
\theoremstyle{thmstyletwo}%

\theoremstyle{thmstylethree}%

\begin{document}

\title[Article Title]{Explore Brain-Inspired Machine Intelligence for Connecting Dots on Graphs Through Holographic Blueprint of Oscillatory Synchronization}


\author[1]{\fnm{Tingting} \sur{Dan}}
\equalcont{These authors contributed equally to this work.}

\author[1,2]{\fnm{Jiaqi} \sur{Ding}}
\equalcont{These authors contributed equally to this work.}

\author*[1,2,3,4]{\fnm{Guorong} \sur{Wu}}\email{grwu@med.unc.edu}

\affil[1]{\orgdiv{Department of Psychiatry}, \orgname{University of North Carolina at Chapel Hill}, \orgaddress{\street{333 S. Columbia Street}, \city{Chapel Hill}, \postcode{27599}, \state{NC}, \country{United States of America}}}

\affil[2]{\orgdiv{Department of Computer Science}, \orgname{University of North Carolina at Chapel Hill}, \orgaddress{\street{201 S. Columbia St}, \city{Chapel Hill}, \postcode{27599}, \state{NC}, \country{United States of America}}}

\affil[3]{\orgdiv{Department of Statistics and Operations Research}, \orgname{University of North Carolina at Chapel Hill}, \orgaddress{\street{318 Hanes Hall, CB 3260}, \city{Chapel Hill}, \postcode{27599}, \state{NC}, \country{United States of America}}}

\affil[4]{\orgdiv{UNC Neuroscience Center}, \orgname{University of North Carolina at Chapel Hill}, \orgaddress{\street{116 Manning Drive}, \city{Chapel Hill}, \postcode{27599}, \state{NC}, \country{United States of America}}}

{\renewcommand{\thefootnote}{}\footnotetext{The version of record of this article, first published in Nature Communications, is available online at Publisher’s website: \url{https://doi.org/10.1038/s41467-025-64471-2}.}}
\abstract{Neural coupling in both neuroscience and AI emerges dynamic oscillatory patterns that encode abstract concepts. To that end, we hypothesize that a deeper understanding of the neural mechanisms that determine brain rhythms could inspire next-generation design principles for machine learning algorithms, leading to greater efficiency and robustness. Following this notion, we first model the evolving brain rhythm by the interference between spontaneously synchronized neural oscillations (termed \textit{HoloBrain}). The success of modeling brain rhythms via an artificial dynamic system of coupled oscillations gives rise to the “first principle” for emerging brain-inspired machine intelligence through the common mechanism of synchronization (termed \textit{HoloGraph}), enabling graph neural networks (GNNs) to move beyond conventional heat diffusion paradigms toward modeling oscillatory synchronization. Our \textit{HoloGraph} not only effectively addresses the over-smoothing issue in GNNs but also manifests the potential of reasoning and solving challenging problems on graphs.}


\keywords{Graph neural networks, \kuramoto{} model, neural synchronization, dynamic system}



\maketitle


\section*{Introduction}
\label{intro}
Since the birth of artificial intelligence (AI), the journey of understanding human intelligence in neuroscience is on the same path as the pursuit of machine intelligence for graph learning in AI. Their similarity lies in both system structure and behavior.
First, human brain is an inter-wired system, where billions of neurons connect through complex networks to facilitate communication and coordination \cite{bassett2017network}. Real-world graph data, such as social networks and citation networks, exhibit similar topological properties, including ``small-world" characteristics \cite{watts1998collective}.
Second, both brain and graph data can be viewed as a complex system characterized by interconnected components that collectively give rise to emergent behavior. The human brain stands out as one of the most intricate dynamical systems, where ubiquitous neural synapses give rise to oscillatory brain rhythms. In parallel, the feature presentation learning process in current graph neural networks (GNNs) is another example of a dynamic system, where graph nodes are interconnected and collaboratively generate optimal features, layer-by-layer, for both individual nodes and the entire graph \cite{xhonneux2020continuous}.

Current machine learning systems excel human at pattern recognition, data analysis, and task-specific performance based on statistical correlations learned from data. However, they lack understanding, reasoning, and consciousness. In contrast, human brain processes information from multiple sensory systems through a highly coordinated and complex network involving sensory integration, parallel processing, and multimodal convergence \cite{ghazanfar2009emergence}. We continuously receive information from different sensory modalities in our everyday life, including sight, sound, and touch. For instance, the sound of the footsteps of a person entering a room affects neural and behavioral processing associated with another sensory modality (e.g., vision), providing a complementary multi-sensory representation of the external environment.

Neural oscillations have frequently been identified as a key mechanism supporting cross-modal influence by promoting the exchange of information between sensory modalities \cite{van2014multisensory}. Deep down at the neuron level, mounting evidence shows that neurons interact via lateral connections \cite{hubel1962receptive}. In this context, neural oscillations reflect the rhythmic fluctuations of excitability (i.e., brain rhythms) in neuronal ensembles related to the dynamics of the circuits in which ensembles are embedded as well as the kinetics of their ionic channels. Neighboring cortical units often exhibit oscillatory synchrony and transiently form assemblies that compete to explain sensory input \cite{gray1989oscillatory}. Such competition induces an information bottleneck and promotes specialization across assemblies \cite{amari1977competition}. The coordination mechanism is synchronization: neurons, similar to fireflies at night, align their activity locally, suppressing inconsistent fluctuations and compressing population states into lower-dimensional, more abstract codes \cite{Notbohm2012}. This perspective is consistent with Kuramoto-style projection dynamics used in oscillatory neural models \cite{miyato2024artificial}.

Revisiting conventional GNN backbones, such as graph convolutional network (GCN) \cite{kipf2016semi}, graph Transformer \cite{Yun2019}, they are powerful but lack the ``intelligence" seen in biological systems. On the flip side, human and animal cognition involves adaptive perception and behavior, driven by large-scale, dynamic oscillatory activity across neural circuits.
In this regard, the crucial link between the human brain and AI motivates us to embrace a more modern dynamical view of graph nodes in GNN as coupled oscillatory neurons in the human brain, which presents an exciting opportunity to inspire the next generation of GNN architectures.

\begin{figure}[h!]
  \centering
  \includegraphics[width=0.9\textwidth]{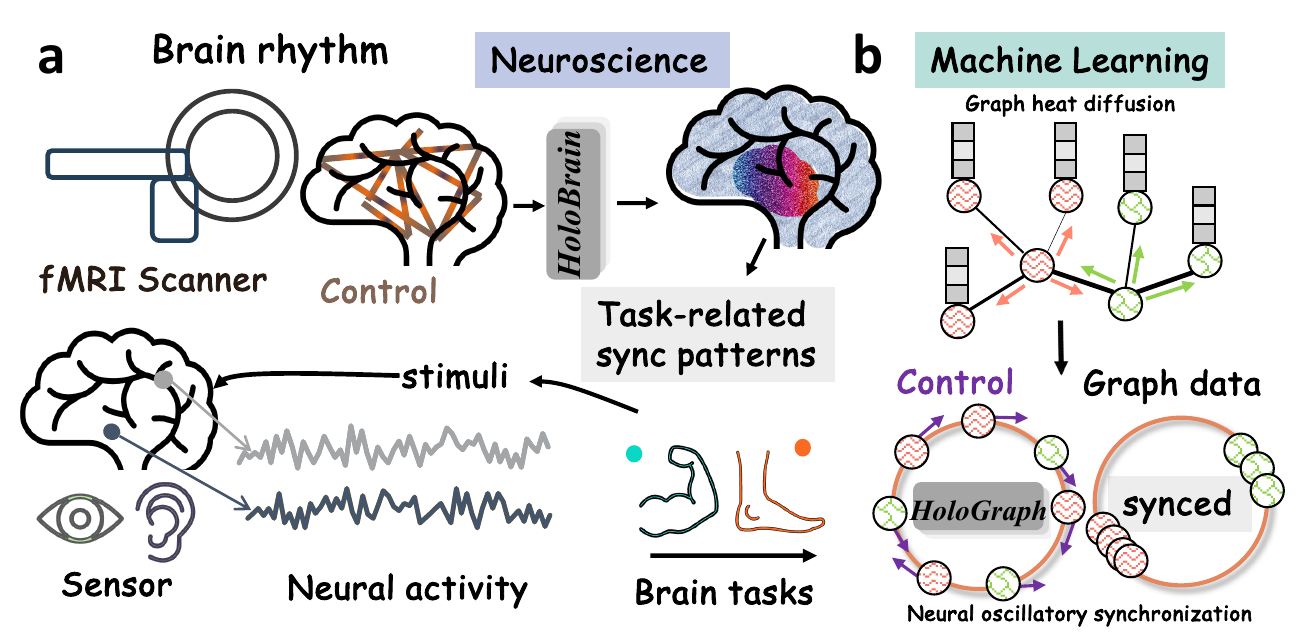}  
  \caption{ \textbf{From \textit{HoloBrain} to \textit{HoloGraph}.} \textbf{a} Neuroscience $\rightarrow$ \textit{HoloBrain}. Task stimuli drive neural activity measured by fMRI. \textit{HoloBrain} models oscillatory coordination on the structural network—via Kuramoto‐style dynamics with attending‐memory (control)—to recover task-evoked synchronization patterns.
 \textbf{b} Machine learning $\rightarrow$ \textit{HoloGraph}. Inspired by this mechanism, \textit{HoloGraph} replaces heat-diffusion message passing with neural-oscillation–based synchronization on graphs, using outcome-specific feedback to align distant nodes into meaningful clusters.}
  \label{fig:overview}
\end{figure}

In this work, we seek to establish a biologically inspired learning mechanism for GNNs in two steps, as illustrated in Fig. \ref{fig:overview}.
First, we propose a harmonic holography technique to describe spontaneous functional fluctuations (coined \modelname{} -- Fig. \ref{fig:overview}a) -- an explainable deep model to uncover latent cross-frequency coupling (CFC) patterns via the interference of self-organized harmonic waves on top of structural brain networks. Specifically, we combine the physics principle in \kuramoto{} model \cite{kuramoto1975self} and the notion of attending memory \cite{hutchinson2012memory} in cognitive neuroscience to uncover the governing equation that gives rise to fluctuating oscillations in human brains. 
To that end, \modelname{} incorporates a biologically inspired learning mechanism grounded in oscillatory synchronization alongside a system‑level attention component driven by historical activity.
Second, we draw inspiration from the neural mechanism of oscillatory synchronization to develop a learning mechanism for graph data, which is also both structurally interconnected and behaviorally dynamic.
As shown in Fig. \ref{fig:overview}b, popular message passing and graph convolution techniques in current GNNs are limited in the clich\'e of graph heat diffusion \cite{chamberlain2021grand}, where excessive information exchange on graph topology often leads to over-smoothing issues \cite{dan2023re}. To ratchet the gear of GNNs another notch forward, we further introduce a biologically inspired graph neural oscillatory synchronization model by regarding each graph node as an oscillator coupled with the others, forming an inter-wired dynamic system on which the evolution of feature representations emulates artificial neural activities.
We name the biologically inspired GNN model \gnnmodelname{} to reflect its shared system dynamics with \modelname{}.
\gnnmodelname{} is trained to steer the pattern formation via oscillatory synchronization, where high-level interference patterns in the spatial domain allow us to perceive that the underlying graph consists of two clusters of `red' and `green' dots, as shown in Fig. \ref{fig:overview}b. Following the notion that the synchronization of neural oscillation enables the brain to process complex information efficiently, the learning mechanism inspired by brain rhythms is expected to revolutionize machine intelligence towards the paradigm of reasoning and solving problems on graphs.

\section*{Results}

\subsection*{Interference patterns from spontaneous functional fluctuations}

\begin{figure}
  \centering
 \includegraphics[width=1.0\textwidth]{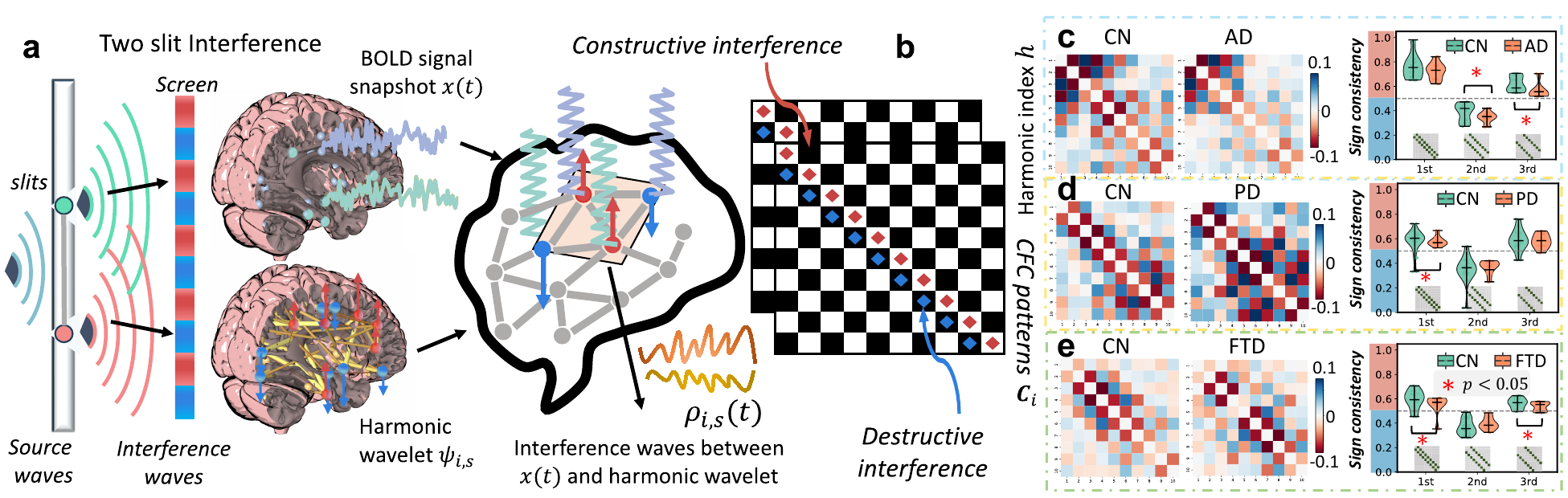}
  \caption{\textbf{From wave interference to CFC patterns: physics-inspired fingerprints for neurodegenerative diseases}. {The physics insight of cross-frequency coupling (building block of \modelname{}) is analogous to the wave interference principle \textbf{a}, which both yields constructive (red) and destructive (blue) interference patterns on the screen and cross-frequency couplings, as shown in \textbf{b}. We present the node-wise averages of CFC patterns along with quantitative measures (sign consistency degree) that highlight disease-specific interference patterns in \textbf{c} AD using ADNI dataset (\href{https://adni.loni.usc.edu/data-samples/adni-data/}{https://adni.loni.usc.edu/data-samples/adni-data/}), \textbf{d} PD using PPMI dataset {\href{https://www.ppmi-info.org/}{https://www.ppmi-info.org/}}, and \textbf{e} FTD using NIFD dataset (\href{https://memory.ucsf.edu/research-trials/research/allftd}{https://memory.ucsf.edu/research-trials/research/allftd}). `$\ast$' denote statistically significant group differences at the levels of $p<0.05$ (two-sided Mann–Whitney U), respectively. The brain images are generated using Surf Ice 
  \cite{rorden2025surfice} \href{https://www.nitrc.org/projects/surfice/}{https://www.nitrc.org/projects/surfice/} (Surf Ice 6), a software for reading surface- and volume-based neuroimaging data.}} 
  \label{fig:young}
\end{figure}

In Young’s double slit experiment conducted in 1801 \cite{young1804bakerian}, the waves from two slits interfere constructively or destructively at the screen, as shown in Fig. \ref{fig:young}a.
In computer vision, holography is a stereo-imaging technique that generates a hologram by superimposing a reference beam on the wavefront of interest \cite{gabor1948new}. The resulting hologram is an interference pattern that can be recorded on a physical medium. Following the concept of wave-to-wave interference, we computationally “record” the CFC of time-evolving interference patterns that are formed by superimposing the neural activities across oscillation frequencies, as shown in Fig. \ref{fig:young}b. Specifically, we first construct subject-specific graph wavelets from the structural connectome data \cite{chen2022characterizing}, which serve as the basis functions of the geometric scattering transform (GST) \cite{gao2019geometric} for each brain region. Applying GST to the fMRI (functional magnetic resonance imaging) data yields a set of ``neural oscillators'', each exhibiting fluctuations governed by its own natural frequency. In each brain region, we further capture the presence of CFC patterns in a frequency-to-frequency correlation matrix based on Pearson's correlation of spontaneous fluctuations between two neural oscillators, as shown in Fig. \ref{fig:young}b.   

We first show the population average of global CFC patterns (i.e., averaging throughout brain regions and across individuals) for healthy vs. Alzheimer's disease (AD), healthy vs. Parkinson's disease (PD), and healthy vs. Frontotemporal dementia (FTD) in Fig. \ref{fig:young}c-e, respectively. There is a clear sign that the population-wise CFC average exhibits remarkable off-diagonal striping patterns, which resemble the constructive and destructive interference phenomena in Young’s double slit shown in Fig. \ref{fig:young}. In light of this, we hypothesize that the consistency of CFC degree along these striping patterns may help explain how brain function gives rise to diverse cognitive and behavioral outcomes. Since the striping patterns are visually distinguishable between healthy and disease cohorts, we further speculate that the coherence of maintaining the striping patterns might be an indicator of how brain function is altered as the disease progresses.
Following this clue, we compute the degree of \textit{sign consistency} for each off-diagonal line by calculating the percentage of values exhibiting the predominant sign across instances. In Fig. \ref{fig:young}c-e, we plot the distribution of sign consistency for the first three off-diagonal lines for each disease cohort. Values above the 0.5 line indicate a predominant distribution of positive CFCs, while values below 0.5 reflect a predominance of negative CFCs. It is apparent that the sign consistency degree manifests strong differences between healthy and disease connectomes at the significance level $p<0.05$. More evidence of CFC patterns in task-fMRI is shown in Supplementary Fig. 1. 

Remarks. The presence of interference patterns in CFC matrices implies that discovering the laws governing neural oscillatory fluctuations in the human brain is a necessary precursor to understanding perceptual and behavioral processes. Such understanding could ultimately inspire the development of biologically grounded learning mechanisms in AI.

\subsection*{Identify governing equation of brain rhythms}

Mounting evidence shows that neural oscillations play a vital role in synchronizing different brain regions, and the Kuramoto model \cite{kuramoto1984chemical} has been widely used to study neural synchronization dynamics in both computational neuroscience and empirical brain data \cite{cabral2011role}. 
Suppose the system consists of $N$ oscillators with their pairwise coupling strength denoted by $K_{ij} (i,j=1,...,N)$. By allowing each oscillator to vibrate at its own natural frequency $\omega_i$, the dynamics of coupled phase oscillator in \kuramoto{} model is formulated as:
\begin{equation}
    \frac{d \theta_i}{d t}=\omega_i+[\sum {_{j = 1}^N} K_{i j} \sin \left(\theta_j-\theta_i\right)],
    \label{Eq.Kuramoto}
\end{equation}
where the evolution of phase $\theta_i \in \mathbb{R}$ associated with $i^\text{th}$ oscillator is synchronized with other coupled oscillators.
Initially, each oscillator evolves independently at its intrinsic frequency $\omega_i$. Over time, the coupling term $[\cdot]$ enables the oscillators to adjust their phases relative to one another, leading to collective synchronization. This process results in oscillators locking their phases and forming coherent groups.

Given the presence of interference patterns in CFC matrices (shown in Fig. \ref{fig:young}), we propose a physics-informed deep neural network to parameterize the Kuramoto model, which allows us to understand the neurobiological mechanism of how coupled oscillatory synchronization emerges underlying cognitive tasks. 
We conceptualize that oscillatory activity across interconnected brain regions together forms a dynamic structure shaped by neural synchronization mechanisms. To that end, we design a neural network architecture that characterizes the whole-brain oscillatory synchronization governed by the Kuramoto model of coupled phase oscillators. Furthermore, we replace the conventional attention module with an optimal control component, inspired by the notion of attending memory in cognitive neuroscience, to focus on task-relevant information while suppressing irrelevant stimuli during dynamic cognitive activities. 
Drawing inspiration from neuroscience and grounded in the principles of holography, we name our method \modelname{}.
In a nutshell, the observed blood oxygenation level dependent (BOLD) signals are first processed using GST, and the resulting output, denoted by $\mathbf{X}^{(0)}$, serves as the input to \modelname{}. The output of \modelname{} includes (1) synchronized neural oscillations $\mathbf{X}^{(L)}$ and (2) the learned control pattern $\mathbf{Y}^{(L)}$ after $L$ layers.
By combining the power of machine learning and an unprecedented amount of public neuroimaging data, our \modelname{} method not only yields promising performance in predicting cognitive tasks but also offers an interpretable perspective of functional brain dynamics.

\textbf{Prediction of cognitive tasks.} 
We evaluate the model performance on the following public datasets. 
    (i) The Lifespan Human Connectome Project Aging (HCP-A) dataset \cite{bookheimer2019lifespan}. The HCP-A dataset provides a comprehensive view of the aging process and is instrumental in task recognition research. It consists of data from 717 subjects, including both fMRI scans (4,849 time series) and diffusion-weighted imaging (DWI) scans (717 in total). The dataset includes four tasks associated with memory: VISMOTOR, CARIT, FACENAME, and REST state. Each fMRI scan comprises 300 time points. In our experiments, these tasks are treated as distinct categories in a four-class classification problem.
    (ii) The Human Connectome Project - Young Adults (HCP-YA) database \cite{van2013wu}. Each scan contains data from seven cognitive tasks related to memory, including Motor, Relational, Social, Working Memory, Language, Emotion, and Gambling. Each task-based fMRI scan contains 175 time points in our experiments. Additionally, the Working Memory task (abbreviated as HCP-WM) involves 2-back and 0-back task conditions with body, place, face, and tool stimuli, interspersed with fixation periods. A resting-state period follows two sequential cognitive task periods in an alternating fashion, resulting in fMRI scans with a total of 405 time points.

For both HCP-A and HCP-YA datasets, the brain is partitioned into 116 regions using the AAL atlas \cite{tzourio2002automated}. The structural connectivity (SC) matrix is a $116 \times 116$ matrix, where each element quantifies the number of fibers linking two brain regions. The fiber counts are normalized by the total fiber count of each subject. Functional connectivity (FC) is computed as the Pearson correlation between BOLD signals of different regions, representing the temporal synchronization of neuronal activity in the brain. The preprocessing pipelines are fully described and accessible through publicly available tools (fMRIPrep: \url{https://fmriprep.org/en/stable/}, 
QSIPrep: \url{https://qsiprep.readthedocs.io/en/latest/}). To evaluate scalability, we employ the Brainnetome Atlas \cite{fan2016human} to partition the whole brain into 246 regions for the HCP-WM data. In our experiments, the HCP-YA tasks are treated as distinct categories in a seven-class classification problem, while the HCP-WM tasks form an eight-class classification problem. For these experiments, we report 5-fold cross-validation results.

Prediction accuracy. We compare against a diverse set of state-of-the-art graph-based methods, including vanilla GCN \cite{kipf2016semi}, graph attention network (GAT) \cite{velivckovic2017graph}, graph isomorphism network (GIN) \cite{xu2018powerful}, GCNII \cite{chen2020simple}, GraphSAGE \cite{hamilton2017inductive}, graph transformer with spectral attention network (SAN) \cite{kreuzer2021rethinking}, graph-coupled oscillator networks (GraphCON) \cite{rusch2022graph} and a kuramoto model-based approach (KuramotoGNN) \cite{nguyen2024coupled}. Fig. \ref{Holobrain}a and Supplementary Table 3 list the quantitative comparison results for three datasets on nine methods. $\modelname{}$ achieves the best performance on all datasets over the existing hand-designed GNN models. 

\begin{figure}
  \centering
  \includegraphics[width=1.0\textwidth]{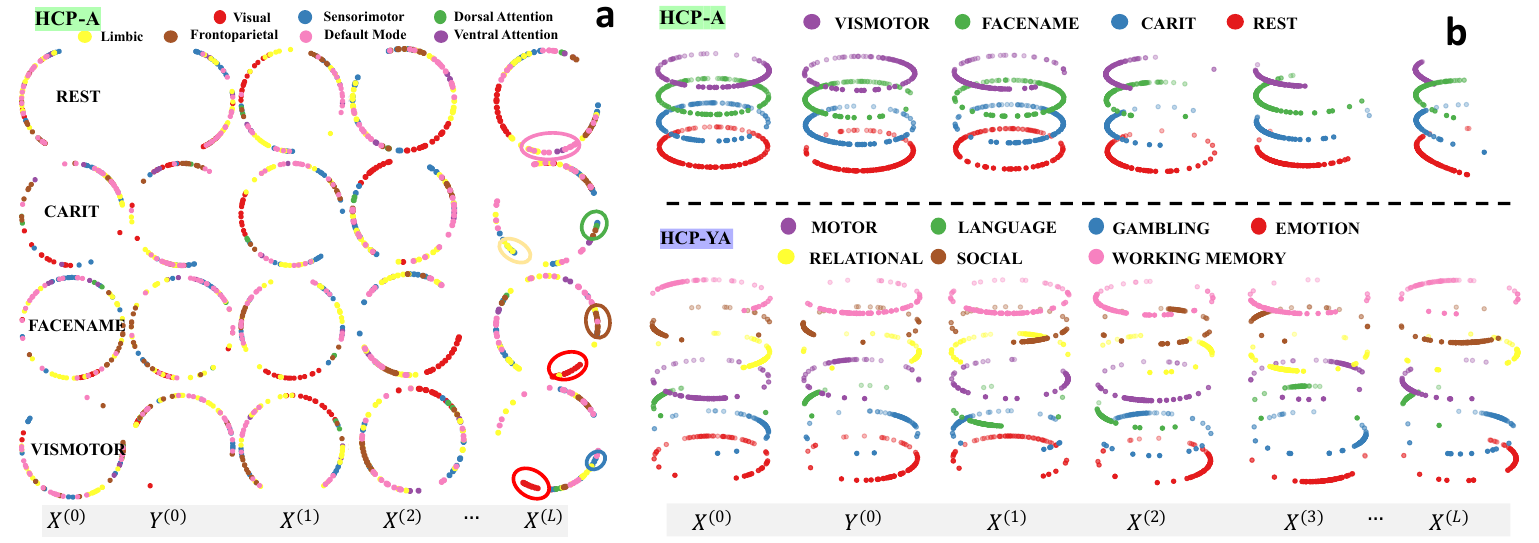}  
\caption{\textbf{Model interpretability: task-specific neural synchronization in phase space.}
Each brain region is mapped to a unit-norm feature and visualized by its phase on the unit circle.
{Color coding:} in panel \textbf{(a)} the colors denote canonical subnetworks;
in panel \textbf{(b)} the colors denote congnitive tasks.
\textbf{a} HCP-A examples. Columns show the evolution through \modelname: $\mathbf{X}^{(0)} \!\rightarrow\! \mathbf{Y}^{(0)} \!\rightarrow\! \mathbf{X}^{(1)}\ldots \mathbf{X}^{(L)}$.
Tighter dots indicate stronger synchrony; e.g., VISMOTOR concentrates in visual/sensorimotor systems, whereas REST shows stronger default-mode synchrony.
\textbf{b} Group-level dynamics. As depth increases, regional phases within the same task collapse toward a task-specific attractor (higher sign-consistency), while different tasks remain separable, revealing distinct synchronization fingerprints for each cognitive state.}
  \label{fig:vis}
\end{figure}

Model interpretability. These results provide compelling evidence that our physics-based oscillation model effectively synchronizes different brain regions, with the resulting synchronization patterns corresponding to distinct brain cognitive states. Especially in HCP-WM, all tasks from the same subject share identical structural connectivity. This design penalizes topology-dependent baselines but favors models that read out temporal coordination. \textit{HoloBrain} explicitly models oscillatory synchronization and therefore separates the eight states more effectively than static GNNs. To examine the relationship between neural synchronization and cognitive tasks, we examine the phase-space representations of various brain regions, as shown in Fig. \ref{fig:vis}a. Our analysis reveals that different brain states exhibit unique synchronization patterns. For instance, during the VISMOTOR task, synchronization is predominantly observed in the visual and sensorimotor regions, reflecting the functional demands of the task. In contrast, during the REST state, the synchronization pattern is more pronounced within the default mode network, aligning with its role in self-referential and intrinsic cognitive processes \cite{raichle2001default}.
In addition, we further investigate task-specific synchronization across all subjects. As neural synchronization progresses (Fig. \ref{fig:vis}b), tasks gradually synchronize across groups, resembling the phenomenon of runners on a track who initially move at their own speeds but gradually form synchronized clusters through mutual interaction. More visualization analyses are shown in Supplementary Fig. 3.

Remarks. The collective synchronization emerged from \modelname{}, akin to patterns observed in natural systems such as fireflies flashing in unison or birds flying in formation, highlights the emergent coordination in neural dynamics. Notably, distinct synchronization patterns emerge between tasks, reinforcing the structured nature of this process.

Emergence of interference patterns from \modelname{}.
Since \modelname{} is inspired by the interference of cross-frequency coupling, we hypothesize that \modelname{} recognizes neural activities by synchronizing oscillations across frequencies and across brain regions, which eventually give rise to interference patterns observed in Fig. \ref{fig:young}.

\begin{figure}
\centering
\includegraphics[width=1.0\textwidth]{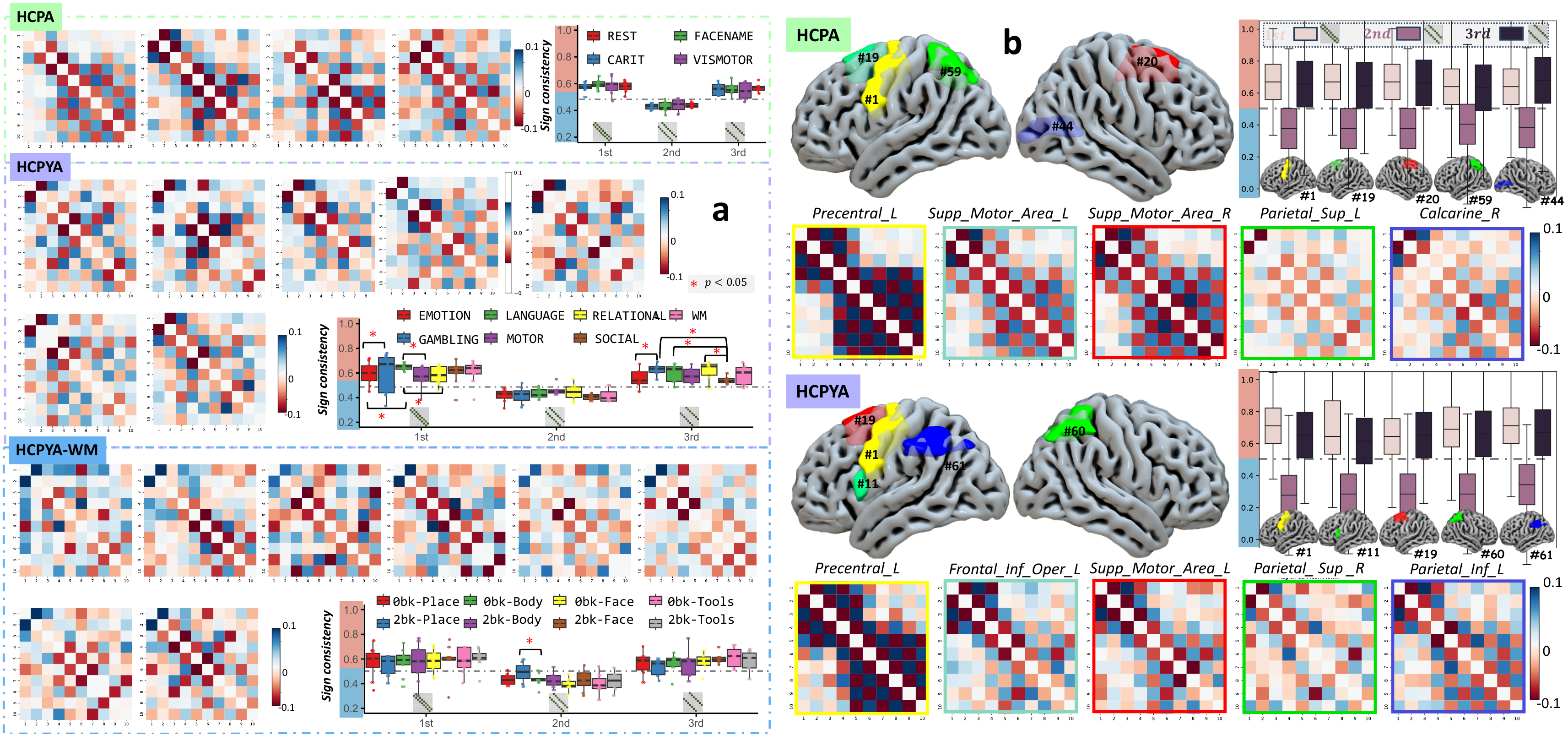}
\caption{\textbf{Whole-brain and regional interference patterns learned by \textit{HoloBrain}.}
\textbf{a} Whole brain. For each cohort/task comparison, we show the group-average CFC matrices (red/blue colors indicate constructive/destructive interference). The box plots quantify the \emph{sign-consistency degree} along the dominant off-diagonal. An asterisk denotes a significant difference between the two tasks (two-sided Mann–Whitney U, $p<0.05$).
\textbf{b} Regional exemplars. Cortical regions with the strongest learned synchrony are highlighted on the brain surface (visualized using Surf Ice \cite{rorden2025surfice} \href{https://www.nitrc.org/projects/surfice/}{https://www.nitrc.org/projects/surfice/}), and their regional CFC matrices are shown below each parcel. These examples illustrate that \modelname{} captures task-specific interference fingerprints both at the whole-brain level and within characteristic regions.}
\label{figcfc_hcp}
\end{figure}

\begin{figure}
\centering
\includegraphics[width=1.0\textwidth, trim=0 0.2cm 0 0, clip]{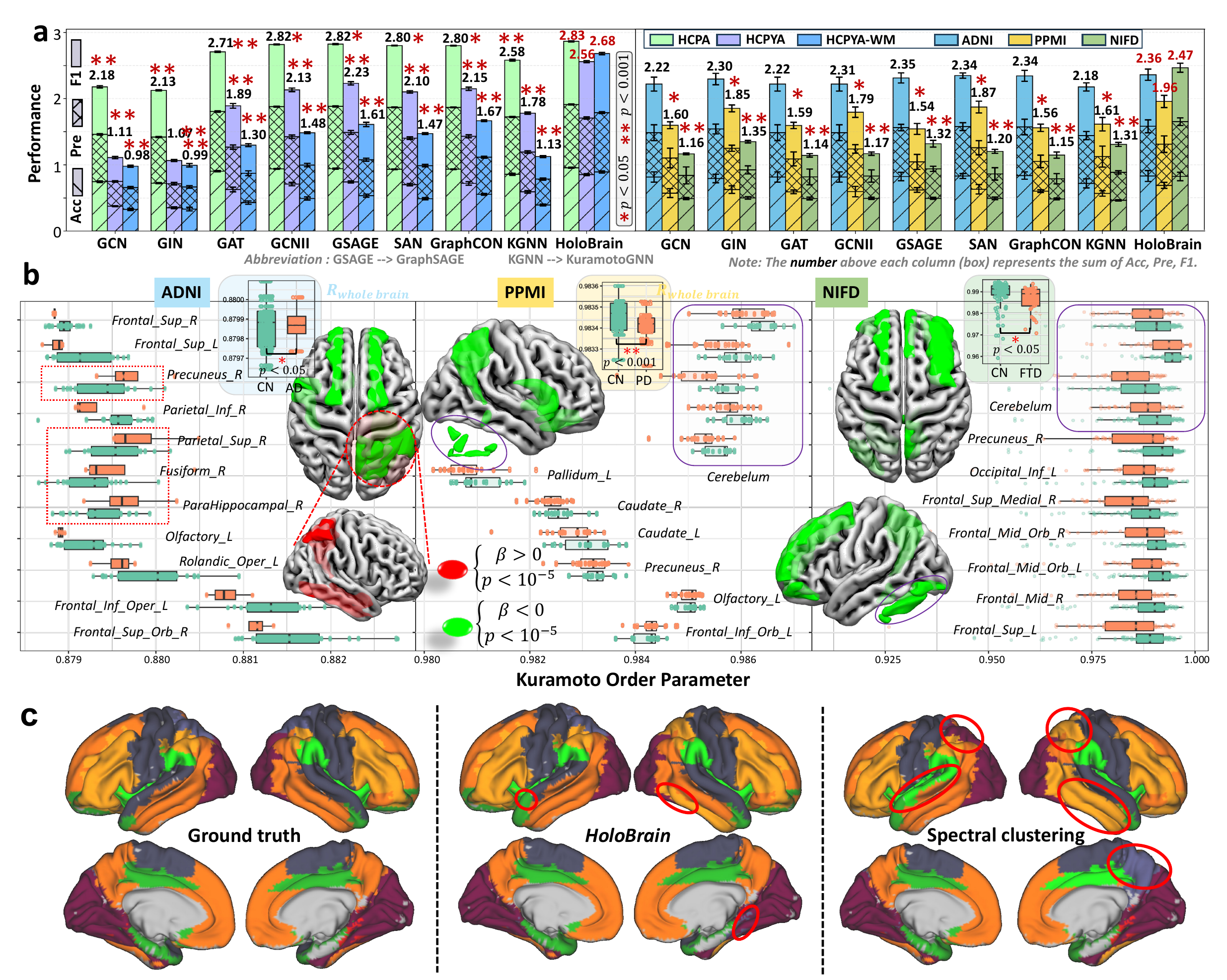}
\caption{\textbf{Overall performance of \modelname{}.} 
\textbf{a} Comparison of nine methods on HCPA, HCPYA, HCPYA-WM, ADNI, PPMI, and NIFD datasets. Bars show accuracy (Acc), precision (Pre), and F1-score (F1), with red asterisks marking cases where \modelname{} significantly outperforms competing methods (paired $t$-test, $p<0.05$ and $p<0.001$). Red fonts denote the best performance. 
\textbf{b} Neural synchronization levels, both at the whole-brain and regional scales, exhibit significant group differences between CN and AD, CN and PD, and CN and FTD. The group comparison results at the whole-brain level are shown in the shaded bounding box. At a significance level of $p<10^{-5}$, we display the brain regions with significant health vs. disease differences as well as the distributions of KOP degree for CN (green) and disease (red) subjects. In addition, brain regions with reduced synchronization levels associated with disease pathology are highlighted in green, while those with increased synchronization are marked in red. The brain images are visualized using Surf Ice \cite{rorden2025surfice} (\href{https://www.nitrc.org/projects/surfice/}{https://www.nitrc.org/projects/surfice/}).
\textbf{c} Clustering results of \modelname{} versus spectral clustering, compared against the ground truth. Red circles highlight discrepancies from the ground truth. Different clusters (colors) are mapped in brain surface using ParaView (v5.10.1) \cite{ahrens2005paraview}  \href{https://www.paraview.org/}{https://www.paraview.org/}.}
\label{Holobrain}
\end{figure}

Emergence of cross-frequency interference patterns associated with cognitive tasks.
In \modelname{}, we predict cognitive tasks by modeling the synchronization of neural oscillatory waves throughout the brain, with each wave oscillating at its own natural frequency. In this context, we compute the $H\times H (H=10)$ CFC matrix on each brain region, where each element in the CFC matrix characterizes the interference between two synchronized oscillatory waves. First, we examine whole-brain CFC patterns associated with underlying cognitive tasks by averaging the $10\times 10$ CFC matrices across brain regions and individuals. As shown in Fig. \ref{figcfc_hcp}a, it is evident that our \modelname{} reproduces the constructive and destructive patterns along the off-diagonal in the CFC matrices, suggesting the emergence of interference patterns during the learning process. Through visual inspection, the population-wise CFC matrices exhibit substantial differences across cognitive tasks. Following the quantitative analysis in Fig. \ref{fig:young}, we calculate the sign consistency degree for the first, second, and third off-diagonal lines of each subject’s whole-brain CFC matrix, where the distribution of sign consistency degree of each cognitive task is shown in the bar-plot in Fig. \ref{figcfc_hcp}a. Significant task-to-task differences at $p<0.05$ are indicated by `$\ast$'.

Emerging knowledge of critical brain regions from the perspective of neural synchronization.
Furthermore, we investigate constructive/destructive interference patterns in each regional CFC matrix. In Fig. \ref{figcfc_hcp}b, we display five regions with the highest sign consistency degrees along the first, second, and third off-diagonal lines in older (HCPA: $58.0\pm 18.5$ years) and young adults (HCPYA: $29.1\pm 3.6$ years). In older adults, left precentral gyrus (\#1), left and right supplementary motor area (\#19–20), right calcarine (\#44), and left superior parietal gyrus (\#59) manifest highest local synchrony, which form a coherent motor–attention–visual compensatory network. Numerous aging studies \cite{Grady2010,Seidler2010} have shown that older adults exhibiting increased $\beta/\gamma$ coupling in primary motor (M1/SMA) circuits have high reserve to preserve motor function. In addition, the observed full synchrony in the primary visual cortex (calcarine) supports prior reports suggesting that increased local coupling in the aging visual system serves as a compensatory strategy for maintaining perception \cite{zhang2022biophysical}.
Similarly, the left superior parietal lobule, a core node of the dorsal attention network, often displays elevated short‑range connectivity to offset declines in spatial attention. Taken together, our findings imply that neural oscillatory synchronization provides a window for understanding critical dynamics of brain function that characterize healthy brain aging. In young adults, the five regions showing the strongest local synchrony are: \#1-left precentral gyrus, \#11-left inferior frontal gyrus (opercular part), \#19-left supplementary motor area, \#60-right superior parietal gyrus, and \#61-left inferior parietal gyrus. 
These brain regions form a tight sensorimotor–attention loop in which primary and supplementary motor regions are anchored to adjacent parietal cortices. This “motor–attentional” network consistently exhibits the highest amplitude‑coupling and local coherence at rest in healthy young adults, reflecting its role in maintaining fine motor control and spatial attention even in the absence of overt movement. This finding has been replicated across multiple HCP analyses \cite{Power2011}, confirming that the sensorimotor network dominates intrinsic neural synchrony in youth.

Remarks. Leveraging the principles of the Kuramoto model, our \modelname{} provides a powerful explanatory framework that captures the majority of variability in the relationship between functional fluctuations and cognitive tasks. The inference patterns underlying cross-frequency coupling have been replicated by synchronizing neural oscillations in our physics-informed neural network, which allows us to investigate criticality through the lens of local synchrony.

\textbf{Clinical value in disease diagnosis.} 
In this experiment, we evaluate the potential of \modelname{} in the early diagnosis of neurodegenerative diseases, using the following three public neuroimaging data cohorts for AD, PD, and FTD. Specifically, we involve (i) Alzheimer’s Disease Neuroimaging Initiative (ADNI): This dataset includes resting-state fMRI data from 135 subjects, comprising individuals diagnosed with Alzheimer's disease (AD) and cognitively normal (CN) controls. It is designed to track brain changes associated with AD progression. (ii) Parkinson’s Progression Markers Initiative (PPMI): A multi-center study that collects neuroimaging data from 173 subjects, including individuals with Parkinson’s disease (PD), scans without evidence of dopaminergic deficit (SWEDD), prodromal PD, and CN. 
(iii) Neuroimaging Initiative for Frontotemporal Lobar Degeneration (NIFD): This dataset focuses on frontotemporal dementia (FTD) and includes resting-state fMRI data from 1,010 subjects. Participants are categorized into CN, logopenic variant of primary progressive aphasia (LVPPA), behavioral variant frontotemporal dementia (BV), progressive non-fluent aphasia (PNFA), and semantic variant (SV) groups. 

Consistent with the preprocessing pipeline applied to the HCP dataset and the comparison methods, we obtain the regional mean BOLD signals and $116 \times 116$ SC matrices based on the AAL atlas \cite{tzourio2002automated}. The graph embeddings of these methods are BOLD signals and the adjacency matrices are SCs. For the disease diagnosis task, the ADNI dataset is formulated as a binary classification problem (AD vs. CN), the PPMI dataset as a four-class classification problem (PD, SWEDD, Prodromal, and CN), and the NIFD dataset as a five-class classification problem (CN, LVPPA, BV, PNFA, and SV). We evaluate model performance using accuracy (ACC), precision (PRE), and F1-score (F1), reporting results based on 5-fold cross-validation.

Diagnosis accuracy.  As shown in Fig. \ref{Holobrain}a and Supplementary Table 3, our \modelname{} method significantly outperforms the counterpart approaches in identifying AD, PD, and FTD subjects, where the improvement is significant (* $p<0.05$, paired t-test) on PPMI and NIFD datasets, demonstrating the effectiveness of physics-informed deep model in disease diagnosis. Compared to other GNN models, the governing equation in our \modelname{} offers a systems-level perspective on functional fluctuations, helping to elucidate the mechanisms underlying dysfunctional syndromes in neurodegenerative diseases \cite{park2013structural}, as detailed below.

Neural synchronization metrics as biomarkers for neurodegenerative disease. We hypothesize that the reduced degree of neural synchronization might be a putative indicator of the neurodegeneration process in aging brains. To validate our hypothesis, we analyze the phase synchronization between different brain regions by computing the Kuramoto Order Parameter (KOP), thus the population-wise KOP for $i^{\text{th}}$ region is defined as:  $R_i = \left| e^{\sqrt{-1} (arg(\mathbf{x}^{(L)}_{i}))}  \right|$, where $arg(\mathbf{x}^{(L)}_{i})$ is the synchronized oscillation (see Supplementary Information for details) in phase space (output of $L^{\text{th}}$ layer of \modelname{}) of $i^{\text{th}}$ brain region.
The population average of whole-brain KOP is also defined as: $R_{whole} = \frac{1}{N} \left| \sum_{i}^N R_i \right|$, where $N$ denotes the number of brain regions. In this context, we examine the group difference of synchronization level for ADNI, PPMI and NIFD datasets (as shown in Fig. \ref{Holobrain}b) to investigate whether and how the neural synchronization mechanism is altered in neurodegenerative diseases.

First, we conduct a $t$-test to compare the whole-brain KOP degree between CN individuals and those with diseases. As illustrated in the shaded panels in Fig.~\ref{Holobrain}b, all disease groups (AD, PD, and FTD) consistently show a significantly lower whole-brain KOP degree compared to the CN group ($p<0.05$).
Second, we scale up the group comparison by expressing the disease outcome via regional KOP degree. At a significance level $p<10^{-5}$, we have identified group differences in 11 brain regions between CN and AD. For CN vs. PD, 5 cerebellar regions and 6 other brain regions showed significant differences. Similarly, the comparison between CN and FTD revealed significant differences in 4 cerebellar regions and 7 additional brain regions. We display these disease-specific brain regions in Fig. \ref{Holobrain}b, along with the distributions of regional KOP degree $R_i$ in CN and disease groups. Specifically, the majority of brain regions show a reduction in KOP degree (effect size $\beta <0$ in the linear regression model) in the disease group (marked in green), indicating that disease-related pathology may disrupt neural synchronization. This pattern is quite consistent in PD and FTD. However, four brain regions—the right precuneus, right superior parietal gyrus, right fusiform gyrus, and right parahippocampal area—exhibit an opposite trend (marked in red), all associated with Alzheimer's disease, where pathological changes appear to enhance synchronization levels. One possible reason is that these regions are located on the spreading pathway of tau aggregates (one of the key pathological hallmarks) at the early stage of AD. As a result, increased synchronization in these areas may facilitate the spread of AD pathology across the brain, thereby contributing to disease progression.

\begin{figure}
\centering
\includegraphics[width=1.0\textwidth]{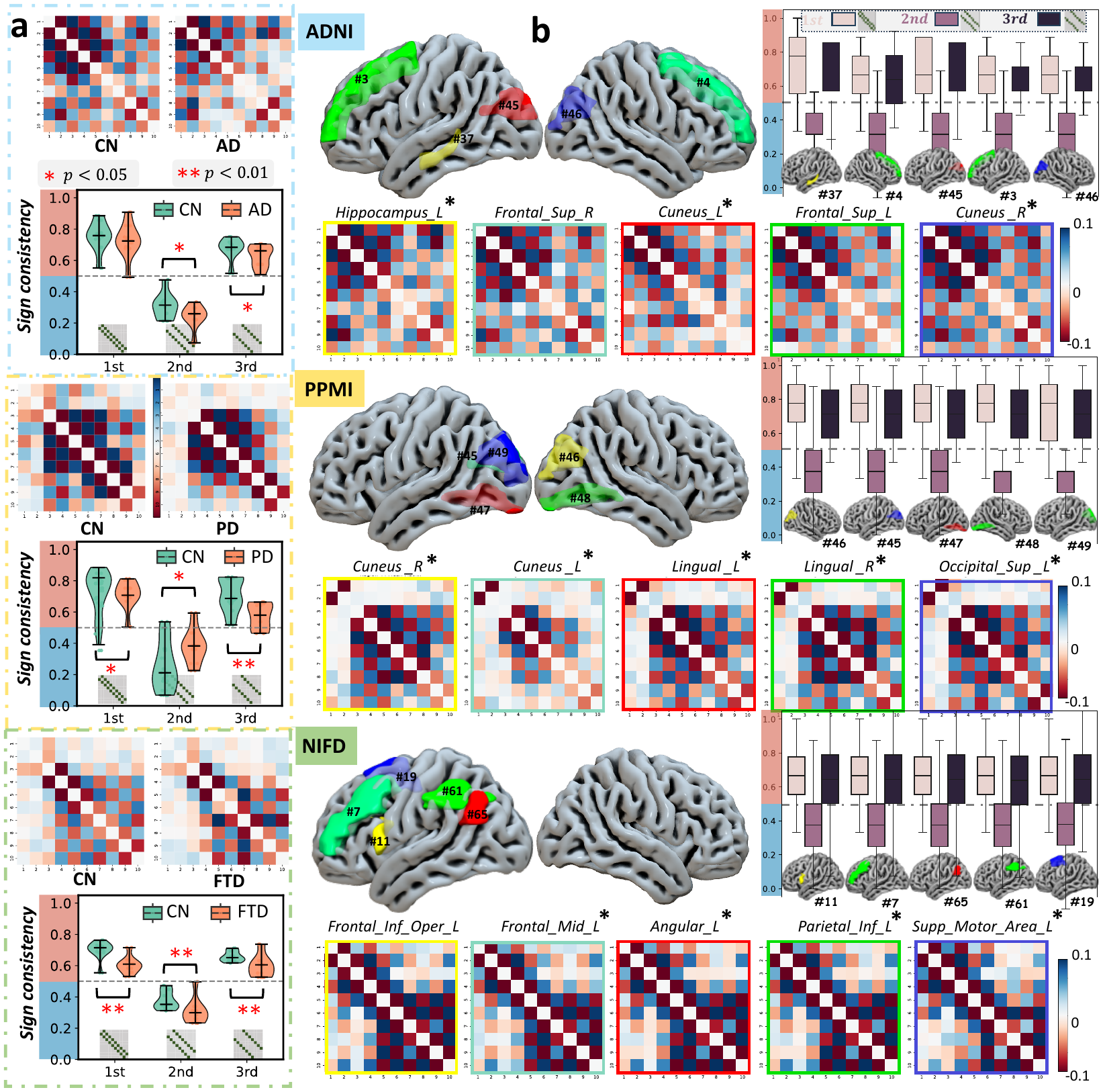}
\caption{\textbf{The interference patterns of CFC emerged from \textit{HoloBrain}.} \textbf{a} Whole-brain comparison. For each dataset (ADNI: AD vs. CN; PPMI: PD vs. CN; NIFD: FTD vs. CN) we show the group-average CFC matrices together with quantitative summaries of the \emph{sign-consistency degree} measured on the 1st, 2nd and 3rd off-diagonal bands (violin plots). Asterisks mark significant group differences (* $p<0.05$, ** $p<0.01$).
\textbf{b} Regional exemplars. The five cortical regions with the strongest learned synchrony per cohort are highlighted on the cortical surface and their regional CFC matrices are displayed below, with accompanying box plots of the sign-consistency degree for the three off-diagonal bands. The prominent striped off-diagonal structure indicates organized cross-frequency interactions that are systematically altered in disease. The brain images are visualized using Surf Ice \cite{rorden2025surfice} (\href{https://www.nitrc.org/projects/surfice/}{https://www.nitrc.org/projects/surfice/}).}
 \label{figcfc_disease}
\end{figure}

Uncover disease-specific interference patterns.
To replicate the hypothesis of wave-to-wave interference (Fig. \ref{fig:young}b), we compute the CFC matrix for each brain region between synchronized oscillations determined by the Kuramoto model in \modelname{} (i.e., the final-layer node representations). By averaging the CFC matrix across regions and individuals, we show the average CFC matrices in Fig. \ref{figcfc_disease}a for AD, PD, and FTD. Compared to the average CFC matrices in Fig. \ref{fig:young}, our \modelname{} reproduces quite similar interference patterns along the off-diagonal lines, while revealing more pronounced group differences. This finding implies that the neurobiological process underlying oscillatory synchronization might be disrupted during the progression of neurodegenerative diseases.

Furthermore, we investigate the hypothesis that the contrast between constructive and destructive interference patterns is linked to neurodegenerative diseases. To do so, we first select five brain regions with the strongest local synchrony, i.e., the total degree of sign consistency along the first, second, and third off-diagonal lines is higher than in other regions, for AD, PD, and FTD subjects. We conduct this step on ADNI, PPMI, and NIFD datasets separately. Note, clinical label information is not used in region selection. 
As shown in Fig. \ref{figcfc_disease}b, left hippocampus, bilateral superior frontal gyrus, and bilateral cuneus exhibit the highest levels of synchrony in older adults with AD, forming a hippocampo–occipital constellation. This pattern recapitulates medial–temporal and occipital short-range hyperconnectivity, previously associated with early tau deposition, memory decline, and visual–attentional symptoms in AD~\cite{Brier2015}.
In PD subjects, the top five regions with the highest level of local synchrony are located in primary/secondary visual cortex—right/left cuneus and lingual gyri plus the left superior occipital gyrus (Fig. \ref{figcfc_disease}b). The resulting “visual belt” mirrors the `$\beta/\gamma$'‑band hypersynchrony and tremor‑related cerebello–thalamo–visual loop repeatedly reported in PPMI cohorts~\cite{weil2016visual}.   
In Fig. \ref{figcfc_disease}b, five regions in the FTD population are strictly peri‑Sylvian and left‑lateralised, comprising the inferior frontal operculum, middle frontal gyrus, angular gyrus, inferior parietal lobule, and the supplementary motor area. These nodes form the executive–speech circuit that shows early structural loss and local FC increases in non‑fluent primary progressive aphasia and behavioral variant FTD~\cite{Mandelli2016,bouzigues2024disruption}.

After that, we perform a group comparison to examine whether the total degree of sign consistency is associated with the clinical label using a linear regression model. At a significance level of $p<0.05$, the synchrony degrees at left hippocampus and left/right cuneus are strongly associated with AD, where cognitively normal subjects manifest higher local synchrony degrees than AD subjects. All of the top five highly synchronous brain regions demonstrate significant differences between CN and PD. Likewise, all regions—except the inferior frontal operculum—show significant differences between CN and FTD. Notably, the group difference patterns observed in PD and FTD are consistent with those in AD, with both conditions exhibiting reduced local synchrony.  

Remarks. The results presented in Fig. \ref{Holobrain}–\ref{figcfc_disease} provide evidence that the neurobiological mechanisms underlying oscillatory synchronization may be disrupted throughout the progression of neurodegenerative diseases.

\subsection*{Brain rhythm–inspired learning mechanism for graphs}
Suppose $\mathbf{X}=[\mathbf{x}_i]_{i=1}^N$ is the initial graph embeddings, where the graph consists of $N$ nodes. 
Current GNNs follow the principle of graph heat diffusion to aggregate features on top of the graph topology by $\frac{\partial \mathbf{X}}{\partial t}+\Delta \mathbf{X}=0$, where $\Delta$ denotes graph Laplacian operator \cite{chamberlain2021grand}. Since the fundamental assumption in current GNNs is the graph smoothness $\sum _{i,j}\mathbf{x}_i^{\intercal}\Delta \mathbf{x}_j$, excessive information exchange often leads to over-smoothing issues \cite{li2021gnns}. 
Inspired by \modelname{} for modeling functional dynamics, we present a solution to address the over-smoothing issues by (1) considering each graph node as an oscillator and (2) compressing the graph feature representations towards partial/full synchronization in the latent oscillatory phase space. In this regard, the oscillatory nature of the brain-inspired system prevents the GNN model from converging to a static and trivial state where all graph embeddings become identical by the end of graph heat diffusion. In a nutshell, we translate the design principle of \modelname{} to the clich\'e of GNN, yielding a GNN model coined \gnnmodelname{}.

Integrate Kuramoto model into \gnnmodelname{}. We treat each graph node as an oscillator, with interactions determined by the graph topology. \gnnmodelname{} takes the graph embedding after GST transform as the initial condition for oscillatory synchronization, denoted by $\mathbf{x}{(0)}$, operating at different harmonic frequencies. 
In this context, the phase dynamics of these oscillators are governed by a vector-valued equation \cite{miyato2024artificial}: 
\begin{equation}
 \frac{d \mathbf{x}_i}{dt}=\omega_i + [\rho\cdot\phi_{{x_i}}(\sum {_{j = 1}^N} K_{ij}\mathbf{x}_j)],   
 \label{eq.kuramoto_vector}
\end{equation}
where $\phi$ represents a projection of the vector $K_{ij}\mathbf{x}_j$ onto the tangent space at point $\mathbf{x}_i$ on the unit sphere and $\rho$ is a scalar hyperparameter.

Integrate optimal control into \gnnmodelname{}. The global nature of dynamics in the Kuramoto model poses challenges to capturing transient patterns of functional fluctuations that are closely associated with cognitive tasks. To address this issue, we introduce the notion of optimal control from cognitive neuroscience \cite{dobbins2002cognitive} into our model by:
\begin{equation}
    \dot{\mathbf{x}}_i=\omega_i + \rho\cdot \phi_{\mathbf{x}_i}(\mathbf{y}_i + \sum\limits_{j\neq i}K_{ij}\mathbf{x}_j),
    \label{eq.kuramoto_control}
\end{equation}
where $\mathbf{y}_i$ is a learnable memory controller of past activity of oscillator $\mathbf{x}_i$. 
Intuitively, each $\mathbf{y}_i$ captures behavior-specific episodic memories and characterizes the mechanistic role of each oscillator engaged in a given cognitive task.
By tracking the progression of cognitive states, we can continuously assess feedback from neural oscillations and adjust the synchronization in Kuramoto system through episodic control.

\begin{figure}
  \centering
  \includegraphics[width=1.0\textwidth]{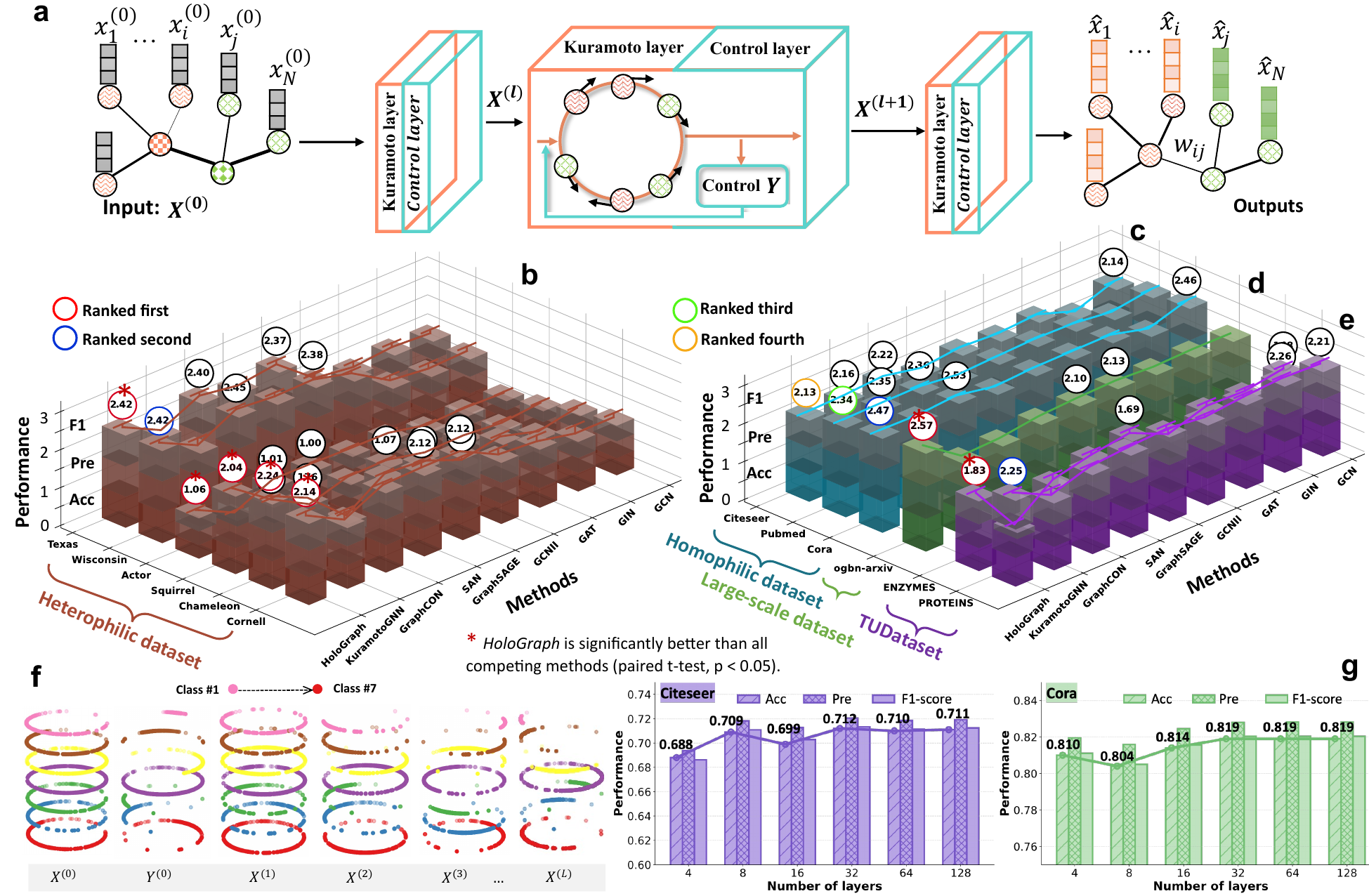}  
  \caption{\textbf{Overview of \gnnmodelname{}.} 
  \textbf{a} Network architecture of \gnnmodelname{}. Each graph node is treated as an oscillator, and feature learning becomes a dynamic synchronization process: oscillators with similar characteristics align on the sphere through alternating Kuramoto and control layers. 
  \textbf{b–d} Node classification performance. Each 3D stacked bar encodes accuracy (Acc), precision (Pre), and F1-score (F1) for one method (x-axis) on one dataset (y-axis); the vertical axis shows the normalized sum $(\text{Acc}+\text{Pre}+\text{F1}) \in [0,3]$. A polyline connects the total score of each method within a dataset; the number in each white circle indicates the total score, and the circle color marks the rank (red=1st, blue=2nd, green=3rd, orange=4th). Red asterisks denote results where \gnnmodelname{} significantly outperforms all competing methods (paired $t$-test, $p<0.05$). 
  \textbf{e} TUDatasets for graph classification (ENZYMES, PROTEINS). 
  \textbf{f} Synchronization trajectories on Cora: $\mathbf{X}^{(0)}, \mathbf{Y}^{(0)}$ are initial states, $\mathbf{X}^{(l)}$ denotes the state after layer $l$; colors indicate classes and arrows show evolution direction. 
  \textbf{g} Over-smoothing resistance: accuracy, precision, and F1 versus layer number on Citeseer (purple) and Cora (green), showing that \gnnmodelname{} maintains stable performance with depth.}
  \label{HoloGraph}
\end{figure}

Taken together, feature learning in \gnnmodelname{} is cast as a dynamic process governed by Kuramoto model. As shown in Fig. \ref{HoloGraph}a, the backbone consists of two layers: the \kuramoto{} layer and the optimal control layer, which together update the feature representations ${\mathbf{X}}^{(l+1)}$ and control patterns ${\mathbf{Y}}^{(l+1)}$.
Kuramoto layer is used to compress the feature $\mathbf{x}_i^{(l)}$ on each oscillator by $\mathbf{x}_i^{(l+1)}=norm(\mathbf{x}_i^{(l)}+\epsilon \cdot \Delta \mathbf{x}_i^{(l)})$, where $\Delta \mathbf{x}_i^{(l)}=\omega_i^{(l)}+\rho\cdot\phi(\mathbf{y}_i^{(l)}+\sum_{j\neq i} w_{ij}^{(l)}\mathbf{x}_j^{(l)})$ and $norm(\cdot)$ denotes the normalization operation that ensures the oscillator stay on the sphere. In parallel with the governing equation in \modelname{}, our \gnnmodelname{} compresses feature representations by aligning the oscillators with similar properties or labels via the \kuramoto{} model, thus preventing the over-smoothing problem in modern GNNs.
Control layer reads out patterns encoded in the oscillators by a mapping function 
${\mathbf{Y}}^{(l+1)}=f_{\varphi}({\mathbf{X}}^{(l+1)})$ with the parameter ${\varphi}$.

Evaluation on graph node classification. We apply our \gnnmodelname{} to both heterophilic and homophilic datasets (sorted by homophily ratio $\hbar$ \cite{zhu2020beyond}): Texas ($\hbar=0.11$), Wisconsin ($\hbar=0.21$), Actor ($\hbar=0.22$), Squirrel ($\hbar=0.22$), Chameleon ($\hbar=0.23$), Cornell ($\hbar=0.3$), Citeseer ($\hbar=0.74$), Pubmed ($\hbar=0.8$) and Cora ($\hbar=0.81$), where $\hbar$ indicate the fraction of edges that connect nodes with the same label. We also evaluate on the large-scale ogbn-arxiv dataset from the Open Graph Benchmark (OGB) \cite{hu2020open}. The detailed data description is shown in Supplementary Table 1. Fig. \ref{HoloGraph}b-c and Supplementary Table 4 present the comparison results across nine datasets using nine different methods. \gnnmodelname{} achieves competitive performance on heterophilic (Fig. \ref{HoloGraph}b, Table Supplementary Table 4) and decent performance on homophilic graphs (Fig. \ref{HoloGraph}c, Table Supplementary Table 4), outperforming existing hand-designed GNN models. Notably, in heterophilic graph data (such as Texas, Wisconsin, Actor, Squirrel, Chameleon, and Cornell), node connections don’t depend on similarity. Particularly, Traditional GNNs, which rely on the assumption that neighboring nodes have similar features (message passing mechanism), struggle with heterophilic graphs where this assumption doesn’t hold. This leads to poor performance. In contrast, our \gnnmodelname{} captures complex interactions through oscillator synchronization. By modeling coupling and phase synchronization, it allows information to be transmitted based on dynamic relationships, not just similarity. This makes it well-suited for heterophilic graphs, as it can propagate information through ``asynchronous synchronization" even when node features are far apart in the initial condition. In short, \gnnmodelname{} effectively handles heterogeneity and improves performance on such graphs. To evaluate model scalability, we further tested \gnnmodelname{} on the large-scale ogbn-arxiv dataset. As shown in Fig. ~\ref{HoloGraph}d and Supplementary Table 5, our model outperforms all standard baselines and achieves best performance, with an accuracy of 0.86. These results highlight the generalization and discriminative power of our oscillatory synchronization mechanism on large graphs. Discussion. These results provide strong evidence that our message aggregation mechanism effectively synchronizes nodes of the same class while adhering to the constraints of the adjacency matrix (as shown in Fig. \ref{HoloGraph}f). This approach is fundamentally distinct from traditional message-passing methods in GNNs, as it does not enforce identical feature representations for nodes within the same class. Consequently, our method mitigates the risk of over-smoothing as the network depth increases. To validate this claim, we evaluated classification performance across multiple network layers, as shown in Fig. \ref{HoloGraph}g. It is clear that the classification performance remains stable even as the number of network layers increases, reaching up to 128 layers. This indicates that our method effectively mitigates the issue of over-smoothing.

Evaluation on graph classification.
We apply \gnnmodelname{} to ENZYMES and PROTEINS (TUDataset \cite{morris2020tudataset}). The detailed data description is shown in Supplementary Table 2.  For homophilic graph data (Cora, Citeseer, Pubmed), we adopt the data-splitting method used in the vanilla GCN \cite{kipf2016semi}, specifically the Public semi-supervised Split manner, which allocates a fixed 20 nodes per class for training, and we report results averaged over 10 runs with different random seeds. For heterophilic graph data and TUDataset, we follow the data-splitting manner introduced in \cite{pei2020geom} and \cite{errica2019fair} respectively, with 10-fold cross-validation and we report the averaged metric on all test folds. Each dataset is evaluated by accuracy, precision, and F1-Score. 
Fig. \ref{HoloGraph}e and Supplementary Table 6 present the quantitative comparison of nine methods across two datasets. \gnnmodelname{} demonstrates decent performance (batch\_size=1) in graph classification. Discussion. The graph classification task shares similarities with brain rhythm identification, as both aim to classify entire graphs rather than individual nodes. Consequently, the synchronization mechanism in our model is designed to exhibit distinct synchronization patterns for different graphs, validating its effectiveness on graph classification tasks.

Additional implementation details and ablation studies are provided in the Supplementary Information (Supplementary Tables 7-10).

\subsection*{Explore the potential of solving problems without ground truth}
In the HCP-A dataset, we use the AAL atlas to divide the brain's 116 regions into eight functional subnetworks: the default mode, frontoparietal, limbic, ventral attention, dorsal attention, sensorimotor, visual subnetworks, and cerebellum. To investigate the potential for identifying population-level functional communities in the absence of ground truth, we train an unsupervised neural oscillatory synchronization model designed to cluster brain regions into distinct regions. This model minimizes the Euclidean distance between $\mathbf{x}_i$ and $\mathbf{x}_j$ through the optimization of the Rayleigh quotient \cite{shaham2018spectralnet}. According to Eq. \ref{energy} in the Method session, our approach effectively identifies population-level functional communities that not only align with established subnetworks but also reveal potential patterns in brain organization. Fig. \ref{Holobrain}c shows the clustering results on our \modelname{} and classic spectral clustering. It is clear that, our proposed \modelname{} closely synchronized related brain regions through oscillation, generating specific functional communities. This demonstrates the potential of our approach to collaboratively enable personalized brain parcellation.

\section*{Discussion}
We introduced a deep model, \modelname{}, to explore the biological mechanisms linking fluctuating brain functions with cognitive states. We then extended the concepts of neural oscillatory synchronization and attentional memory to graph learning, yielding a GNN architecture \gnnmodelname{}. 

Related works. (i) System identification. 
Tremendous efforts have been made to find a mechanistic explanation for the brain’s control in processing information and making decisions \cite{botvinick2014computational}. Since the neuroscience notion of cognitive control is analogous to the system control used in engineering, many computational models formulate the dynamic neural process into a linear system \cite{gu2015controllability,mcgowan2022controllability}, where the hidden states of the complex neural system are modulated by energetic simulations. By doing so, we are able to examine the control mechanism over functional dynamics through the lens of well-studied system control principles \cite{medaglia2017brain}. Recently, the research paradigm has shifted toward integrating recurrent neural networks with continuous-time hidden states governed by physics models \cite{chen2018neuralode,hasani2021liquid,dan2022neuro}. Specifically, the Kuramoto model has been used to model the coupling between brain structure and function \cite{cabral2011role,capouskova2022modes}, where resting-state neural activity is found to arise from the interaction between local neural dynamics and the brain's large-scale structure. Following these prior works, we propose the \modelname{} model to parameterize the governing equation of functional fluctuations using deep learning techniques. (ii) \kuramoto{} model in machine learning. Most relevant work to our \gnnmodelname{} is the recent work of artificial \kuramoto{} oscillatory neurons \cite{miyato2024artificial} for unsupervised object detection, which also uses the concept of oscillatory synchronization to compress the visual feature representations. Although both works are inspired by the neural mechanism of oscillation, we put the spotlight on bridging the reciprocal relationship between neuroscience and AI through the lens of governing equations in the dynamical system. 

Our study design has two milestones: (1) Existing ML technique $\rightarrow$ Improved understanding of neural mechanism. We present a deep model \modelname{} to uncover the biological mechanism that links fluctuating brain function and cognitive states. (2) Biologically inspired ML mechanism $\leftarrow$ Neural mechanism. We transplant the concept of neural oscillatory synchronization and attending memory to graph learning, with the application of identifying large-scale functional networks, which is a proof-of-concept exploration for reasoning and solving problems in the absence of ground truth.

Impact of our work. Our major contribution is a graph learning mechanism via neural oscillatory synchronization, yielding a computational model (\modelname{}) for system identification and a GNN architecture (\gnnmodelname{}) with competitive performance on real-world graph datasets. Our experiments showed that this innovative graph learning mechanism not only addresses the existing issues typically encountered in GNNs but also demonstrates its potential for tackling challenging learning problems on graphs.
By conceptualizing brain oscillations as coupled systems of synchronized graph nodes, we aim to inspire the next generation of machine learning techniques that are more efficient and robust. The proposed \modelname{} model provides a deeper understanding of neural synchronization and offers a powerful framework for graph learning (\gnnmodelname{}), addressing key challenges such as the over-smoothing issue in GNNs. Beyond theoretical contributions, the impact of this work could extend to real-world applications, particularly in advancing AI systems that emulate human cognitive processes, ultimately benefiting fields such as healthcare, robotics, and cognitive computing.

\section*{Methods}
In our work, we hypothesize that both human brains and graph data form a dynamical system where the dynamics are characterized by \kuramoto{} model. Suppose the system consists of $N$ oscillators with their pairwise coupling strength denoted by $K_{ij} (i,j=1,...,N)$. By allowing each oscillator to vibrate at its own natural frequency $\omega_i$, the dynamics of coupled phase oscillators in the \kuramoto{} model is formulated in Eq. \ref{Eq.Kuramoto}.
Initially, each oscillator evolves independently at its intrinsic frequency $\omega_i$. Over time, the coupling term $[\cdot]$ in Eq. \ref{Eq.Kuramoto} enables the oscillators to adjust their phases relative to one another, leading to collective synchronization. This process results in oscillators locking their phases and forming coherent groups. 
Our hypothesis is supported by the interference patterns of CFC in the human brain (shown in Fig. \ref{fig:young}), captured by the following computational approach. 

\subsection*{Geometric scattering transform for capturing CFC patterns}
\begin{figure}
  \centering
  \includegraphics[width=1.0\textwidth]{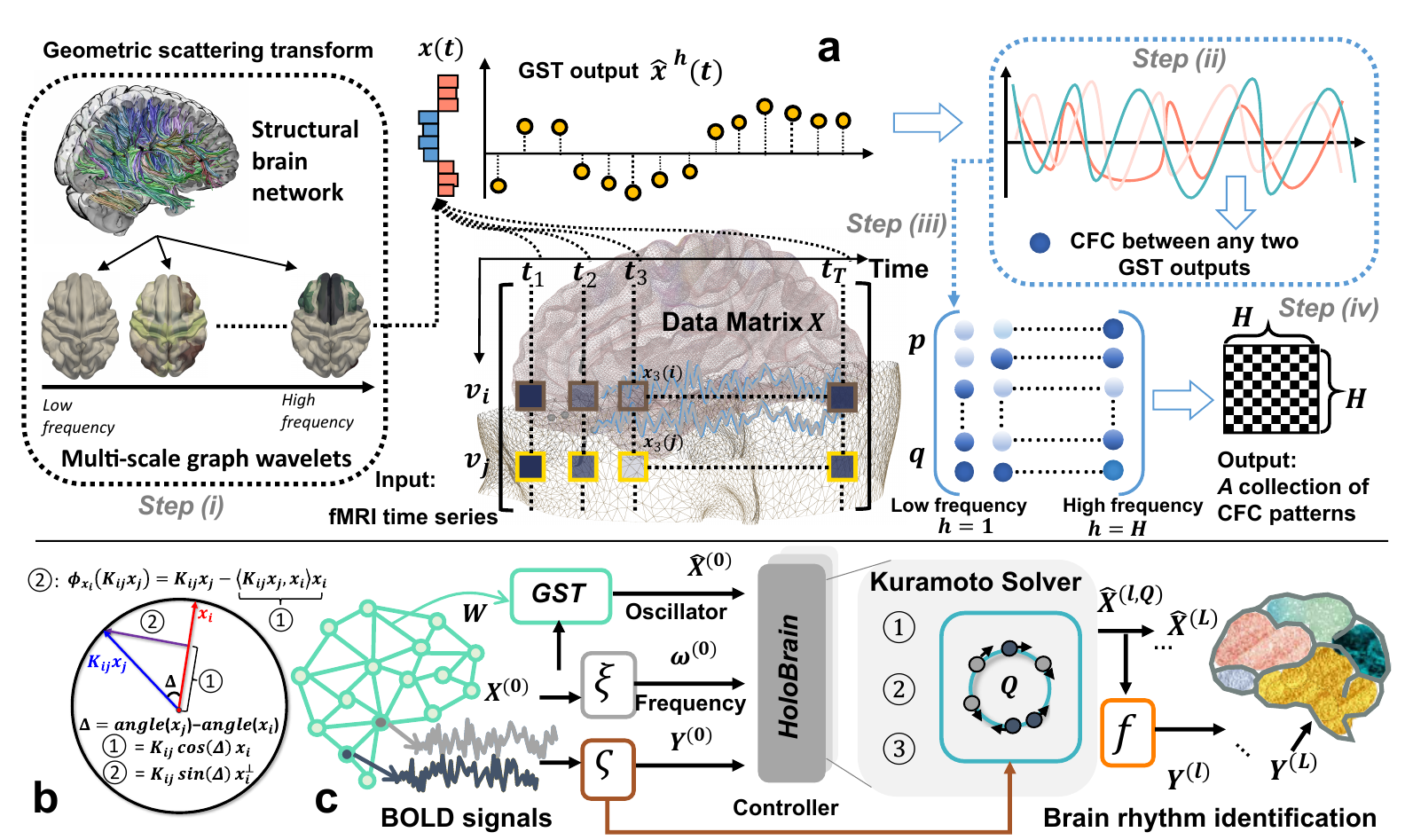}  
  \caption{ \textbf{\modelname{}: Physics-informed modeling of cross-crequency coupling in the human brain.} {\textbf{a} Conceptual overview of CFC patterns in the human brain. Step (i): Each graph wavelet, associated with each region $v_i$ and specific frequency $h$, is used to lift the frequency of the BOLD signal. Step (ii): Each lifted BOLD signal is considered as a spatial oscillator operating in a harmonic frequency. Step (iii): We construct a CFC matrix to capture the synchronization of oscillators. Step (iv): Our model is expected to generate synchronized interference patterns at each brain region.} {\textbf{b} Intuition of project operator $\phi$.  Suppose $N=2$, and all oscillators synchronize on a 2D sphere. The projection $\phi$ consists of two steps: \textcircled{1} project the signal from oscillator $\mathbf{x}_j$ to the direction of underlying oscillator $\mathbf{x}_i$ by $K_{ij}\cos{(\Delta)}\mathbf{x}_i$, and \textcircled{2} synchronize $\mathbf{x}_i$ w.r.t. $\mathbf{x}_j$ in the tangent space by $K_{ij}\sin{(\Delta)}\mathbf{x}_i^\perp$. $\Delta =angle(\mathbf{x}_j)-angle(\mathbf{x}_i)$ denotes the angular difference between $\mathbf{x}_i$ and $\mathbf{x}_j$ on the circle.} {\textbf{c} The network architecture of our \modelname{}. The input to \modelname{} is the lifted BOLD signals by GST, which are considered as initial oscillators $\mathbf{X}^{(0)}=\{\mathbf{x}_i\}_{i=1}^N$. The backbone of \modelname{} is a cascade of Kuramotor solvers, repeated $Q$ times, which consists of \textcircled{1} compute oscillator-to-oscillator interaction, \textcircled{2} phase project, and \textcircled{3} phase update for $Q$ times. By cascading $L$ \modelname{}, the synchronized oscillations $\mathbf{\hat X}^{(L)}$ will be used to predict cognitive state using identified brain rhythms. Note: the frequency maps in Step (i) are rendered in ParaView \cite{ahrens2005paraview} (v5.10.1; \href{https://www.paraview.org/}{https://www.paraview.org/}); all other brain visualizations are produced with Surf Ice \cite{rorden2025surfice} (Surf Ice 6; \href{https://www.nitrc.org/projects/surfice/}{https://www.nitrc.org/projects/surfice/}). }} 
  \label{fig:cfc}
\end{figure}

Suppose the brain network is represented as a graph structure $\mathcal{G} = (\mathbf{V}, \mathbf{W})$, where $\mathbf{V}$ denotes the set of $N$ nodes and $\mathbf{W} = [w_{ij}]_{i,j=1}^N$ is the $N \times N$ weighted adjacency matrix. Each element $w_{ij}$ represents the coupling strength measured from neuroimaging data (Note: Inherent coupling degree $w_{ij}$ can be normalized count of neuronal fibers connecting nodes $v_i$ and $v_j$ in a structural brain network (structural connectivity, SC) or co-activation of neural activities between nodes $v_i$ and $v_j$ in a functional brain network (functional connectivity, FC)). 
Additionally, we use $\mathbf{X} = \{\mathbf{x}_i|\mathbf{x}_i(t)\in \mathbb{R}, t=1,...,T\}$ to denote the BOLD (blood-oxygen-level-dependent) signals of $T$ time points measured at region $v_i$.
To enrich the feature representation from a single degree $\mathbf{x}_i(t)$, we employ geometric scattering transform (GST) \cite{gao2019geometric} to capture the multi-frequency representation using a set of harmonic wavelets that are derived from the underlying graph Laplacian matrix. 
Specifically, we first define the lazy random walk matrix \cite{min2020scattering} as $\mathbf{P}=\frac{1}{2}{(\mathbf{I}_N+\mathbf{W}\mathbf{D}^{\rm  -1})}$ where ${\mathbf{I}}_N$ is an $N\times N$ identity matrix and $\mathbf{D}$ is a diagonal matrix of connectivity degree associated with $\mathbf{W}$. In this context, we recursively compute the multi-scale graph wavelets $\{{\Psi^h\}}_{h=0}^{H-1}$ as:

\begin{equation}
\begin{aligned}
{\Psi}^0&:= {\mathbf{I}}_N-{\mathbf{P}}, \quad h=0\\
{\Psi}^h&:= {\mathbf{P}}^{2^{h-1}}-{\mathbf{P}}^{2^h},\quad 1\leq h < H
\label{eq.wavelets}
\end{aligned}
\end{equation}
Since each row in $\Psi^h$ is considered as the graph wavelets associated with the corresponding brain region, we map the row vector of $\Psi^h$ into the brain as shown in Fig. \ref{fig:cfc}a (Step (i)), where red and blue arrows indicate the positive and negative oscillations, respectively. 

Suppose $\mathbf{x}(t) \in \mathbb{R}^N$ is the snapshot of BOLD at time $t$. For each scale $h$, the computation of GST output $\hat{\mathbf{x}}^h(t)\in \mathbb{R}^N$ involves three steps: (1) apply the wavelet transform $({\Psi}^h,\mathbf{x}(t))$, (2) compute element-wise absolute value, and (3) apply a low-pass filter ${\Phi}={\mathbf{P}}^{2^H}$, denoted as: $\hat{\mathbf{x}}^h(t)={\Phi}|({\Psi}^h,\mathbf{x}(t))|$. At each brain region $v_i$, we, therefore, obtain a set of GST-lifted BOLD signals $\{\mathbf{x}_i^h(t)\}$. As shown in Fig. \ref{fig:cfc}a, it is straightforward to construct a $H\times H$ CFC matrix at each brain region, where each element reflects the cross-frequency coupling between two GST-lifted BOLD signals.

\subsection*{Brain rhythm identification by \modelname{}}
Given the presence of interference patterns in the human brain, we conceptualize the behavior of dynamic CFC can be effectively described using Kuramoto model. To that end, we present the physics-informed deep model to link neural activities and outcomes (cognitive states). 

Vectorized \kuramoto{} model for oscillatory synchronization across brain regions.
After GST, each brain region $v_i$ has collected a set of time-dependent partial observations of BOLD signal $\hat{\mathbf{x}}_i(t)=[\hat{\mathbf{x}}_i^h(t)]_{h=1}^H$ across frequency scale $h$.
The phase dynamics of these oscillators are governed by the following vector-valued equation \cite{miyato2024artificial}: 

\begin{equation}
\label{kuramotor}
\frac{d \hat{\mathbf{x}}_i}{dt}=\omega_i + [\rho\cdot\phi_{{\hat{\mathbf{x}}_i}}(\sum {_{j = 1}^N} K_{ij}\hat{\mathbf{x}}_j)],
\end{equation}
where $\phi$ represents a projection of the vector $K_{ij}\hat{\mathbf{x}}_j$ onto the tangent space at point $\hat{\mathbf{x}}_i$ on the unit sphere and $\rho$ is a scalar hyperparameter. The projection ensures the neural synchronization lies within the tangent space of the underlying oscillator $\hat{\mathbf{x}}_i$. Herein, $K_{ij}$ is derived from the adjacency matrix $\mathbf{W}$, scaled by a symmetric learnable matrix $\mathbf{A}$, such that $K_{ij} = a_{ij} \cdot w_{ij}$.
$\hat{\mathbf{x}}_i \in \mathbb{R}^{H \times T}$ is defined as an oscillator residing on a hypersphere. The intuition of $\phi$ is shown in Fig. \ref{fig:cfc}b. The key to understanding dynamic neural synchronization in the brain lies in achieving optimal synchronization of oscillators while minimizing the associated costs. Therefore, our objective is to develop an oscillator representation and a control mechanism that aligns closely with fundamental neuroscientific principles.

\kuramoto{} model with attending memory.
The global nature of dynamics in Kuramoto model in Eq. \ref{kuramotor} poses challenges to capturing transient patterns of functional fluctuations that are closely associated with cognitive tasks. In contrast, human brains continuously decide ``what to remember" and ``what to ignore" during dynamic cognitive activities, which involve focusing attention on task-relevant information while suppressing irrelevant stimuli. Thus, it is a natural choice to leverage the neural mechanism of attending memory \cite{hutchinson2012memory} to guide \kuramoto{} model to make optimal choices based on past events. To that end, we introduce a global optimal control into our model by: 
\begin{equation}
\label{brick}
\frac{d\hat{\mathbf{x}}_i}{dt}=\omega_i + [\rho\cdot\phi_{{\hat{\mathbf{x}}_i}}(\mathbf{y}_i+\sum {_{j = 1}^N} K_{ij}\hat{\mathbf{x}}_j)],
\end{equation}
where $\mathbf{y}_i$ is a feedback control of attending memory driven by the observed neural activities. Intuitively, each $\mathbf{y}_i$ captures population-level, behavior-specific attending memories and characterizes the mechanistic role of each oscillator engaged in a given cognitive task. By tracking the progression of cognitive states, we can continuously assess feedback from neural oscillations and adjust the synchronization in \kuramoto{} system through feedback control.

Intuition of \kuramoto{} model with attending memory.
Assume each brain region $v_i$ is equally important. Since the coupling matrix $\mathbf{K}$ is symmetric, the \kuramoto{} model in Eq. \ref{brick} is the gradient flow of minimizing the energy functional $E$ \cite{aoyagi1995network,miyato2024artificial}:

\begin{equation}
\label{energy}
E=-\sum_{i, j} {\hat{\mathbf{x}}}_i^{\intercal} {K}_{i j} {\hat{\mathbf{x}}}_j-\sum_i {\mathbf{y}}_i^{\intercal} {\hat{\mathbf{x}}}_i
\end{equation}

The intuitions behind Eq. \ref{brick} include:
\begin{itemize}[left=0cm]
    \setlength{\itemsep}{0pt} 
    \setlength{\parskip}{2pt} 
    \item The energy function $E$ is a ``natural” Lyapunov function for the governing equation Eq. \ref{brick} as $\frac{dE(\hat{\mathbf{x}}(t))}{dt}\leq 0$.
    \item The first term in $E$ promotes synchronization by encouraging oscillators $\hat{\mathbf{x}}_i$ and $\hat{\mathbf{x}}_j$ to operate at the same phase if they are strongly coupled. The second term of $E$ constrains the alignment between oscillatory synchronization and behavior-specific control patterns.
    \item The spatial pattern of feedback controls $\mathbf{Y}=\{\mathbf{y}_i|i=1,...,N\}$ (aka. control patterns) provides an alternative attention mechanism associated with downstream tasks (such as cognitive state recognition). In contrast to the conventional graph attention mechanism \cite{velivckovic2017graph}, control patterns allow us to understand the system controllability \cite{liu2011controllability} of each brain region in the dynamical system of evolving functional fluctuations. 
    \item The energy function $E$ offers insight into effective information processing without labels, where biologically-meaningful patterns emerging from the neural mechanism inform the development of more capable reasoning strategies for graph learning (lead to \gnnmodelname{} model).
\end{itemize}

Physics-informed deep model for brain rhythm identification.
Taking together, we present a deep model to predict the cognitive tasks based on neural synchronization results. 
The network architecture of our proposed \modelname{} model is shown in the solid blue box of Fig. \ref{fig:cfc}c.  Specifically, we employ two neural networks to parameterize the function $\xi_{\sigma}(\hat{\mathbf{x}}_i)$ for determining the natural frequency $\omega_i$ and the function $\zeta_\mu(\hat{\mathbf{x}}_i)$ for generating controller $\mathbf{y}_i$ in the governing equation Eq. \ref{brick}, where $\sigma$ and $\mu$ are network parameters.
Given $\omega_i$ and $\mathbf{y}_i$, we gradually synchronize oscillators in three steps: \textcircled{1} Compute influence from other oscillators. For each oscillator $\hat{\mathbf{x}}_i$, we define the total influence from other coupled oscillators by $\mathbf{z}_i=\mathbf{y}_i+\sum {_{j = 1}^N} (a_{ij}\cdot w_{ij})\hat{\mathbf{x}}_j$, where $\mathbf{A}=[a_{ij}]_{i,j=1}^N$ is a symmetrical learnable weighting parameter to adaptively weight the inherent coupling strength $\mathbf{W}$. The output of this step is a collection of oscillator-to-oscillator interactions $\mathbf{Z}=[\mathbf{z}_i]_{i=1}^N$.
\textcircled{2} Projection. For each oscillator $\hat{\mathbf{x}}_i$, the analytic formulation of projection is defined as $\phi_{\hat{\mathbf{x}}_i}(\mathbf{z}_i)={\mathbf{z}}_i-\left\langle{\mathbf{z}}_i, {\hat{\mathbf{z}}}_i\right\rangle {\hat{\mathbf{z}}}_i$. This project operation plays a crucial role in Riemannian optimization on the sphere, ensuring that the updated direction remains within the tangent space at the point $\hat{\mathbf{x}}_i$ on the sphere \cite{chandra2019continuous}.
\textcircled{3} Update phase information for each oscillator. We update each oscillatory via the forward difference as shown in Eq. \ref{brick}. Further, we employ a mapping operation $f_{\varphi}$ with parameter $\varphi$ to update the controller $\mathbf{y}_{i}$ in the current $l^{\text{th}}$ layer, ensuring that $\mathbf{y}_i$ resides in the same space as the oscillators, i.e., onto the (unit) hyper-sphere, to capture phase-invariant
patterns.
The implementation detail is shown in Algorithm 1 (see Supplementary Information).

The loss function $\mathcal{L}$ of the downstream task uses a cross-entropy loss between the ground truth event of the cognitive task and the predicted event. Note, the total loss function in \modelname{} model is eventually the combination of downstream task loss $\mathcal{L}$ and the energy function oscillatory synchronization $E$.
For the unseen new subject, our \modelname{} model synchronizes a large population of oscillators $\mathbf{X}$ via the learned model parameters $\sigma$ and $\mu$. The model output includes not only the predicted outcome (e.g., cognitive status) but also an associated task-specific control pattern $\mathbf{Y}$.  

\subsection*{\gnnmodelname{}: Graph learning via neural oscillation}
\label{Sec.BIGNOS}

\gnnmodelname{}: A learning mechanism of GNN.
Inspired by \modelname{} for the human brain, we seek a solution to address the over-smoothing issues by compressing the graph feature representations in the latent oscillatory phase space rather than diffusing on the graph domain. 
The oscillatory nature of the system, the coupling dynamics, and the intrinsic frequencies prevent the Kuramoto model from converging to a static, trivial state as all identical features in the graph heat diffusion model \cite{strogatz2000kuramoto}. Instead, \kuramoto{} model achieves dynamic behaviors towards partial or full synchronization \cite{acebron2005kuramoto}.

In addition, graph attention mechanism in current GNNs is essentially dynamically assigning weights to neighboring nodes based on their importance to the underlying node in information aggregation \cite{velivckovic2017graph}. As shown in Eq. \ref{brick}, we introduce the notion of control pattern $\mathbf{Y}$, which acts as the outcome-specific feedback control in the dynamical process of graph feature representation learning. In this context, our control patterns provide a window to investigate the system controllability of graph learning using well-studied complex system analysis approaches. 

The comparison between the learning mechanisms in current GNNs and the proposed \gnnmodelname{} model inspired by brain rhythms is summarized in Table \ref{com}. We also provide a comparative analysis of the dynamics exhibited by diffusion-based GNNs, wave-like systems, and our proposed \gnnmodelname{}, as shown in Supplementary Fig. 2.

Network architecture. In general, we discretize the \modelname{} for the human brain into a \gnnmodelname{} for graph data. Specifically, we model each graph node as an oscillator, with interactions determined by the underlying graph topology. \gnnmodelname{} takes the graph embedding vectors ${\mathbf{X}}^{(0)}$ as input, which serves as the initial conditions for oscillatory synchronization. The control term $\mathbf{Y}^{(0)}$ is initialized using an encoder, such as GCN layer, i.e., $\mathbf{Y}^{(0)}=\zeta(\mathbf{X}^{(0)})\gets GCN(\mathbf{X}^{(0)},\mathbf{W})$ . As shown in Fig. \ref{HoloGraph}a, feature learning in \gnnmodelname{} is cast as a dynamic process governed by Kuramoto model, as described in Algorithm 1 (see Supplementary Methods).

\section*{Data Availability}

The HCP-A data used in this study are available in the Human Connectome Project database under accession link \url{https://www.humanconnectome.org/study/hcp-lifespan-aging}. The HCP-A data are available under restricted access for compliance with data use terms; access can be obtained by registering for a ConnectomeDB account and agreeing to the Open Access Data Use Terms. The HCP-YA data used in this study are available in the Human Connectome Project database under accession link \url{https://www.humanconnectome.org/study/hcp-young-adult/data-releases}. The HCP-YA data are available under restricted access for compliance with data use policies; access can be obtained by registering and agreeing to the Open Access Data Use Terms on ConnectomeDB. The ADNI, NIFD, and PPMI datasets used in this study are available through the Image and Data Archive (IDA) platform. Access to these datasets is restricted for participant privacy protection; access can be obtained by creating an IDA account and submitting a data use application at \url{https://ida.loni.usc.edu}. The raw imaging data are protected and are not publicly available due to data privacy regulations from the original studies. The processed data used for model input and evaluation are available upon reasonable request from the corresponding author. The preprocessing pipelines are fully described and accessible through publicly available tools (fMRIPrep: \url{https://fmriprep.org/en/stable/}, 
QSIPrep: \url{https://qsiprep.readthedocs.io/en/latest/}). No new datasets were generated in this study. All data used were obtained from publicly available repositories as noted above.

\section*{Code Availability}

The code used in this study has been deposited in a GitHub repository at \url{https://github.com/acmlab/HoloBrain_HoloGraph} and archived on Zenodo at \url{https://doi.org/10.5281/zenodo.17095726}. The version used corresponds to Release v1.0. Detailed configurations of parameters are available in Supplementary Table 9.

\backmatter


\section*{Acknowledgements}

This work was supported by the National Institutes of Health (AG091653, AG068399, AG084375) and the Foundation of Hope, all awarded to G. W.

\section*{Author Contributions Statement}
T. D. designed the experimental framework, performed the experimental studies, wrote the paper, prepared the figures and interpreted of results. J. D. performed the data analyses, implemented the computational models. G. W. supervised and coordinated the overall study, provided domain expertise, contributed to study design, wrote the paper, and critically reviewed the manuscript.

\section*{Competing Interests Statement} 
The authors declare no competing interests.

\clearpage
\section*{Tables}

\begin{table}[h!]
\caption{Comparison between graph-based models and \gnnmodelname{}}
\label{com}
\setlength{\tabcolsep}{8pt} 
\renewcommand{\arraystretch}{1.2} 
\begin{tabular}{l c c}
\hline
 & Current GNN & \gnnmodelname{} \\
\hline
Governing equation   & $\frac{\partial X}{\partial t}=\nabla\cdot (\nabla X)$ 
                     & $\frac{dx_i}{dt}=\omega_i + [\rho\cdot\phi_{x_i}(y_i+\sum K x_i)]$ \\
Learning mechanism    & Heat diffusion & Oscillatory synchronization \\
Attention mechanism   & Weighting matrix & Feedback control \\
\hline
\end{tabular}
\end{table}

\section*{Figure Captions}

\clearpage
\newpage
\section*{Supplementary Figures}

\begin{figure}[!h]
\centering
\includegraphics[width=0.94\textwidth]{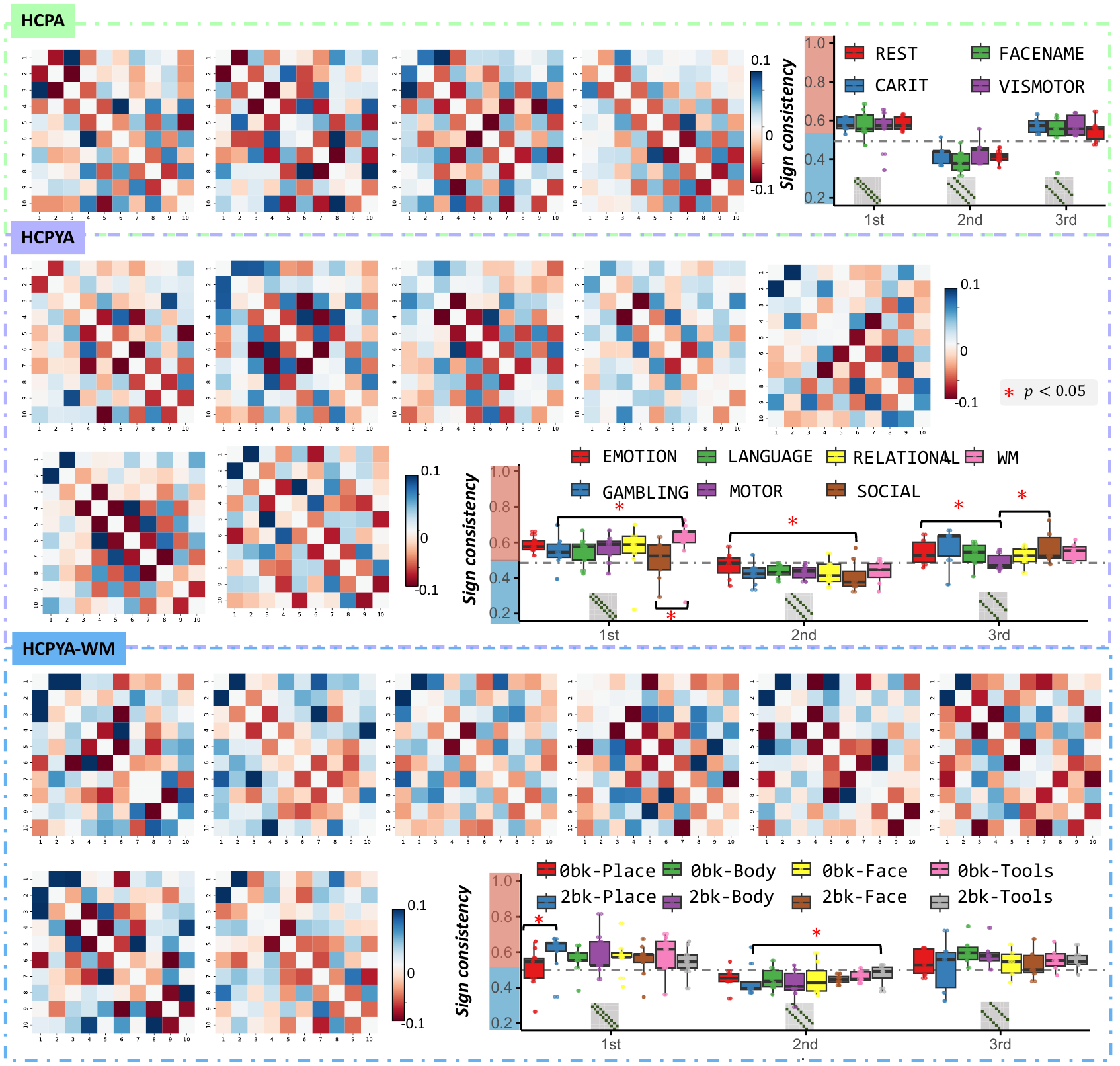}
\caption{\textbf{Statistical power of interference patterns in CFC matrices derived from GST across HCP-A, HCP-YA, and HCP-YA-WM.} 
Each $10 \times 10$ matrix represents frequency–frequency coupling at a brain region, computed using graph spectral transform (GST) with $H=10$ wavelet filters. For each cognitive task, matrices are averaged across all brain regions and participants to yield population-level CFC patterns. Prominent off-diagonal striping—resembling constructive/destructive wave interference (e.g., Young’s double slit)—emerges consistently across tasks. These patterns are quantified via sign consistency along off-diagonal lines, with statistical tests ($p{<}0.05$; `$\ast$') revealing moderate but significant discriminative power across tasks. The findings suggest that interference-like CFC patterns may encode functional signatures of cognition. Compared to Fig. 4 in the main text, these GST-only results establish a baseline. \modelname{} achieves improved statistical power in recognizing cognitive tasks, indicating the effectiveness of the governing equation in capturing dynamic functional fluctuations.}
\label{fig:cfc_ori}
\end{figure}

\begin{figure}[h]
  \centering
  \includegraphics[width=0.6\textwidth]{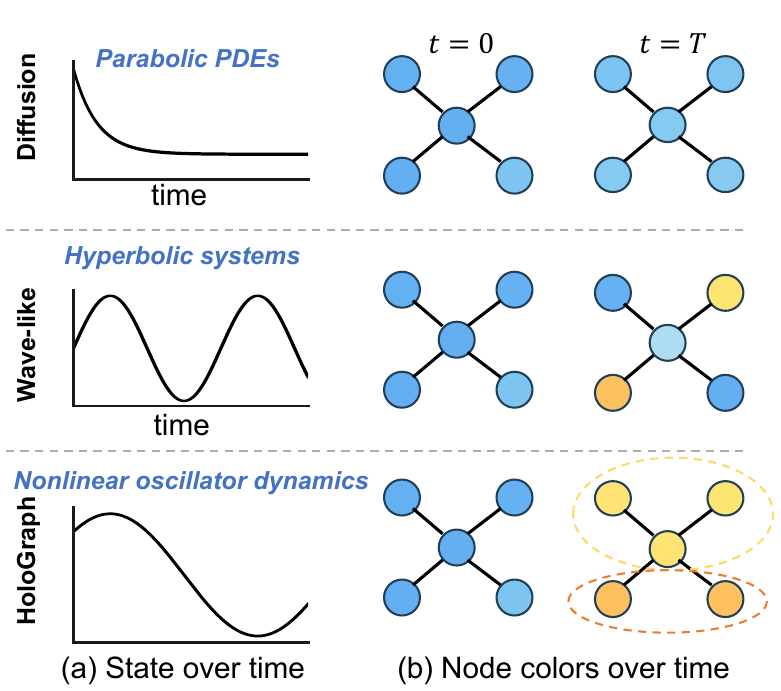}  
  \caption{\textbf{Comparative dynamics of diffusion-based GNNs, wave-like systems, and our proposed \gnnmodelname{}.} 
  We model graph representation learning as a dynamic process governed by physical principles, aligning with the emerging paradigm of physics-informed neural networks. Classical graph neural networks (GNNs) often rely on diffusion-like message passing, which corresponds to parabolic partial differential equations (PDEs) such as the heat equation. These operations smooth node representations over time, often suppressing informative but non-smooth local variations. In contrast, our proposed \gnnmodelname{} is grounded in oscillator synchronization dynamics, inspired by the Kuramoto model in neuroscience. This model governs nonlinear phase-based interactions among coupled nodes and can retain local structure without averaging, closely resembling the dynamics of wave equations (hyperbolic PDEs) that support structured propagation and local preservation.
  Panel (a) illustrates how a node's state evolves under three paradigms. Diffusion-based GNNs exhibit rapid smoothing that erodes distinctions between nodes. Wave-like systems maintain oscillatory structure over time, and \gnnmodelname{} introduces nonlinear synchronization dynamics that preserve both temporal structure and topological distinctions. Panel (b) visualizes toy graphs at initial ($t=0$) and final ($t=T$) states. Diffusion dynamics lead to homogenized node states that obscure community structure, while wave-like systems and \gnnmodelname{} preserve more discriminative spatial patterns. Notably, \gnnmodelname{} maintains phase-coherent, non-averaged interactions within subnetworks (highlighted with dashed circles), preserving modular structure in a biologically plausible fashion.
  Although \gnnmodelname{} is not formally hyperbolic, its synchronization dynamics share core properties with wave equations—namely, the ability to propagate structured activity and preserve sharp local variations. This offers theoretical and empirical advantages in modeling neural-inspired graph learning mechanisms, bringing us a step closer to uncovering first principles behind machine intelligence.
  }
  \label{fig:toy}
\end{figure}

\begin{figure}[h!]
    \centering
    \includegraphics[width=1\linewidth]{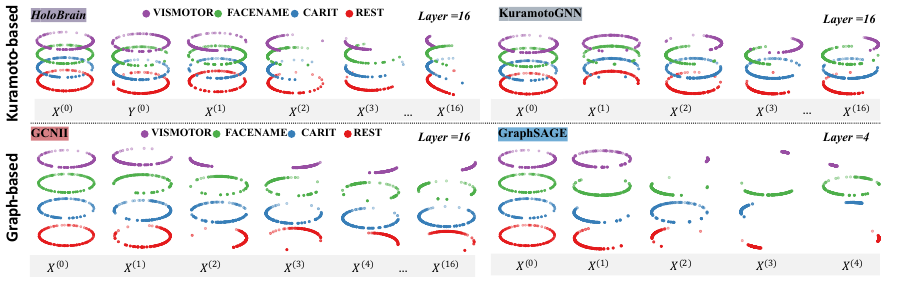}
\caption{\textbf{Comparison of phase-space synchronization trajectories across four models—\textit{HoloBrain}, KuramotoGNN, GCNII, and GraphSAGE—on the HCP-A dataset.} 
We visualize the layerwise evolution of brain region representations across four graph models by projecting their states onto a shared 2D unit circle. Each dot represents a brain region, and its angular position reflects its instantaneous phase $\theta_i^{(l)}$ at layer $l$. Over layers, \textit{HoloBrain} exhibits progressive synchronization, with phase clusters becoming increasingly compact and organized by cognitive task. This locking is measurable via the Kuramoto order parameter $R^{(l)}$, which monotonically increases across layers, indicating growing coherence among regions. In contrast, standard graph models (GCNII, GraphSAGE) yield latent features with no inherent oscillatory meaning. To enable visualization, we apply an ad hoc phase projection (e.g., $\theta_i=\angle(\Theta {x}_i)$) using a learned linear readout $\Theta$. However, these projections are not physically grounded, and the resulting phase dynamics are inconsistent across layers—points jump, arcs fragment, and clusters fluctuate chaotically. Compared to KuramotoGNN, which uses the same base oscillator but lacks a top-down task controller, \textit{HoloBrain} introduces an attending-memory field $\mathbf{Y} = \{\mathbf{y}_i\}$ that injects task-specific information. The resulting dynamics minimize a Lyapunov energy functional (Eq.~7 in the main text), ensuring convergence toward task-consistent, stable phase-locked manifolds. This memory-based modulation induces interpretable control patterns (the $\mathbf{y}_i$ maps) and structured interference in CFC matrices—such as the emergence of off-diagonal stripes associated with cross-frequency phase interactions. What this figure reveals is a clear contrast in the nature of cluster formation: \textit{HoloBrain} produces smooth, directed trajectories in phase space—rings tighten, angular spreads shrink, and each cognitive task stably occupies a phase sector. KuramotoGNN locks partially, but without task control, trajectories drift or fragment. Graph baselines appear to cluster, but these visual separations are artifacts of projection, not outcomes of meaningful neural dynamics. Only \textit{HoloBrain} offers a biologically interpretable synchronization pathway, with measurable order $R^{(l)}$, controller-driven energy descent, and structured cross-frequency coupling—all grounded in neuro-inspired physics.}
\label{fig:sync_hcpa}
\end{figure}

\clearpage
\newpage
\section*{Supplementary Figures}

\begin{figure}[!h]
\centering
\includegraphics[width=0.94\textwidth]{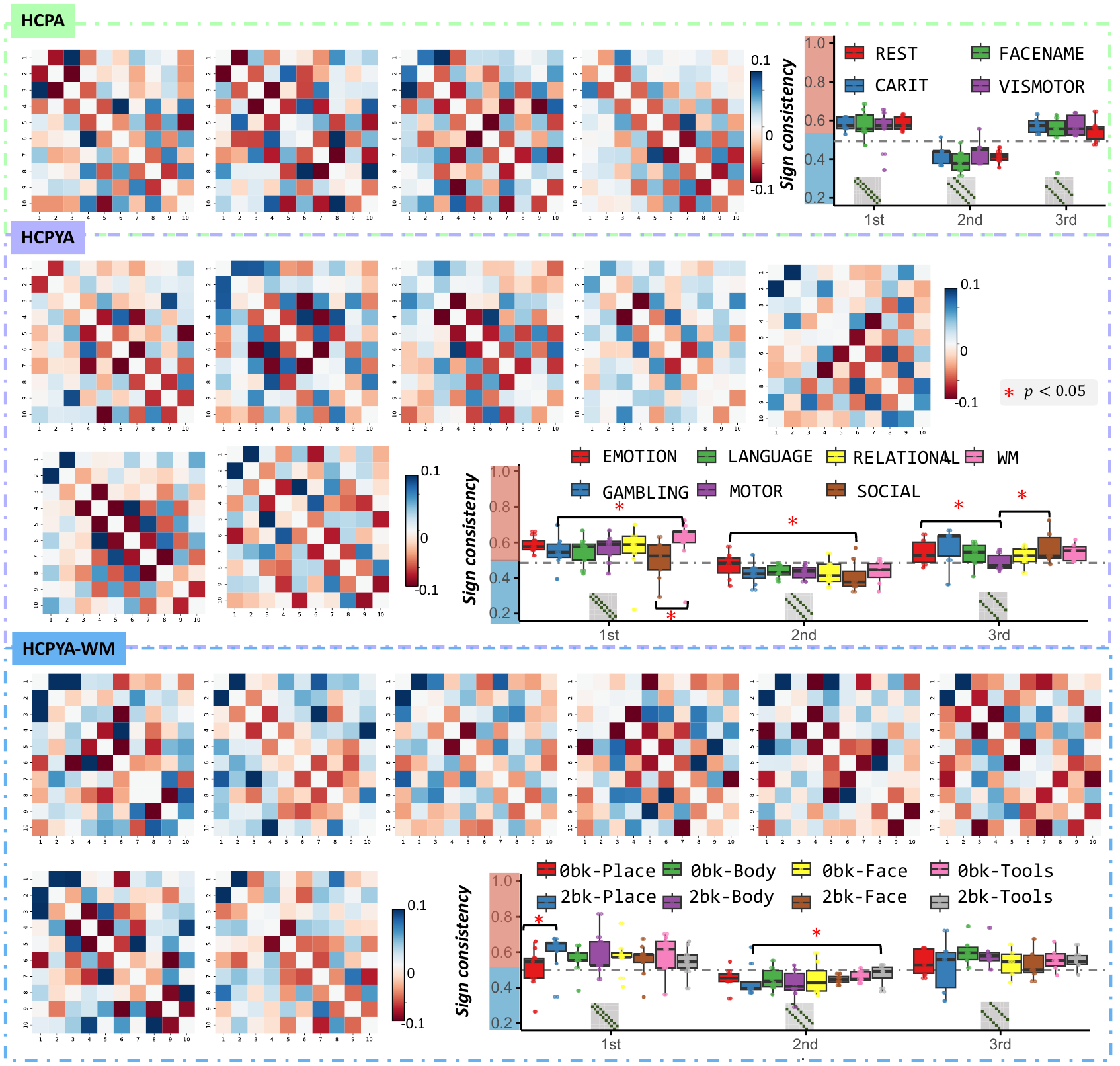}
\caption{\textbf{Statistical power of interference patterns in CFC matrices derived from GST across HCP-A, HCP-YA, and HCP-YA-WM.} 
Each $10 \times 10$ matrix represents frequency–frequency coupling at a brain region, computed using graph spectral transform (GST) with $H=10$ wavelet filters. For each cognitive task, matrices are averaged across all brain regions and participants to yield population-level CFC patterns. Prominent off-diagonal striping—resembling constructive/destructive wave interference (e.g., Young’s double slit)—emerges consistently across tasks. These patterns are quantified via sign consistency along off-diagonal lines, with statistical tests ($p{<}0.05$; `$\ast$') revealing moderate but significant discriminative power across tasks. The findings suggest that interference-like CFC patterns may encode functional signatures of cognition. Compared to Fig. 4 in the main text, these GST-only results establish a baseline. \modelname{} achieves improved statistical power in recognizing cognitive tasks, indicating the effectiveness of the governing equation in capturing dynamic functional fluctuations.}
\label{fig:cfc_ori}
\end{figure}

\begin{figure}[h]
  \centering
  \includegraphics[width=0.6\textwidth]{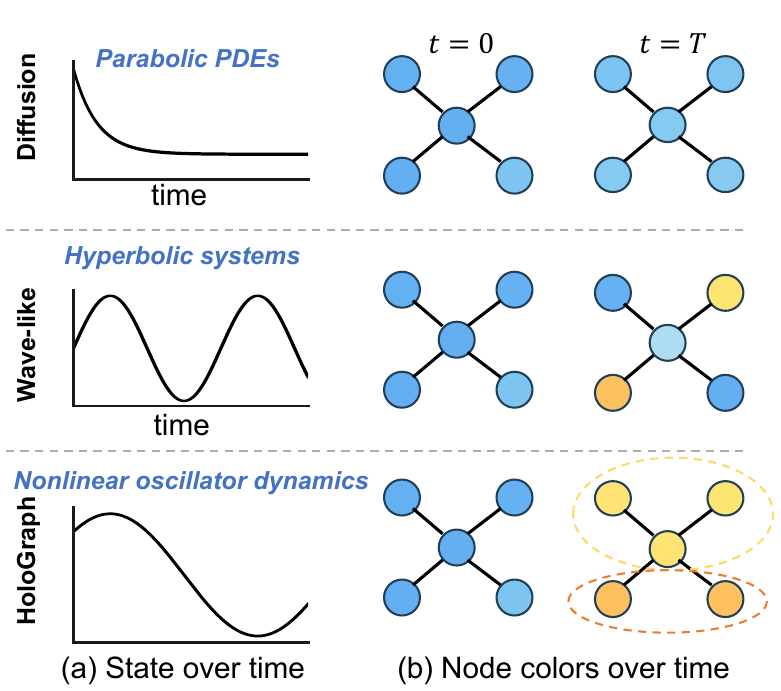}  
  \caption{\textbf{Comparative dynamics of diffusion-based GNNs, wave-like systems, and our proposed \gnnmodelname{}.} 
  We model graph representation learning as a dynamic process governed by physical principles, aligning with the emerging paradigm of physics-informed neural networks. Classical graph neural networks (GNNs) often rely on diffusion-like message passing, which corresponds to parabolic partial differential equations (PDEs) such as the heat equation. These operations smooth node representations over time, often suppressing informative but non-smooth local variations. In contrast, our proposed \gnnmodelname{} is grounded in oscillator synchronization dynamics, inspired by the Kuramoto model in neuroscience. This model governs nonlinear phase-based interactions among coupled nodes and can retain local structure without averaging, closely resembling the dynamics of wave equations (hyperbolic PDEs) that support structured propagation and local preservation.
  Panel (a) illustrates how a node's state evolves under three paradigms. Diffusion-based GNNs exhibit rapid smoothing that erodes distinctions between nodes. Wave-like systems maintain oscillatory structure over time, and \gnnmodelname{} introduces nonlinear synchronization dynamics that preserve both temporal structure and topological distinctions. Panel (b) visualizes toy graphs at initial ($t=0$) and final ($t=T$) states. Diffusion dynamics lead to homogenized node states that obscure community structure, while wave-like systems and \gnnmodelname{} preserve more discriminative spatial patterns. Notably, \gnnmodelname{} maintains phase-coherent, non-averaged interactions within subnetworks (highlighted with dashed circles), preserving modular structure in a biologically plausible fashion.
  Although \gnnmodelname{} is not formally hyperbolic, its synchronization dynamics share core properties with wave equations—namely, the ability to propagate structured activity and preserve sharp local variations. This offers theoretical and empirical advantages in modeling neural-inspired graph learning mechanisms, bringing us a step closer to uncovering first principles behind machine intelligence.
  }
  \label{fig:toy}
\end{figure}

\begin{figure}[h!]
    \centering
    \includegraphics[width=1\linewidth]{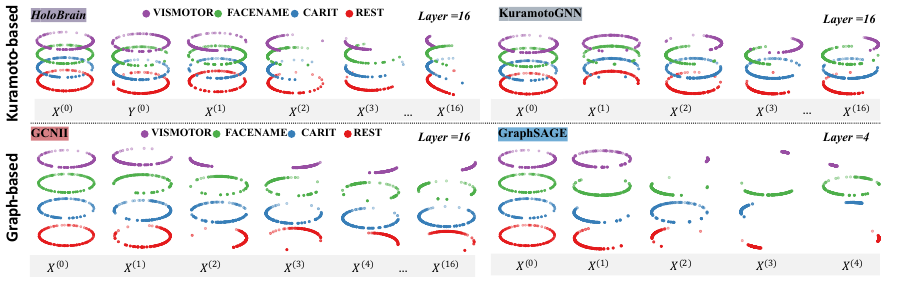}
\caption{\textbf{Comparison of phase-space synchronization trajectories across four models—\textit{HoloBrain}, KuramotoGNN, GCNII, and GraphSAGE—on the HCP-A dataset.} 
We visualize the layerwise evolution of brain region representations across four graph models by projecting their states onto a shared 2D unit circle. Each dot represents a brain region, and its angular position reflects its instantaneous phase $\theta_i^{(l)}$ at layer $l$. Over layers, \textit{HoloBrain} exhibits progressive synchronization, with phase clusters becoming increasingly compact and organized by cognitive task. This locking is measurable via the Kuramoto order parameter $R^{(l)}$, which monotonically increases across layers, indicating growing coherence among regions. In contrast, standard graph models (GCNII, GraphSAGE) yield latent features with no inherent oscillatory meaning. To enable visualization, we apply an ad hoc phase projection (e.g., $\theta_i=\angle(\Theta {x}_i)$) using a learned linear readout $\Theta$. However, these projections are not physically grounded, and the resulting phase dynamics are inconsistent across layers—points jump, arcs fragment, and clusters fluctuate chaotically. Compared to KuramotoGNN, which uses the same base oscillator but lacks a top-down task controller, \textit{HoloBrain} introduces an attending-memory field $\mathbf{Y} = \{\mathbf{y}_i\}$ that injects task-specific information. The resulting dynamics minimize a Lyapunov energy functional (Eq.~7 in the main text), ensuring convergence toward task-consistent, stable phase-locked manifolds. This memory-based modulation induces interpretable control patterns (the $\mathbf{y}_i$ maps) and structured interference in CFC matrices—such as the emergence of off-diagonal stripes associated with cross-frequency phase interactions. What this figure reveals is a clear contrast in the nature of cluster formation: \textit{HoloBrain} produces smooth, directed trajectories in phase space—rings tighten, angular spreads shrink, and each cognitive task stably occupies a phase sector. KuramotoGNN locks partially, but without task control, trajectories drift or fragment. Graph baselines appear to cluster, but these visual separations are artifacts of projection, not outcomes of meaningful neural dynamics. Only \textit{HoloBrain} offers a biologically interpretable synchronization pathway, with measurable order $R^{(l)}$, controller-driven energy descent, and structured cross-frequency coupling—all grounded in neuro-inspired physics.}
\label{fig:sync_hcpa}
\end{figure}

\clearpage

\section*{Supplementary Tables}

\begin{table*}[h!]
\centering
\caption{Dataset description for node classification.}
\label{tab:node_desc}
\small
\scalebox{0.65}{%
\begin{tabular}{lcccccccccc}
\toprule
 & Texas & Wisconsin & Actor & Squirrel & Chameleon & Cornell & Citeseer & Pubmed & Cora & ogbn-arxiv\\
\midrule
Hom.\ ratio $h$ & 0.11 & 0.21 & 0.22 & 0.22 & 0.23 & 0.30 & 0.57 & 0.74 & 0.81 & -- \\
\# Nodes & 183 & 251 & 7{,}600 & 5{,}201 & 2{,}277 & 183 & 3{,}327 & 19{,}717 & 2{,}708 & 169{,}343\\
\# Edges & 295 & 466 & 26{,}752 & 198{,}493 & 31{,}421 & 280 & 4{,}676 & 44{,}327 & 5{,}278 & 1{,}166{,}243\\
\# Classes & 5 & 5 & 5 & 5 & 5 & 5 & 7 & 3 & 6 & 40\\
\bottomrule
\end{tabular}}
\end{table*}

\begin{table*}[!h]
\centering
\caption{Dataset description for graph classification (TUDataset).}
\label{tab:tudesc}
\small
\begin{tabular}{lcc}
\toprule
 & ENZYMES & PROTEINS\\
\midrule
Avg \# Nodes & 32.63 & 39.36\\
Avg \# Edges & 62.14 & 39.06\\
\# Classes & 6 & 2\\
\# Graphs & 600 & 1{,}113\\
\bottomrule
\end{tabular}
\end{table*}

\begin{table*}[!h]
 \caption{Performance metrics (\%) on HCPA, HCPYA, HCPYA-WM, ADNI, PPMI, and NIFD. Each entry reports mean\,$\pm$\,SD across 5-fold cross-validation; higher is better. Within each row, \first{red} marks the best value and \second{blue} marks the second best. Superscript $^*$ indicates that \modelname{} is significantly better than all baselines on that dataset/metric (paired $t$-test, $p<0.05$).}
  \label{holobrain_exp_result_si}
  \small
   \scalebox{0.55}{\begin{tabular}{ccccccccccc}
   \toprule
     \textbf{Dataset} & \textbf{Metric} & {GCN} & {GIN} & {GAT} & {GCNII} & {GraphSAGE} & {SAN} &GraphCON & KuramotoGNN& \textit{HoloBrain} \\
    \toprule
     & Acc & $74.57_{\pm1.25}$ & $72.28_{\pm0.85}$ & $90.45_{\pm0.97}$ & $94.00_{\pm0.65}$ & $94.20_{\pm0.84}$ & ${93.34_{\pm0.51}}$ & $93.40_{\pm0.53}$ &$85.73_{\pm1.60}$ & {\first{$95.55_{\pm0.77}$}} \\
\textbf{HCP-A}     & Pre & $71.48_{\pm1.21}$ & $69.74_{\pm0.80}$ & $90.19_{\pm1.00}$ & $93.95_{\pm0.62}$ & $94.15_{\pm0.89}$ & $93.54_{\pm0.50}$ & $93.45_{\pm0.46}$  & $86.70_{\pm0.87}$& {\first{$95.57_{\pm0.77}$}} \\
                   & F1  & $71.78_{\pm1.81}$ & $70.49_{\pm0.85}$ & $90.12_{\pm1.03}$ & $93.92_{\pm0.65}$ & $94.12_{\pm0.86}$ & $93.37_{\pm0.48}$ & $93.30_{\pm0.55}$  & $85.68_{\pm1.40}$& {\first{$95.50_{\pm0.78}$}} \\
    \midrule
     & Acc & $37.54_{\pm0.70}$ & $35.17_{\pm1.55}$ & $62.70_{\pm3.53}$ & $70.89_{\pm2.56}$ & $74.16_{\pm2.43}$ & $69.98_{\pm1.88}$ & $71.92_{\pm2.67}$ &$59.18_{\pm3.88}$ & {\first{$85.20_{\pm1.60}$}} \\
\textbf{HCP-YA}  & Pre & $37.40_{\pm1.10}$ & $36.34_{\pm1.88}$ & $64.08_{\pm4.20}$ & $71.52_{\pm2.66}$ & $74.87_{\pm2.30}$ & $70.34_{\pm1.96}$ & $71.23_{\pm2.32}$ &$60.15_{\pm2.95}$ & {\first{$85.53_{\pm1.72}$}} \\
& F1  & $36.09_{\pm0.78}$ & $35.02_{\pm1.49}$ & $62.36_{\pm3.56}$ & $70.84_{\pm2.70}$ & $74.14_{\pm2.47}$ & $69.85_{\pm1.99}$ & $71.75_{\pm2.65}$ &$58.77_{\pm3.67}$ & {\first{$85.18_{\pm1.65}$}} \\
    \midrule
     & Acc & $32.65_{\pm1.63}$ & $33.16_{\pm2.67}$ & $42.94_{\pm2.48}$ & $49.26_{\pm3.99}$ & $53.46_{\pm3.26}$ & $48.82_{\pm1.60}$ & $55.51_{\pm1.29}$ &$35.37_{\pm4.19}$ & {\first{$89.22_{\pm1.71}$}} \\
\textbf{HCP-WM}      & Pre & $33.01_{\pm1.78}$ & $33.40_{\pm3.01}$ & $44.34_{\pm3.63}$ & $50.44_{\pm4.25}$ & $54.22_{\pm3.42}$ & $50.01_{\pm0.84}$ & $56.03_{\pm1.65}$ & $39.09_{\pm1.56}$& {\first{$89.66_{\pm1.49}$}} \\
 & F1  & $32.25_{\pm1.53}$ & $32.87_{\pm2.61}$ & $42.54_{\pm2.26}$ & $48.79_{\pm4.10}$ & $53.34_{\pm3.26}$ & $48.41_{\pm1.33}$ & $55.06_{\pm1.27}$ &$34.06_{\pm4.76}$ & {\first{$89.18_{\pm1.77}$}} \\
    \bottomrule
    \bottomrule
     & Acc & ${81.48_{\pm7.77}}$ & $79.26_{\pm6.46}$ & $81.48_{\pm7.77}$ & $81.48_{\pm7.77}$ & ${82.22_{\pm6.37}}$ & ${82.96_{\pm3.78}}$ & {\first{$83.70_{\pm5.54}$}} &  ${71.85_{\pm6.87}}$& {\second{$83.05_{\pm8.28}$}} \\
\textbf{ADNI}     & Pre & ${67.00_{\pm11.96}}$ & $74.76_{\pm7.32}$ & $67.00_{\pm11.96}$ & ${67.00_{\pm11.96}}$  & ${74.00_{\pm4.84}}$ & $74.68_{\pm5.89}$ & ${73.39_{\pm11.74}}$ &  ${72.26_{\pm6.19}}$& {\first{$75.06_{\pm9.71}$}} \\
                   & F1  & ${73.38_{\pm10.56}}$ & $75.49_{\pm8.64}$ & $73.38_{\pm10.56}$ & ${73.38_{\pm10.56}}$ & ${77.10_{\pm3.98}}$ & $77.10_{\pm3.98}$ &  ${72.28_{\pm8.25}}$&  ${73.48_{\pm6.14}}$ & {\first{$77.94_{\pm8.52}$}} \\
    \midrule
     & Acc & ${57.14_{\pm6.72}}$ & $62.39_{\pm5.71}$ & $58.96_{\pm2.80}$ & $59.46_{\pm7.88}$ & ${61.83_{\pm3.50}}$ & ${62.99_{\pm7.16}}$ &$60.12_{\pm2.04}$ &$56.64_{\pm4.18}$& {\first{$68.83_{\pm4.53}$}} \\
\textbf{PPMI}     & Pre & ${53.07_{\pm14.79}}$ & $62.53_{\pm4.57}$ & $49.85_{\pm10.51}$ & $64.41_{\pm6.45}$ & ${42.50_{\pm11.29}}$ & $62.75_{\pm10.66}$ &$44.98_{\pm9.41}$ &$55.71_{\pm15.82}$ & {\first{$62.86_{\pm11.98}$}} \\
                   & F1  & ${49.51_{\pm5.19}}$ & $60.27_{\pm5.20}$ & $50.56_{\pm4.78}$ & $55.5_{\pm7.84}$ & ${49.82_{\pm8.25}}$ & $61.47_{\pm8.92}$  &$50.72_{\pm5.52}$ &$48.95_{\pm10.17}$ & {\first{$64.62_{\pm9.22}$}} \\
    \midrule
     & Acc & ${48.81_{\pm1.55}}$ & $49.90_{\pm1.97}$ & $48.91_{\pm2.06}$ & $49.21_{\pm1.70}$ & ${49.21_{\pm1.70}}$ & ${49.21_{\pm1.99}}$ &${48.31_{\pm1.35}}$ &${46.24_{\pm0.86}}$ & {\first{$68.28_{\pm2.98}$}} \\
\textbf{NIFD}     & Pre & ${35.12_{\pm12.83}}$ & $42.33_{\pm6.13}$ & $32.70_{\pm11.88}$ & $44.63_{\pm3.87}$ & ${44.63_{\pm3.87}}$ & $36.67_{\pm9.34}$ &${30.49_{\pm7.77}}$ & ${42.29_{\pm2.47}}$& {\first{$66.72_{\pm3.85}$}} \\
                   & F1  & ${35.52_{\pm1.33}}$ & $42.64_{\pm1.96}$ & $32.57_{\pm2.67}$ & $37.96_{\pm5.10}$ & ${37.96_{\pm5.10}}$ & $34.05_{\pm3.01}$ & ${35.73_{\pm4.67}}$&${42.20_{\pm3.01}}$ & {\first{$66.80_{\pm3.53}$}} \\
    \bottomrule
   \end{tabular}}
\end{table*}

\begin{table*}[!h]
  \caption{Test performance (\%) on nine graph networks for node classification. Significance based on 10 runs with different random seeds, $^*$ denotes $p<0.05$ vs. the second-best method. Colors indicate within-row ranks: \first{1st (red)}, \second{2nd (blue)}, \third{3rd (green)}, \fourth{4th (orange)}.}
  \label{more_results_si}
  \small
      \scalebox{0.58}{\begin{tabular}{ccccccccccc}
    \toprule
     \textbf{Dataset}&Metric&{GCN} &{GIN} &{GAT} &{GCNII} &{GraphSAGE} &{SAN} & GraphCON & KuramotoGNN &\gnnmodelname{} \\
    \midrule
\multirow{3}{*}{\textbf{Texas}}
&Acc&$58.65_{\pm3.64}$&$57.03_{\pm5.98}$&$56.49_{\pm5.85}$&$57.30_{\pm4.56}$ & \third{$78.92_{\pm5.51}$}&\fourth{$73.24_{\pm7.40}$} & \second{$81.08_{\pm4.52}$}  &  $69.46_{\pm 9.90}$&$\first{81.35_{\pm3.51}}$\\
&Pre&$41.80_{\pm7.01}$&$46.86_{\pm7.10}$&$41.44_{\pm10.41}$&$49.02_{\pm11.50}$& \first{$80.68_{\pm6.30}$}&{$77.54_{\pm7.17}$}   &\second{$80.33_{\pm 5.02}$}  &  $61.79_{\pm 7.37}$  &\third{$79.75_{\pm5.54}$}\\
&F1 &$47.46_{\pm4.76}$&$50.10_{\pm5.32}$&$46.33_{\pm7.50}$&$51.70_{\pm8.37}$ & \third{$77.50_{\pm6.31}$}&{$73.08_{\pm7.53}$} &\second{$78.40_{\pm 5.74}$} &$60.51_{\pm 7.95}$ &\first{$81.20_{\pm5.00}^*$}\\
\hline
\multirow{3}{*}{\textbf{Wisconsin}}
&Acc& $52.75_{\pm6.35}$ & $47.84_{\pm5.20}$ & $53.53_{\pm7.60}$ & $70.20_{\pm3.90}$ & \third{$79.61_{\pm7.55}$} & {$78.63_{\pm6.17}$}  &\first{$83.33_{\pm 2.36}$} & $69.61_{\pm 9.04}$& \second{$82.16_{\pm3.56}$}  \\
&Pre& $43.93_{\pm6.97}$ & $46.13_{\pm10.84}$ & $48.07_{\pm7.58}$ & $72.47_{\pm4.16}$ & \third{$79.48_{\pm8.02}$} & {$79.47_{\pm5.19}$}  &\second{$80.49_{\pm 4.31}$} & $63.01_{\pm 9.26}$& \first{$80.52_{\pm4.04}$}  \\
&F1& $45.11_{\pm6.95}$ & $44.21_{\pm7.74}$ & $48.35_{\pm6.92}$ & $68.79_{\pm4.00}$ & \third{$78.43_{\pm7.92}$} & {$77.78_{\pm5.92}$}  &\first{$81.07_{\pm 2.72}$} &$59.95_{\pm 13.53}$ & \second{$79.68_{\pm4.20}$}  \\
\hline
\multirow{3}{*}{\textbf{Actor}}
&Acc& $28.38_{\pm0.96}$ & $25.91_{\pm1.01}$ & $29.05_{\pm0.80}$ & {$33.80_{\pm1.43}$} & \third{$34.88_{\pm1.19}$}  & $32.94_{\pm0.92}$ &\second{$35.66_{\pm 1.18}$} &$25.62_{\pm 1.35}$ &  \first{$35.80_{\pm0.60}$} \\
&Pre& $20.77_{\pm1.65}$ & $25.33_{\pm1.07}$ & $20.83_{\pm3.25}$ & $30.17_{\pm2.35}$ & {$30.89_{\pm2.78}$}  & \second{$34.18_{\pm2.16}$} &\third{$34.08_{\pm 1.42}$} &$16.20_{\pm 4.00}$ &  \first{$36.05_{\pm1.79}$} \\
&F1& $23.27_{\pm1.55}$ & $24.62_{\pm0.82}$ & $23.43_{\pm2.25}$ & {$31.00_{\pm2.19}$} & $30.89_{\pm2.78}$  & \second{$32.67_{\pm1.01}$} &\third{$31.14_{\pm 1.77}$}  &$14.28_{\pm 2.94}$ &  \first{$33.86_{\pm1.60}$} \\
\hline
\multirow{3}{*}{\textbf{Squirrel}}
&Acc & $28.87_{\pm1.55}$ & $25.16_{\pm2.17}$  & $30.08_{\pm1.03}$  & $28.41_{\pm1.15}$  &  {$36.90_{\pm1.03}$} &  \third{$37.00_{\pm1.28}$}  & $27.70_{\pm 2.45}$ &$\second{43.17_{\pm 1.08}}$ & \first{$68.06_{\pm1.65}^*$}  \\
&Pre & $29.92_{\pm1.36}$ & $25.97_{\pm5.51}$  & $28.45_{\pm2.65}$  & $30.01_{\pm2.22}$  &  \third{$35.36_{\pm1.00}$} &  {$35.22_{\pm1.58}$}  & $25.69_{\pm 3.35}$ &\second{$48.05_{\pm 2.23}$} & \first{$67.88_{\pm1.60}^*$}  \\
&F1 & $27.93_{\pm1.99}$ & $25.97_{\pm2.42}$  & $27.96_{\pm3.95}$  & $26.05_{\pm1.70}$  &  \third{$34.78_{\pm1.19}$} &  {$34.00_{\pm2.00}$}  & $22.74_{\pm 2.19}$ &\second{$40.42_{\pm 1.11}$} & \first{$67.80_{\pm1.63}^*$}  \\
\hline
\multirow{3}{*}{\textbf{Chameleon}}
&Acc & $39.36_{\pm1.93}$ & $32.17_{\pm1.75}$ & $43.16_{\pm1.56}$ & $37.41_{\pm2.54}$ & {$48.03_{\pm2.22}$}  &  \third{$51.80_{\pm1.95}$} & $34.23_{\pm 2.46}$ & \second{$56.93_{\pm 1.65}$}& \first{$74.50_{\pm1.13}^*$}  \\
&Pre& $39.08_{\pm1.93}$ & $32.04_{\pm2.05}$ & $40.85_{\pm4.80}$ & $36.09_{\pm6.51}$ & {$49.76_{\pm3.01}$}  &  \third{$52.25_{\pm2.04}$} & $26.69_{\pm 7.86}$ &\second{$62.25_{\pm 2.25}$}  & \first{$74.78_{\pm1.15}^*$}  \\
&F1& $36.15_{\pm2.59}$ & $31.18_{\pm1.84}$ & $40.23_{\pm2.87}$ & $32.05_{\pm4.64}$ & {$48.05_{\pm2.27}$}  &  \third{$50.70_{\pm1.78}$} & $26.35_{\pm 4.30}$ &\second{$57.26_{\pm 1.82}$} & \first{$74.46_{\pm1.09}^*$}  \\
\hline
\multirow{3}{*}{\textbf{Cornell}}
&Acc & $45.14_{\pm4.84}$ & $48.65_{\pm10.61}$ & $51.62_{\pm4.75}$ & $53.51_{\pm7.91}$ & \third{$70.81_{\pm3.15}$} & {$69.46_{\pm6.05}$}  &\second{$71.89_{\pm 3.67}$} & $62.43_{\pm 7.30}$&  \first{$73.24_{\pm4.75}$} \\
&Pre & $44.24_{\pm8.28}$ & $39.74_{\pm14.52}$ & $51.91_{\pm12.06}$ & {$55.46_{\pm10.90}$} & \second{$72.27_{\pm6.22}$}  & \first{$73.10_{\pm8.96}$}  & $68.23_{\pm 4.41}$  & $49.89_{\pm 18.93}$ & \third{$70.14_{\pm6.85}^*$} \\
&F1 & $40.86_{\pm5.11}$ & $41.82_{\pm12.30}$ & $48.76_{\pm8.74}$ & $51.52_{\pm8.21}$ & \third{$68.68_{\pm3.54}$} & \second{$69.03_{\pm6.82}$}  &{$68.46_{\pm 2.90}$} &$47.40_{\pm 15.49}$ &  \first{$70.15_{\pm6.22}$} \\
\hline
\hline
\multirow{3}{*}{\textbf{Citeseer}}
&Acc &$70.62_{\pm0.48}$&$58.28_{\pm3.09}$&$70.38_{\pm0.84}$&$68.08_{\pm0.89}$&$70.44_{\pm0.23}$&$66.30_{\pm0.88}$ & \first{$74.19_{\pm 2.68}$}&\second{$72.90_{\pm 1.75}$} &  \third{$70.63_{\pm0.36}$} \\
&Pre &\third{$72.32_{\pm0.53}$}&$61.75_{\pm2.13}$&{$72.19_{\pm0.44}$}&$69.99_{\pm1.40}$&$71.32_{\pm0.28}$&$68.47_{\pm0.52}$ & \first{$74.28_{\pm 2.36}$}& \second{$72.61_{\pm 2.06}$}&  ${71.47_{\pm0.40}}$ \\
&F1 &\second{$70.98_{\pm0.46}$}&$59.02_{\pm2.75}$&\third{$70.79_{\pm0.67}$}&$68.44_{\pm1.14}$&$70.62_{\pm0.26}$&$66.66_{\pm0.83}$ & \first{$73.78_{\pm 2.58}$} & $70.02_{\pm 2.32}$& \fourth{$70.66_{\pm0.38}$} \\
\hline

\multirow{3}{*}{\textbf{Pubmed}}
&Acc &$77.76_{\pm0.50}$  &$71.40_{\pm2.10}$  & $76.80_{\pm0.75}$ &$77.62_{\pm1.07}$ &$76.36_{\pm0.29}$    &  $74.68_{\pm0.81}$  &\second{$78.69_{\pm 2.17}$} &\first{$78.77_{\pm 2.57}$} &\third{$77.80_{\pm0.23}$}   \\
&Pre &$77.84_{\pm0.50}$  &$72.04_{\pm1.85}$  & $76.95_{\pm0.56}$ &$77.75_{\pm1.02}$ &$76.45_{\pm0.37}$    &  $75.12_{\pm0.94}$ &\second{$78.73_{\pm 2.09}$} & \first{$78.98_{\pm 2.74}$}&\third{$78.25_{\pm0.32}$}   \\
&F1 &\third{$77.77_{\pm0.49}$}  &$74.90_{\pm2.01}$  & $76.82_{\pm0.72}$ &{$77.65_{\pm1.06}$} &$76.34_{\pm0.27}$    &  $74.70_{\pm0.81}$ & \first{$78.58_{\pm 2.17}$}& $77.64_{\pm 3.47}$&\second{$77.79_{\pm0.23}$}   \\
\hline

\multirow{3}{*}{\textbf{Cora}}
&Acc &\third{$81.66_{\pm0.50}$}&$70.76_{\pm2.66}$&$80.14_{\pm1.13}$&$79.92_{\pm1.09}$ &$79.58_{\pm0.37}$  & $77.98_{\pm1.31}$   &\first{$83.80_{\pm 0.86}$} &{$80.45_{\pm 1.59}$} &\second{$81.86_{\pm0.25}$}   \\
&Pre &{$82.61_{\pm0.52}$}&$73.19_{\pm2.47}$&$80.88_{\pm1.37}$&$80.77_{\pm1.09}$ &$80.52_{\pm0.44}$  & $77.80_{\pm0.65}$  &\first{$84.97_{\pm 0.60}$} &\second{$83.42_{\pm 0.99}$} &\third{$82.72_{\pm0.27}$}   \\
&F1 &\third{$81.83_{\pm0.51}$}&$71.04_{\pm2.74}$&$80.18_{\pm1.18}$&$80.00_{\pm1.08}$ &$79.70_{\pm0.36}$  & $77.96_{\pm1.33}$  &\first{$83.98_{\pm 0.76}$} &\fourth{$80.78_{\pm 1.69}$}&\second{$82.00_{\pm0.23}$}   \\
    \bottomrule
  \end{tabular}}
\end{table*}

\begin{table*}[!h]
  \centering
  \caption{Performance (\%) on the ogbn-arxiv dataset. \first{Red} indicates the best results.}
  \label{tab:ogb}
  \small
    \scalebox{0.67}{
  \begin{tabular}{ccccccccccc}
    \toprule
    \textbf{Dataset} & \textbf{Metric} & GCN & GIN & GAT & GCNII & GraphSAGE & SAN & GraphCON & KuramotoGNN & {\gnnmodelname{}} \\
    \midrule
     & Acc & 70.13 & 70.52 & 69.97 & 71.22 & 70.51 & 68.72 & 65.50  & 64.26  & \first{85.84} \\
\textbf{ogbn-arxiv} & Pre & 69.71 & 69.87 & 68.94 & 71.36 & 70.20 & 68.25 & 63.14 & 62.49  & \first{85.52} \\
                    & F1  & 68.97 & 69.26 & 68.00 & 70.23 & 69.25 & 67.51  & 61.90  & 59.94 & \first{85.47} \\
    \bottomrule
  \end{tabular}}
\end{table*}

\begin{table*}[!h]
  \caption{Performance (\%) on TUDataset datasets.}
  \label{tudata_si}
  \small
 \scalebox{0.58}{\begin{tabular}{ccccccccccc}
    \toprule
     \textbf{Dataset} & \textbf{Metric} & {GCN} & {GIN} & {GAT} & {GCNII} & {GraphSAGE} & {SAN} &GraphCON &KuramotoGNN & {\gnnmodelname{}} \\
    \midrule
     & Acc & ${55.67_{\pm4.96}}$ & $50.33_{\pm4.58}$ & $51.15_{\pm6.39}$ & $42.00_{\pm6.89}$ & ${55.67_{\pm5.69}}$ & ${51.50_{\pm5.70}}$ & $43.83_{\pm8.21}$ & $24.17_{\pm1.86}$& ${\first{60.00_{\pm4.28}^*}}$ \\
\textbf{ENZYMES}     & Pre & ${57.10_{\pm4.93}}$ & $51.58_{\pm5.09}$ & $53.18_{\pm6.04}$ & $41.61_{\pm7.58}$ & ${58.70_{\pm6.54}}$ & $52.51_{\pm5.92}$ & $45.69_{\pm8.75}$ &$17.48_{\pm5.32}$ & ${\first{63.13_{\pm4.98}^*}}$ \\
                   & F1  & ${54.96_{\pm4.89}}$ & $49.79_{\pm4.76}$ & $50.72_{\pm6.28}$ & $40.25_{\pm7.75}$ & ${55.04_{\pm5.48}}$ & $51.01_{\pm5.99}$ & $42.92_{\pm8.13}$ & $15.38_{\pm1.55}$& ${\first{59.79_{\pm4.49}^*}}$ \\
    \midrule
     & Acc & ${78.64_{\pm1.88}}$ & $\first{81.11_{\pm1.75}}$ & $73.32_{\pm3.85}$ & $69.99_{\pm4.11}$ & $73.32_{\pm5.70}$ & $71.97_{\pm3.28}$ & $69.45_{\pm3.53}$ &$61.73_{\pm 5.80}$ & $\second{75.02_{\pm2.61}}$ \\
\textbf{PROTEINS}      & Pre & $71.61_{\pm4.13}$ & $74.12_{\pm3.89}$ & $73.41_{\pm3.90}$ & $71.02_{\pm5.41}$ & $73.51_{\pm5.94}$ & $71.75_{\pm3.33}$ & $69.84_{\pm4.57}$ &$57.75_{\pm 13.66}$ & $\first{75.18_{\pm2.56}^*}$ \\
 & F1  & $71.15_{\pm4.01}$ & $73.74_{\pm3.75}$ & $72.82_{\pm3.84}$ & $68.23_{\pm3.85}$ & $72.78_{\pm5.65}$ & $71.73_{\pm3.27}$ & $67.87_{\pm3.65}$ &$56.06_{\pm 8.49}$ & $\first{74.68_{\pm2.43}^*}$ \\
    \bottomrule
   \end{tabular}}
\end{table*}


\begin{table*}[!h]
    \centering
    \caption{Inference time (ms/subject) on three human brain datasets. All the experiments are conducted on four NVIDIA RTX 6000 Ada GPUs.}
    \label{runtime_si}
    \small
    \scalebox{0.8}{%
    \begin{tabular}{lccccccccc}
        \toprule
        & {GCN} & {GIN} & {GAT} & {GCNII} & {GraphSAGE} & {SAN} &{GraphCON} & {KuramotoGNN} & {\textit{HoloBrain}}  \\
        \midrule
        {HCP-A} & 0.66 & 0.58 & 0.86 & 0.82 & 0.57 & 1.02 &0.85 & 12.48 & 1.02   \\
        {HCP-YA}   & 0.63 & 0.55 & 0.81 & 0.70 & 0.54 & 1.02 & 0.99&19.55 & 1.17  \\
        {HCP-WM}  & 1.41 & 1.09 & 2.24 & 1.74 & 1.05 & 2.81 &1.89 &13.28 & 1.56  \\
        \bottomrule
    \end{tabular}%
    }
\end{table*}

\begin{table*}[!h]
\centering
\caption{\textit{HoloGraph} inference time and accuracy under different $L,Q$ on Cora dataset. We empirically examine the model's parameter size and inference time under varying depths. In practice, we generally set $L < Q$, as control signals are updated less frequently, typically once per full oscillation cycle governed by $Q$. }
\label{holograph_hyperparameters_si}
\small
\scalebox{0.60}{\begin{tabular}{c|c|c|c|c|c|c|c|c|c|c|c|c|c|c|c|c|c}
\hline
{$L$} & \multicolumn{6}{c|}{1 (params=41.2M)} & \multicolumn{5}{c|}{2 (params=50.7M)} & \multicolumn{4}{c|}{4 (params=69.5M)} & \multicolumn{2}{c}{8 (params=107.3M)} \\
\hline
{$Q$} & 4 & 8 & 16 & 32 & 64 & 128 & 4 & 8 & 16 & 32 & 64 & 4 & 8 & 16 & 32 & 8 & 16 \\
\hline
{layers} & 4 & 8 & 16 & 32 & 64 & 128 & 8 & 16 & 32 & 64 & 128 & 16 & 32 & 64 & 128 & 64 & 128 \\
\hline
{Inference Time (ms)} & 39.7 & 44.4 & 46.9 & 48.1 & 53.9 & 76.3 & 41.5 & 49.7 & 51.8 & 63.8 & 85.1 & 46.4 & 53.5 & 74.9 & 94.7 & 75.5 & 99.3 \\
\hline
{Acc (\%)} & 81.0 & 80.4 & 81.4 & 81.9 & 81.9 & 81.9 & 80.4 & 81.4 & 81.9  &81.9  & 81.9 & 81.4 & 81.9 & 81.9 & 81.9 & 81.9 & 81.9 \\
\bottomrule
\end{tabular}}
\end{table*}

\begin{table*}[!h]
\centering
\caption{Hyperparameter settings for different models of \textit{HoloBrain} on HCP-YA. To ensure faithful and fair comparisons, we closely followed the \_official implementation\_ and \_default configurations\_ provided by the original papers. In KuramotoGNN, setting the parameter to 16 means that the ODE solver integrates from $t=0$ to $t=16.0$. A larger value corresponds to deeper or longer propagation of information across the graph, serving a role analogous to increasing the number of layers in a standard GCN.}
\label{tab:model_hyperparams_si}
\small
\scalebox{0.65}{%
\begin{tabular}{l|ccccccccc}
\hline
\textbf{Model} & GCN & GIN & GAT & GCNII & GraphSAGE &SAN  & GraphCON & KuramotoGNN & \textit{HoloBrain}\\
\hline
Optimizer & Adam & Adam & Adam & Adam & Adam & Adam & Adam & Adam & Adam  \\
Learning rate & $5 \times 10^{-4}$  & $5 \times 10^{-4}$  & $5 \times 10^{-4}$ & $5 \times 10^{-4}$ & $5 \times 10^{-4}$ & $5 \times 10^{-4}$  & $10^{-3}$ & $10^{-3}$ & $10^{-3}$\\
Weight decay & $5 \times 10^{-3}$ & $5 \times 10^{-3}$ & $5 \times 10^{-3}$ & $5 \times 10^{-3}$ & $5 \times 10^{-3}$ & $5 \times 10^{-3}$ & $5 \times 5^{-4}$ & $5 \times 10^{-4}$ & $10^{-4}$\\
Batch size & 64 & 64 & 64 & 64 & 64 & 64 & 32 & 32 & 64 \\
Epochs & 200 & 200 & 200 & 200 & 200 & 200 & 200 & 200 & 200  \\
Hidden dim & 256 & 256 & 256 & 256 & 256 & 256 & 64 & 64 & 256 \\
Network layer & 2 & 2 & 2 & 4 & 2 & 2 & 4 & 16 & 16 \\
\hline
\end{tabular}}
\end{table*}

\begin{table*}[!h]
\centering
\caption{Ablation of GST level/order on HCP-A (fixed split). We vary the number of frequency channels extracted from the GST to evaluate its effect on node classification.
Each GST configuration is denoted by {level} (scaling depth) and {order} (harmonic orders used).
Higher orders and deeper levels allow richer spectral–temporal representation, though at greater computational cost.
We observe that {level=1, order=[0,1,2]} provides a favorable tradeoff, achieving the highest performance with moderate model size.
This configuration is adopted as the default in all downstream experiments.}
\label{tab:gst_ablation_si}
\small
\scalebox{0.95}{\begin{tabular}{ccccc}
\toprule
{Level} & {Order} & {Acc} (\%) & {Pre} (\%) & {F1}  (\%)\\
\midrule
1 & {[0]}         & 94.82 & 94.70 & 94.73 \\
1 & {[0, 1]}      & 95.02 & 94.96 & 94.86 \\
1 & {[0, 1, 2]}   & 95.55 & 95.57 & 95.50 \\
\midrule
2 & {[0]}         & 95.02 & 94.86 & 94.85 \\
2 & {[0, 1]}      & 95.41 & 95.63 & 95.45 \\
\bottomrule
\end{tabular}}
\end{table*}

\clearpage

\section*{Supplementary Methods}
\label{sec:sup_methods}

\subsection*{Presence of cross-frequency coupling in cognitive tasks}
We set $H{=}10$ in the geometric scattering transform (GST) to obtain a $10{\times}10$ CFC matrix at each brain region. Fig. \ref{fig:cfc_ori} shows population averages of global CFC patterns (averaged across regions and subjects) for HCP-A, HCP-YA (single tasks), and HCP-YA working memory. Population-level CFC exhibits clear off-diagonal striping, reminiscent of constructive/destructive interference. We further compute sign-consistency along the first three off-diagonals and test for between-task differences. At $p{<}0.05$, the interference patterns derived from GST show moderate discriminative power across tasks (asterisks in Fig. \ref{fig:cfc_ori}). These findings motivate our hypothesis that the CFC striping consistency carries task-relevant information about functional dynamics.

\subsection*{Graph data description}
Node- and graph-level datasets are summarized in Table  \ref{tab:node_desc} and Table  \ref{tab:tudesc}. Additional numerical results appear in Table  \ref{holobrain_exp_result_si}–Table  \ref{tudata_si}.

\subsection*{Hyperparameters and runtime}
For deep baselines we use hidden dimension 256, depth 2 (GCNII: 4), batch size 64; \textit{HoloBrain} uses depth 16, batch size 256; see Table  \ref{tab:model_hyperparams_si}. Inference time comparisons on human datasets are given in Table  \ref{runtime_si}. For \textit{HoloGraph}, outer iterations $L$ (controller updates) and inner steps $Q$ (oscillator evolution) govern depth/runtime; Table  \ref{holograph_hyperparameters_si} reports Cora results under various $(L,Q)$. To ensure faithful and fair comparisons, we closely followed the \_official implementations\_ and \_default configurations\_ provided by the original papers (comparison methods). 

\subsection*{Ablation study}
We ablate the GST configuration by varying its \texttt{level} and \texttt{order} on the HCP-A dataset, as summarized in Table  \ref{tab:gst_ablation_si}. Results show that incorporating richer multiscale spectral features improves performance, with the configuration {level=1, order=[0,1,2]} achieving the best tradeoff between accuracy and model complexity. We further investigate the role of the coupling matrix $\mathbf{K}$ by substituting structural connectivity (SC) weights with functional connectivity (FC) weights. This substitution results in a consistent performance drop—accuracy $93.73 \pm 0.75$\%, precision $93.71 \pm 0.77$\%, and F1-score $93.49 \pm 0.85$\%—highlighting that SC-based couplings more effectively guide synchronization dynamics and phase alignment.

\subsection*{Visualization notes}
Layer-wise phase-space trajectories on HCP-A are shown in Fig. \ref{fig:sync_hcpa}, contrasting graph baselines and Kuramoto-type models. Fig. \ref{fig:toy} illustrates why diffusion tends to homogenize features whereas oscillatory synchronization preserves structured variation. Note: Each region is represented by a \emph{unit} oscillator state
$\mathbf{x}_i(t)=[\cos\theta_i(t),\;\sin\theta_i(t)]^\top$ that lives on the circle,
so the \emph{instantaneous phase} is the angle
$\theta_i(t)=\mathrm{atan2}(\mathbf{x}_{i,2}(t),\mathbf{x}_{i,1}(t))=\arg(\mathbf{x}_i(t))$ \footnote{$\arg(\mathbf{x}_i)$ denotes the phase (principal argument) of the oscillator state vector for region $i$.
For the 2-D case $\mathbf{x}_i=[\mathbf{x}_i^{(c)},\,\mathbf{x}_i^{(s)}]^\top$ with $\|\mathbf{x}_i\|=1$,
$\arg(\mathbf{x}_i)=\mathrm{atan2}\!\big(\mathbf{x}_i^{(s)},\,\mathbf{x}_i^{(c)}\big)\in(-\pi,\pi].$
For a higher-dimensional vector $\mathbf{x}_i\in\mathbb{R}^d$, we first project it onto the model’s
oscillatory 2-D subspace spanned by the cosine–sine basis $(u,v)$ and define
$\arg(\mathbf{x}_i)=\mathrm{atan2}\!\big(\langle v,\mathbf{x}_i\rangle,\;\langle u,\mathbf{x}_i\rangle\big).$
This phase is invariant to amplitude scaling and is used to compute phase differences
$\Delta=\arg(\mathbf{x}_j)-\arg(\mathbf{x}_i)$ and the tangent-space projection $\phi_{\mathbf{x}_i}(\cdot)$.}.
Updates are projected to the tangent space by the geometry–aware project operator
$\phi_{\mathbf{x}_i}(\mathbf{z})=(\mathbf{I}-\mathbf{x}_i \mathbf{x}_i^\top)\mathbf{z}$, which preserves $\|\mathbf{x}_i(t)\|=1$ and turns
feature interactions into phase changes. The resulting dynamics
$
\frac{d \mathbf{x}_i}{dt}
\;=\;
\omega_i\,\mathbf{x}_i^\perp
\;+\;
\rho\,\phi_{\mathbf{x}_i}\!\Big(\sum_{j=1}^{N} K_{ij}\,\mathbf{x}_j\Big)
\quad(\text{with }\mathbf{x}_i^\perp=[-\sin\theta_i,\,\cos\theta_i]^\top)
$
are equivalent to the Kuramoto–type phase ODE
$
\dot\theta_i
=
\omega_i
+
\rho\sum_{j=1}^N K_{ij}\sin(\theta_j-\theta_i),
$
so the phase $\theta_i(t)$ is a genuine \emph{state variable} that evolves by
construction. Consequently, the rings in Fig.~\ref{fig:sync_hcpa} trace true phase
trajectories $\{\theta_i^{(l)}\}$ across layers. By contrast, standard graph-based models produce generic feature vectors $\mathbf{x}_i^{(l)}$ with no phase semantics; to visualize them in the same space one must apply an \emph{ad hoc} projection (e.g., $\theta_i=\angle(\Theta{\mathbf{x}}_i)$). Thus, while any model can be \emph{mapped} onto a circle, our \textit{HoloBrain} \emph{evolves} on that circle 

\subsection*{Algorithm details}


  
\begin{algorithm}[H]
  \caption{Kuramoto model (Eq.~\ref{brick}) Solver}
  \label{brick_a}
  \begin{algorithmic}[1]
    \Require BOLD signal $\mathbf{X}^{(0)}$, adjacency matrix $\mathbf{W}$
    \Ensure Terminal oscillators $\hat{\mathbf{X}}^{(L)}$ and updated controller $\mathbf{Y}^{(L)}$

    \State Initialize oscillator state $\hat{\mathbf{X}}^{(0)} \gets \text{GST}(\mathbf{X}^{(0)})$

    \For{$l = 1 \ldots L$}
      \For{$q = 1 \ldots Q$}
        \State Parameterize natural frequency $\omega \gets \xi_{\sigma}(\hat{\mathbf{X}}^{(q)})$
        \State Construct projection function $\phi_{\hat{\mathbf{X}}}(\mathbf{Z}) \gets \mathbf{Z} - \langle \mathbf{Z}, \hat{\mathbf{X}} \rangle \hat{\mathbf{X}}$
        \State Optimize coupling matrix $\mathbf{K} \gets \mathbf{A} \odot \mathbf{W}$
        \State Compute oscillator change $\Delta \hat{\mathbf{X}}^{(q)}$ using Eq.~6
        \State Update oscillators $\hat{\mathbf{X}}^{(q+1)} \gets \hat{\mathbf{X}}^{(q)} + \epsilon \cdot \Delta \hat{\mathbf{X}}^{(q)}$
        \State Normalize $\hat{\mathbf{X}}^{(q+1)}$
      \EndFor
      \State Update controller $\mathbf{Y}^{(l+1)} \gets f_{\varphi}(\hat{\mathbf{X}}^{(l+1)})$
    \EndFor
  \end{algorithmic}
\end{algorithm}

\subsection*{Limitations and solutions}
The gains on TUDataset (Table \ref{tudata_si}) are smaller than those on brain data and node classification (Tables \ref{holobrain_exp_result_si}, \ref{more_results_si}). The reason is structural: \gnnmodelname{} operates on a coupling matrix $\mathbf{A}$ (the adjacency). In TUDataset, graphs have different sizes, so batching requires padding; padding introduces singular operators and degenerates the geometric scattering transform. We therefore train with batch size 1, which limits optimization. Even so, our results are competitive with state-of-the-art methods, indicating strong potential. In future work, we will develop size-agnostic operators / loaders or integrate with PyTorch Geometric to enable efficient batching.


\begin{thebibliography}{65}
\ifx \bisbn   \undefined \def \bisbn  #1{ISBN #1}\fi
\ifx \binits  \undefined \def \binits#1{#1}\fi
\ifx \bauthor  \undefined \def \bauthor#1{#1}\fi
\ifx \batitle  \undefined \def \batitle#1{#1}\fi
\ifx \bjtitle  \undefined \def \bjtitle#1{#1}\fi
\ifx \bvolume  \undefined \def \bvolume#1{\textbf{#1}}\fi
\ifx \byear  \undefined \def \byear#1{#1}\fi
\ifx \bissue  \undefined \def \bissue#1{#1}\fi
\ifx \bfpage  \undefined \def \bfpage#1{#1}\fi
\ifx \blpage  \undefined \def \blpage #1{#1}\fi
\ifx \burl  \undefined \def \burl#1{\textsf{#1}}\fi
\ifx \doiurl  \undefined \def \doiurl#1{\url{https://doi.org/#1}}\fi
\ifx \betal  \undefined \def \betal{\textit{et al.}}\fi
\ifx \binstitute  \undefined \def \binstitute#1{#1}\fi
\ifx \binstitutionaled  \undefined \def \binstitutionaled#1{#1}\fi
\ifx \bctitle  \undefined \def \bctitle#1{#1}\fi
\ifx \beditor  \undefined \def \beditor#1{#1}\fi
\ifx \bpublisher  \undefined \def \bpublisher#1{#1}\fi
\ifx \bbtitle  \undefined \def \bbtitle#1{#1}\fi
\ifx \bedition  \undefined \def \bedition#1{#1}\fi
\ifx \bseriesno  \undefined \def \bseriesno#1{#1}\fi
\ifx \blocation  \undefined \def \blocation#1{#1}\fi
\ifx \bsertitle  \undefined \def \bsertitle#1{#1}\fi
\ifx \bsnm \undefined \def \bsnm#1{#1}\fi
\ifx \bsuffix \undefined \def \bsuffix#1{#1}\fi
\ifx \bparticle \undefined \def \bparticle#1{#1}\fi
\ifx \barticle \undefined \def \barticle#1{#1}\fi
\bibcommenthead
\ifx \bconfdate \undefined \def \bconfdate #1{#1}\fi
\ifx \botherref \undefined \def \botherref #1{#1}\fi
\providecommand{\url}[1]{\textsf{#1}}
\ifx \bchapter \undefined \def \bchapter#1{#1}\fi
\ifx \bbook \undefined \def \bbook#1{#1}\fi
\ifx \bcomment \undefined \def \bcomment#1{#1}\fi
\ifx \oauthor \undefined \def \oauthor#1{#1}\fi
\ifx \citeauthoryear \undefined \def \citeauthoryear#1{#1}\fi
\ifx \endbibitem  \undefined \def \endbibitem {}\fi
\ifx \bconflocation  \undefined \def \bconflocation#1{#1}\fi
\ifx \arxivurl  \undefined \def \arxivurl#1{\textsf{#1}}\fi
\csname PreBibitemsHook\endcsname

\bibitem[\protect\citeauthoryear{Bassett and Sporns}{2017}]{bassett2017network}
\begin{barticle}
\bauthor{\bsnm{Bassett}, \binits{D.S.}},
\bauthor{\bsnm{Sporns}, \binits{O.}}:
\batitle{Network neuroscience}.
\bjtitle{Nature Neuroscience}
\bvolume{20}(\bissue{3}),
\bfpage{353}--\blpage{364}
(\byear{2017})
\end{barticle}
\endbibitem

\bibitem[\protect\citeauthoryear{Watts and Strogatz}{1998}]{watts1998collective}
\begin{barticle}
\bauthor{\bsnm{Watts}, \binits{D.J.}},
\bauthor{\bsnm{Strogatz}, \binits{S.H.}}:
\batitle{Collective dynamics of ‘small-world’networks}.
\bjtitle{Nature}
\bvolume{393}(\bissue{6684}),
\bfpage{440}--\blpage{442}
(\byear{1998})
\end{barticle}
\endbibitem

\bibitem[\protect\citeauthoryear{Xhonneux et~al.}{2020}]{xhonneux2020continuous}
\begin{bchapter}
\bauthor{\bsnm{Xhonneux}, \binits{L.-P.}},
\bauthor{\bsnm{Defferrard}, \binits{M.}},
\bauthor{\bsnm{Vandergheynst}, \binits{P.}}:
\bctitle{Continuous graph neural networks}.
In: \bbtitle{Proceedings of the 37th International Conference on Machine Learning},
pp. \bfpage{10432}--\blpage{10441}
(\byear{2020})
\end{bchapter}
\endbibitem

\bibitem[\protect\citeauthoryear{Ghazanfar and Lewkowicz}{2009}]{ghazanfar2009emergence}
\begin{barticle}
\bauthor{\bsnm{Ghazanfar}, \binits{A.A.}},
\bauthor{\bsnm{Lewkowicz}, \binits{D.J.}}:
\batitle{The emergence of multisensory systems through perceptual narrowing}.
\bjtitle{Trends in Cognitive Sciences}
\bvolume{13}(\bissue{11}),
\bfpage{470}--\blpage{478}
(\byear{2009})
\doiurl{10.1016/j.tics.2009.08.004}
\end{barticle}
\endbibitem

\bibitem[\protect\citeauthoryear{Van~Atteveldt et~al.}{2014}]{van2014multisensory}
\begin{barticle}
\bauthor{\bsnm{Van~Atteveldt}, \binits{N.}},
\bauthor{\bsnm{Murray}, \binits{M.M.}},
\bauthor{\bsnm{Thut}, \binits{G.}},
\bauthor{\bsnm{Schroeder}, \binits{C.E.}}:
\batitle{Multisensory integration: flexible use of general operations}.
\bjtitle{Neuron}
\bvolume{81}(\bissue{6}),
\bfpage{1240}--\blpage{1253}
(\byear{2014})
\end{barticle}
\endbibitem

\bibitem[\protect\citeauthoryear{Hubel and Wiesel}{1962}]{hubel1962receptive}
\begin{barticle}
\bauthor{\bsnm{Hubel}, \binits{D.H.}},
\bauthor{\bsnm{Wiesel}, \binits{T.N.}}:
\batitle{Receptive fields, binocular interaction and functional architecture in the cat's visual cortex}.
\bjtitle{The Journal of Physiology}
\bvolume{160}(\bissue{1}),
\bfpage{106}--\blpage{154}
(\byear{1962})
\end{barticle}
\endbibitem

\bibitem[\protect\citeauthoryear{Gray et~al.}{1989}]{gray1989oscillatory}
\begin{barticle}
\bauthor{\bsnm{Gray}, \binits{C.M.}},
\bauthor{\bsnm{K{\"o}nig}, \binits{P.}},
\bauthor{\bsnm{Engel}, \binits{A.K.}},
\bauthor{\bsnm{Singer}, \binits{W.}}:
\batitle{Oscillatory responses in cat visual cortex exhibit inter-columnar synchronization which reflects global stimulus properties}.
\bjtitle{Nature}
\bvolume{338}(\bissue{6213}),
\bfpage{334}--\blpage{337}
(\byear{1989})
\end{barticle}
\endbibitem

\bibitem[\protect\citeauthoryear{Amari and Arbib}{1977}]{amari1977competition}
\begin{botherref}
\oauthor{\bsnm{Amari}, \binits{S.-I.}},
\oauthor{\bsnm{Arbib}, \binits{M.A.}}:
Competition and cooperation in neural nets.
Systems neuroscience,
119--165
(1977)
\end{botherref}
\endbibitem


\bibitem[\protect\citeauthoryear{Burgess}{2012}]{Notbohm2012}
\begin{barticle}
\bauthor{\bsnm{Burgess}, \binits{A.P.}}:
\batitle{Towards a unified understanding of event-related changes in the EEG:
the firefly model of synchronization through cross-frequency phase modulation}.
\bjtitle{PLOS ONE}
\bvolume{7}(\bissue{9}),
\bfpage{e45630}
(\byear{2012})
\doiurl{10.1371/journal.pone.0045630}
\end{barticle}
\endbibitem

\bibitem[\protect\citeauthoryear{Miyato et~al.}{2025}]{miyato2024artificial}
\begin{bchapter}
\bauthor{\bsnm{Miyato}, \binits{T.}},
\bauthor{\bsnm{L{\"o}we}, \binits{S.}},
\bauthor{\bsnm{Geiger}, \binits{A.}},
\bauthor{\bsnm{Welling}, \binits{M.}}:
\bctitle{Artificial kuramoto oscillatory neurons}.
In: \bbtitle{The Thirteenth International Conference on Learning Representations}
(\byear{2025}).
\burl{https://openreview.net/forum?id=nwDRD4AMoN}
\end{bchapter}
\endbibitem

\bibitem[\protect\citeauthoryear{Kipf and Welling}{2017}]{kipf2016semi}
\begin{bchapter}
\bauthor{\bsnm{Kipf}, \binits{T.N.}},
\bauthor{\bsnm{Welling}, \binits{M.}}:
\bctitle{Semi-supervised classification with graph convolutional networks}.
In: \bbtitle{International Conference on Learning Representations}
(\byear{2017}).
\burl{https://openreview.net/forum?id=SJU4ayYgl}
\end{bchapter}
\endbibitem

\bibitem[\protect\citeauthoryear{Yun et~al.}{2019}]{Yun2019}
\begin{botherref}
\oauthor{\bsnm{Yun}, \binits{S.}},
\oauthor{\bsnm{Jeong}, \binits{M.}},
\oauthor{\bsnm{Kim}, \binits{R.}},
\oauthor{\bsnm{Kang}, \binits{J.}},
\oauthor{\bsnm{Kim}, \binits{H.J.}}:
Graph transformer networks.
Advances in neural information processing systems
\textbf{32}
(2019)
\end{botherref}
\endbibitem

\bibitem[\protect\citeauthoryear{Kuramoto}{1975}]{kuramoto1975self}
\begin{bchapter}
\bauthor{\bsnm{Kuramoto}, \binits{Y.}}:
\bctitle{Self-entrainment of a population of coupled non-linear oscillators}.
In: \bbtitle{International Symposium on Mathematical Problems in Theoretical Physics: January 23--29, 1975, Kyoto University, Kyoto/Japan},
pp. \bfpage{420}--\blpage{422}
(\byear{2005})
\end{bchapter}
\endbibitem

\bibitem[\protect\citeauthoryear{Hutchinson and Turk-Browne}{2012}]{hutchinson2012memory}
\begin{barticle}
\bauthor{\bsnm{Hutchinson}, \binits{J.B.}},
\bauthor{\bsnm{Turk-Browne}, \binits{N.B.}}:
\batitle{Memory-guided attention: control from multiple memory systems}.
\bjtitle{Trends in Cognitive Sciences}
\bvolume{16}(\bissue{12}),
\bfpage{576}--\blpage{579}
(\byear{2012})
\end{barticle}
\endbibitem

\bibitem[\protect\citeauthoryear{Chamberlain et~al.}{2021}]{chamberlain2021grand}
\begin{bchapter}
\bauthor{\bsnm{Chamberlain}, \binits{B.}},
\bauthor{\bsnm{Rowbottom}, \binits{J.}},
\bauthor{\bsnm{Gorinova}, \binits{M.I.}},
\bauthor{\bsnm{Bronstein}, \binits{M.}},
\bauthor{\bsnm{Webb}, \binits{S.}},
\bauthor{\bsnm{Rossi}, \binits{E.}}:
\bctitle{Grand: Graph neural diffusion}.
In: \bbtitle{International Conference on Machine Learning},
pp. \bfpage{1407}--\blpage{1418}
(\byear{2021})
\end{bchapter}
\endbibitem

\bibitem[\protect\citeauthoryear{Dan et~al.}{2023}]{dan2023re}
\begin{barticle}
\bauthor{\bsnm{Dan}, \binits{T.}},
\bauthor{\bsnm{Ding}, \binits{J.}},
\bauthor{\bsnm{Wei}, \binits{Z.}},
\bauthor{\bsnm{Kovalsky}, \binits{S.}},
\bauthor{\bsnm{Kim}, \binits{M.}},
\bauthor{\bsnm{Kim}, \binits{W.H.}},
\bauthor{\bsnm{Wu}, \binits{G.}}:
\batitle{Re-think and re-design graph neural networks in spaces of continuous graph diffusion functionals}.
\bjtitle{Advances in Neural Information Processing Systems}
\bvolume{36},
\bfpage{59375}--\blpage{59387}
(\byear{2023})
\end{barticle}
\endbibitem

\bibitem[\protect\citeauthoryear{Young}{1804}]{young1804bakerian}
\begin{barticle}
\bauthor{\bsnm{Young}, \binits{T.I.}}:
\batitle{The Bakerian Lecture. experiments and calculations relative to physical optics}.
\bjtitle{Philosophical Transactions of the Royal Society of London}
\bvolume{94},
\bfpage{1}--\blpage{16}
(\byear{1804})
\end{barticle}
\endbibitem

\bibitem[\protect\citeauthoryear{Gabor}{1948}]{gabor1948new}
\begin{barticle}
\bauthor{\bsnm{Gabor}, \binits{D.}}:
\batitle{A new microscopic principle}.
\bjtitle{Nature}
\bvolume{161},
\bfpage{777}
(\byear{1948})
\end{barticle}
\endbibitem

\bibitem[\protect\citeauthoryear{Chen et~al.}{2022}]{chen2022characterizing}
\begin{barticle}
\bauthor{\bsnm{Chen}, \binits{J.}},
\bauthor{\bsnm{Cai}, \binits{H.}},
\bauthor{\bsnm{Yang}, \binits{D.}},
\bauthor{\bsnm{Styner}, \binits{M.}},
\bauthor{\bsnm{Wu}, \binits{G.}}, \betal:
\batitle{Characterizing the propagation pathway of neuropathological events of alzheimer's disease using harmonic wavelet analysis}.
\bjtitle{Medical Image Analysis}
\bvolume{79},
\bfpage{102446}
(\byear{2022})
\end{barticle}
\endbibitem

\bibitem[\protect\citeauthoryear{Gao et~al.}{2019}]{gao2019geometric}
\begin{bchapter}
\bauthor{\bsnm{Gao}, \binits{F.}},
\bauthor{\bsnm{Wolf}, \binits{G.}},
\bauthor{\bsnm{Hirn}, \binits{M.}}:
\bctitle{Geometric scattering for graph data analysis}.
In: \bbtitle{International Conference on Machine Learning},
pp. \bfpage{2122}--\blpage{2131}
(\byear{2019})
\end{bchapter}
\endbibitem


\bibitem[\protect\citeauthoryear{Kuramoto}{1984}]{kuramoto1984chemical}
\begin{bchapter}
\bauthor{\bsnm{Kuramoto}, \binits{Y.}}:
\batitle{Chemical turbulence}.
In:
\bbtitle{Chemical oscillations, waves, and turbulence}
\bfpage{111--140}
(\byear{1984})
\end{bchapter}
\endbibitem

\bibitem[\protect\citeauthoryear{Cabral et~al.}{2011}]{cabral2011role}
\begin{barticle}
\bauthor{\bsnm{Cabral}, \binits{J.}},
\bauthor{\bsnm{Hugues}, \binits{E.}},
\bauthor{\bsnm{Sporns}, \binits{O.}},
\bauthor{\bsnm{Deco}, \binits{G.}}:
\batitle{Role of local network oscillations in resting-state functional connectivity}.
\bjtitle{NeuroImage}
\bvolume{57}(\bissue{1}),
\bfpage{130}--\blpage{139}
(\byear{2011})
\end{barticle}
\endbibitem

\bibitem[\protect\citeauthoryear{Bookheimer et~al.}{2019}]{bookheimer2019lifespan}
\begin{barticle}
\bauthor{\bsnm{Bookheimer}, \binits{S.Y.}},
\bauthor{\bsnm{Salat}, \binits{D.H.}},
\bauthor{\bsnm{Terpstra}, \binits{M.}},
\bauthor{\bsnm{Ances}, \binits{B.M.}},
\bauthor{\bsnm{Barch}, \binits{D.M.}},
\bauthor{\bsnm{Buckner}, \binits{R.L.}},
\bauthor{\bsnm{Burgess}, \binits{G.C.}},
\bauthor{\bsnm{Curtiss}, \binits{S.W.}},
\bauthor{\bsnm{Diaz-Santos}, \binits{M.}},
\bauthor{\bsnm{Elam}, \binits{J.S.}}, \betal:
\batitle{The lifespan human connectome project in aging: an overview}.
\bjtitle{NeuroImage}
\bvolume{185},
\bfpage{335}--\blpage{348}
(\byear{2019})
\end{barticle}
\endbibitem

\bibitem[\protect\citeauthoryear{Van~Essen et~al.}{2013}]{van2013wu}
\begin{barticle}
\bauthor{\bsnm{Van~Essen}, \binits{D.C.}},
\bauthor{\bsnm{Smith}, \binits{S.M.}},
\bauthor{\bsnm{Barch}, \binits{D.M.}},
\bauthor{\bsnm{Behrens}, \binits{T.E.}},
\bauthor{\bsnm{Yacoub}, \binits{E.}},
\bauthor{\bsnm{Ugurbil}, \binits{K.}}:
\batitle{The wu-minn human connectome project: An overview}.
\bjtitle{NeuroImage}
\bvolume{80},
\bfpage{62}--\blpage{79}
(\byear{2013})
\end{barticle}
\endbibitem

\bibitem[\protect\citeauthoryear{Tzourio-Mazoyer et~al.}{2002}]{tzourio2002automated}
\begin{barticle}
\bauthor{\bsnm{Tzourio-Mazoyer}, \binits{N.}},
\bauthor{\bsnm{Landeau}, \binits{B.}},
\bauthor{\bsnm{Papathanassiou}, \binits{D.}},
\bauthor{\bsnm{Crivello}, \binits{F.}},
\bauthor{\bsnm{Etard}, \binits{O.}},
\bauthor{\bsnm{Delcroix}, \binits{N.}},
\bauthor{\bsnm{Mazoyer}, \binits{B.}},
\bauthor{\bsnm{Joliot}, \binits{M.}}:
\batitle{Automated anatomical labeling of activations in spm using a macroscopic anatomical parcellation of the MNI MRI single-subject brain}.
\bjtitle{NeuroImage}
\bvolume{15}(\bissue{1}),
\bfpage{273}--\blpage{289}
(\byear{2002})
\end{barticle}
\endbibitem

\bibitem[\protect\citeauthoryear{Fan et~al.}{2016}]{fan2016human}
\begin{barticle}
\bauthor{\bsnm{Fan}, \binits{L.}},
\bauthor{\bsnm{Li}, \binits{H.}},
\bauthor{\bsnm{Zhuo}, \binits{J.}},
\bauthor{\bsnm{Zhang}, \binits{Y.}},
\bauthor{\bsnm{Wang}, \binits{J.}},
\bauthor{\bsnm{Chen}, \binits{L.}},
\bauthor{\bsnm{Yang}, \binits{Z.}},
\bauthor{\bsnm{Chu}, \binits{C.}},
\bauthor{\bsnm{Xie}, \binits{S.}},
\bauthor{\bsnm{Laird}, \binits{A.R.}}, \betal:
\batitle{The human brainnetome atlas: a new brain atlas based on connectional architecture}.
\bjtitle{Cerebral Cortex}
\bvolume{26}(\bissue{8}),
\bfpage{3508}--\blpage{3526}
(\byear{2016})
\end{barticle}
\endbibitem

\bibitem[\protect\citeauthoryear{Veličković et~al.}{2018}]{velivckovic2017graph}
\begin{bchapter}
\bauthor{\bsnm{Veličković}, \binits{P.}},
\bauthor{\bsnm{Cucurull}, \binits{G.}},
\bauthor{\bsnm{Casanova}, \binits{A.}},
\bauthor{\bsnm{Romero}, \binits{A.}},
\bauthor{\bsnm{Liò}, \binits{P.}},
\bauthor{\bsnm{Bengio}, \binits{Y.}}:
\bctitle{Graph attention networks}.
In: \bbtitle{International Conference on Learning Representations}
(\byear{2018}).
\burl{https://openreview.net/forum?id=rJXMpikCZ}
\end{bchapter}
\endbibitem

\bibitem[\protect\citeauthoryear{Xu et~al.}{2019}]{xu2018powerful}
\begin{bchapter}
\bauthor{\bsnm{Xu}, \binits{K.}},
\bauthor{\bsnm{Hu}, \binits{W.}},
\bauthor{\bsnm{Leskovec}, \binits{J.}},
\bauthor{\bsnm{Jegelka}, \binits{S.}}:
\bctitle{How powerful are graph neural networks?}
In: \bbtitle{International Conference on Learning Representations}
(\byear{2019}).
\burl{https://openreview.net/forum?id=ryGs6iA5Km}
\end{bchapter}
\endbibitem

\bibitem[\protect\citeauthoryear{Chen et~al.}{2020}]{chen2020simple}
\begin{bchapter}
\bauthor{\bsnm{Chen}, \binits{M.}},
\bauthor{\bsnm{Wei}, \binits{Z.}},
\bauthor{\bsnm{Huang}, \binits{Z.}},
\bauthor{\bsnm{Ding}, \binits{B.}},
\bauthor{\bsnm{Li}, \binits{Y.}}:
\bctitle{Simple and deep graph convolutional networks}.
In: \bbtitle{International Conference on Machine Learning},
pp. \bfpage{1725}--\blpage{1735}
(\byear{2020})
\end{bchapter}
\endbibitem

\bibitem[\protect\citeauthoryear{Hamilton et~al.}{2017}]{hamilton2017inductive}
\begin{bchapter}
\bauthor{\bsnm{Hamilton}, \binits{W.L.}},
\bauthor{\bsnm{Ying}, \binits{Z.}},
\bauthor{\bsnm{Leskovec}, \binits{J.}}:
\bctitle{Inductive representation learning on large graphs}.
In: \bbtitle{Advances in Neural Information Processing Systems},
pp. \bfpage{1025}--\blpage{1035}
(\byear{2017}).
\burl{http://papers.nips.cc/paper/6703-inductive-representation-learning-on-large-graphs}
\end{bchapter}
\endbibitem

\bibitem[\protect\citeauthoryear{Kreuzer et~al.}{2021}]{kreuzer2021rethinking}
\begin{barticle}
\bauthor{\bsnm{Kreuzer}, \binits{D.}},
\bauthor{\bsnm{Beaini}, \binits{D.}},
\bauthor{\bsnm{Hamilton}, \binits{W.}},
\bauthor{\bsnm{L{\'e}tourneau}, \binits{V.}},
\bauthor{\bsnm{Tossou}, \binits{P.}}:
\batitle{Rethinking graph transformers with spectral attention}.
\bjtitle{Advances in Neural Information Processing Systems}
\bvolume{34},
\bfpage{21618}--\blpage{21629}
(\byear{2021})
\end{barticle}
\endbibitem

\bibitem[\protect\citeauthoryear{Rusch et~al.}{2022}]{rusch2022graph}
\begin{bchapter}
\bauthor{\bsnm{Rusch}, \binits{T.K.}},
\bauthor{\bsnm{Chamberlain}, \binits{B.}},
\bauthor{\bsnm{Rowbottom}, \binits{J.}},
\bauthor{\bsnm{Mishra}, \binits{S.}},
\bauthor{\bsnm{Bronstein}, \binits{M.}}:
\bctitle{Graph-coupled oscillator networks}.
In: \bbtitle{International Conference on Machine Learning},
pp. \bfpage{18888}--\blpage{18909}
(\byear{2022})
\end{bchapter}
\endbibitem

\bibitem[\protect\citeauthoryear{Nguyen et~al.}{2024}]{nguyen2024coupled}
\begin{bchapter}
\bauthor{\bsnm{Nguyen}, \binits{T.}},
\bauthor{\bsnm{Honda}, \binits{H.}},
\bauthor{\bsnm{Sano}, \binits{T.}},
\bauthor{\bsnm{Nguyen}, \binits{V.}},
\bauthor{\bsnm{Nakamura}, \binits{S.}},
\bauthor{\bsnm{Nguyen}, \binits{T.M.}}:
\bctitle{From coupled oscillators to graph neural networks: Reducing over-smoothing via a kuramoto model-based approach}.
In: \bbtitle{International Conference on Artificial Intelligence and Statistics},
pp. \bfpage{2710}--\blpage{2718}
(\byear{2024})
\end{bchapter}
\endbibitem

\bibitem[\protect\citeauthoryear{Raichle et~al.}{2001}]{raichle2001default}
\begin{barticle}
\bauthor{\bsnm{Raichle}, \binits{M.E.}},
\bauthor{\bsnm{MacLeod}, \binits{A.M.}},
\bauthor{\bsnm{Snyder}, \binits{A.Z.}},
\bauthor{\bsnm{Powers}, \binits{W.J.}},
\bauthor{\bsnm{Gusnard}, \binits{D.A.}},
\bauthor{\bsnm{Shulman}, \binits{G.L.}}:
\batitle{A default mode of brain function}.
\bjtitle{Proceedings of the National Academy of Sciences}
\bvolume{98}(\bissue{2}),
\bfpage{676}--\blpage{682}
(\byear{2001})
\end{barticle}
\endbibitem


\bibitem[\protect\citeauthoryear{Grady et~al.}{2003}]{Grady2010}
\begin{barticle}
\bauthor{\bsnm{Grady}, \binits{C.L.}},
\bauthor{\bsnm{McIntosh}, \binits{A.R.}},
\bauthor{\bsnm{Craik}, \binits{F.I.M.}}:
\batitle{Age-related differences in the functional connectivity of the hippocampus during memory encoding}.
\bjtitle{Hippocampus}
\bvolume{13}(\bissue{5}),
\bfpage{572--586}
(\byear{2003})
\doiurl{10.1002/hipo.10114}
\end{barticle}
\endbibitem


\bibitem[\protect\citeauthoryear{Seidler et~al.}{2010}]{Seidler2010}
\begin{barticle}
\bauthor{\bsnm{Seidler}, \binits{R.D.}},
\bauthor{\bsnm{Bernard}, \binits{J.A.}},
\bauthor{\bsnm{Burutolu}, \binits{T.B.}},
\bauthor{\bsnm{Fling}, \binits{B.W.}},
\bauthor{\bsnm{Gordon}, \binits{M.T.}},
\bauthor{\bsnm{Gwin}, \binits{J.T.}},
\bauthor{\bsnm{Kwak}, \binits{Y.}},
\bauthor{\bsnm{Lipps}, \binits{D.B.}}:
\batitle{Motor control and aging: Links to age‐related brain structural, functional, and biochemical effects}.
\bjtitle{Neuroscience \& Biobehavioral Reviews}
\bvolume{34}(\bissue{5}),
\bfpage{721}--\blpage{733}
(\byear{2010})
\doiurl{10.1016/j.neubiorev.2009.10.005}
\end{barticle}
\endbibitem


\bibitem[\protect\citeauthoryear{Pathak et~al.}{2022}]{zhang2022biophysical}
\begin{barticle}
\bauthor{\bsnm{Pathak}, \binits{A.}},
\bauthor{\bsnm{Sharma}, \binits{V.}},
\bauthor{\bsnm{Roy}, \binits{D.}},
\bauthor{\bsnm{Banerjee}, \binits{A.}}:
\batitle{Biophysical mechanism underlying compensatory preservation of neural synchrony over the adult lifespan}.
\bjtitle{Communications Biology}
\bvolume{5}(\bissue{1}),
\bfpage{567}
(\byear{2022})
\end{barticle}
\endbibitem


\bibitem[\protect\citeauthoryear{Power et~al.}{2011}]{Power2011}
\begin{barticle}
\bauthor{\bsnm{Power}, \binits{J.D.}},
\bauthor{\bsnm{Cohen}, \binits{A.L.}},
\bauthor{\bsnm{Nelson}, \binits{S.M.}},
\bauthor{\bsnm{Wig}, \binits{G.S.}},
\bauthor{\bsnm{Barnes}, \binits{K.A.}},
\bauthor{\bsnm{Church}, \binits{J.A.}},
\bauthor{\bsnm{Vogel}, \binits{A.C.}},
\bauthor{\bsnm{Laumann}, \binits{T.O.}},
\bauthor{\bsnm{Miezin}, \binits{F.M.}},
\bauthor{\bsnm{Schlaggar}, \binits{B.L.}},
\bauthor{\bsnm{Petersen}, \binits{S.E.}}:
\batitle{Functional network organization of the human brain}.
\bjtitle{Neuron}
\bvolume{72}(\bissue{4}),
\bfpage{665}--\blpage{678}
(\byear{2011})
\doiurl{10.1016/j.neuron.2011.09.006}
\end{barticle}
\endbibitem

\bibitem[\protect\citeauthoryear{Park and Friston}{2013}]{park2013structural}
\begin{barticle}
\bauthor{\bsnm{Park}, \binits{H.-J.}},
\bauthor{\bsnm{Friston}, \binits{K.}}:
\batitle{Structural and functional brain networks: from connections to cognition}.
\bjtitle{Science}
\bvolume{342}(\bissue{6158}),
\bfpage{1238411}
(\byear{2013})
\end{barticle}
\endbibitem

\bibitem[\protect\citeauthoryear{Brier et~al.}{2016}]{Brier2015}
\begin{barticle}
\bauthor{\bsnm{Brier}, \binits{M.R.}},
\bauthor{\bsnm{Gordon}, \binits{B.}},
\bauthor{\bsnm{Friedrichsen}, \binits{K.}},
\bauthor{\bsnm{McCarthy}, \binits{J.}},
\bauthor{\bsnm{Stern}, \binits{A.}},
\bauthor{\bsnm{Christensen}, \binits{J.}},
\bauthor{\bsnm{Owen}, \binits{C.}},
\bauthor{\bsnm{Aldea}, \binits{P.}},
\bauthor{\bsnm{Su}, \binits{Y.}},
\bauthor{\bsnm{Hassenstab}, \binits{J.}}, \betal:
\batitle{Tau and a$\beta$ imaging, csf measures, and cognition in alzheimer’s disease}.
\bjtitle{Science Translational Medicine}
\bvolume{8}(\bissue{338}),
\bfpage{338}--\blpage{6633866}
(\byear{2016})
\end{barticle}
\endbibitem

\bibitem[\protect\citeauthoryear{Weil et~al.}{2016}]{weil2016visual}
\begin{barticle}
\bauthor{\bsnm{Weil}, \binits{R.S.}},
\bauthor{\bsnm{Schrag}, \binits{A.E.}},
\bauthor{\bsnm{Warren}, \binits{J.D.}},
\bauthor{\bsnm{Crutch}, \binits{S.J.}},
\bauthor{\bsnm{Lees}, \binits{A.J.}},
\bauthor{\bsnm{Morris}, \binits{H.R.}}:
\batitle{Visual dysfunction in parkinson’s disease}.
\bjtitle{Brain}
\bvolume{139}(\bissue{11}),
\bfpage{2827}--\blpage{2843}
(\byear{2016})
\end{barticle}
\endbibitem

\bibitem[\protect\citeauthoryear{Mandelli et~al.}{2016}]{Mandelli2016}
\begin{barticle}
\bauthor{\bsnm{Mandelli}, \binits{M.L.}},
\bauthor{\bsnm{Vilaplana}, \binits{E.}},
\bauthor{\bsnm{Brown}, \binits{J.A.}},
\bauthor{\bsnm{Hubbard}, \binits{H.I.}},
\bauthor{\bsnm{Binney}, \binits{R.J.}},
\bauthor{\bsnm{Attygalle}, \binits{S.}},
\bauthor{\bsnm{Santos-Santos}, \binits{M.A.}},
\bauthor{\bsnm{Miller}, \binits{Z.A.}},
\bauthor{\bsnm{Pakvasa}, \binits{M.}},
\bauthor{\bsnm{Henry}, \binits{M.L.}}, \betal:
\batitle{Healthy brain connectivity predicts atrophy progression in non-fluent variant of primary progressive aphasia}.
\bjtitle{Brain}
\bvolume{139}(\bissue{10}),
\bfpage{2778}--\blpage{2791}
(\byear{2016})
\end{barticle}
\endbibitem


\bibitem[\protect\citeauthoryear{Bouzigues et~al.}{2025}]{bouzigues2024disruption}
\begin{barticle}
\bauthor{\bsnm{Bouzigues}, \binits{A.}},
\bauthor{\bsnm{Godefroy}, \binits{V.}},
\bauthor{\bsnm{Le Du}, \binits{V.}},
\bauthor{\bsnm{Russell}, \binits{L.L.}},
\bauthor{\bsnm{Houot}, \binits{M.}},
\bauthor{\bsnm{Le Ber}, \binits{I.}},
\bauthor{\bsnm{Batrancourt}, \binits{B.}},
\bauthor{\bsnm{Levy}, \binits{R.}},
\bauthor{\bsnm{Warren}, \binits{J.D.}},
\bauthor{\bsnm{Rohrer}, \binits{J.D.}}:
\batitle{Disruption of macroscale functional network organisation in patients with frontotemporal dementia}.
\bjtitle{Molecular Psychiatry}
\bvolume{30}(\bissue{6}),
\bfpage{2436--2447}
(\byear{2025})
\end{barticle}
\endbibitem


\bibitem[\protect\citeauthoryear{Li et~al.}{2018}]{li2021gnns}
\begin{barticle}
\bauthor{\bsnm{Li}, \binits{Q.}},
\bauthor{\bsnm{Han}, \binits{Z.}},
\bauthor{\bsnm{Wu}, \binits{X.-M.}}:
\batitle{Deeper insights into graph convolutional networks for semi-supervised learning}.
\bjtitle{Proceedings of the AAAI Conference on Artificial Intelligence}
\bvolume{32}(\bissue{1})
(\byear{2018})
\end{barticle}
\endbibitem



\bibitem[\protect\citeauthoryear{Buckner and Wheeler}{2001}]{dobbins2002cognitive}
\begin{barticle}
\bauthor{\bsnm{Buckner}, \binits{R.L.}},
\bauthor{\bsnm{Wheeler}, \binits{M.E.}}:
\batitle{The cognitive neuroscience of remembering}.
\bjtitle{Nature Reviews Neuroscience}
\bvolume{2}(\bissue{9}),
\bfpage{624--634}
(\byear{2001})
\end{barticle}
\endbibitem


\bibitem[\protect\citeauthoryear{Zhu et~al.}{2020}]{zhu2020beyond}
\begin{barticle}
\bauthor{\bsnm{Zhu}, \binits{J.}},
\bauthor{\bsnm{Yan}, \binits{Y.}},
\bauthor{\bsnm{Zhao}, \binits{L.}},
\bauthor{\bsnm{Heimann}, \binits{M.}},
\bauthor{\bsnm{Akoglu}, \binits{L.}},
\bauthor{\bsnm{Koutra}, \binits{D.}}:
\batitle{Beyond homophily in graph neural networks: Current limitations and effective designs}.
\bjtitle{Advances in Neural Information Processing Systems}
\bvolume{33},
\bfpage{7793}--\blpage{7804}
(\byear{2020})
\end{barticle}
\endbibitem

\bibitem[\protect\citeauthoryear{Hu et~al.}{2020}]{hu2020open}
\begin{barticle}
\bauthor{\bsnm{Hu}, \binits{W.}},
\bauthor{\bsnm{Fey}, \binits{M.}},
\bauthor{\bsnm{Zitnik}, \binits{M.}},
\bauthor{\bsnm{Dong}, \binits{Y.}},
\bauthor{\bsnm{Ren}, \binits{H.}},
\bauthor{\bsnm{Liu}, \binits{B.}},
\bauthor{\bsnm{Catasta}, \binits{M.}},
\bauthor{\bsnm{Leskovec}, \binits{J.}}:
\batitle{Open graph benchmark: Datasets for machine learning on graphs}.
\bjtitle{Advances in Neural Information Processing Systems}
\bvolume{33},
\bfpage{22118}--\blpage{22133}
(\byear{2020})
\end{barticle}
\endbibitem

\bibitem[\protect\citeauthoryear{Morris et~al.}{2020}]{morris2020tudataset}
\begin{bchapter}
\bauthor{\bsnm{Morris}, \binits{C.}},
\bauthor{\bsnm{Kriege}, \binits{N.M.}},
\bauthor{\bsnm{Bause}, \binits{F.}},
\bauthor{\bsnm{Kersting}, \binits{K.}},
\bauthor{\bsnm{Mutzel}, \binits{P.}},
\bauthor{\bsnm{Neumann}, \binits{M.}}:
\bctitle{Tudataset: A collection of benchmark datasets for learning with graphs}.
In: \bbtitle{ICML 2020 Workshop on Graph Representation Learning and Beyond (GRL+ 2020)}
(\byear{2020}).
\burl{https://arxiv.org/abs/2007.08663}
\end{bchapter}
\endbibitem

\bibitem[\protect\citeauthoryear{Pei et~al.}{2020}]{pei2020geom}
\begin{bchapter}
\bauthor{\bsnm{Pei}, \binits{H.}},
\bauthor{\bsnm{Wei}, \binits{B.}},
\bauthor{\bsnm{Chang}, \binits{K.C.-C.}},
\bauthor{\bsnm{Lei}, \binits{Y.}},
\bauthor{\bsnm{Yang}, \binits{B.}}:
\bctitle{Geom-gcn: Geometric graph convolutional networks}.
In: \bbtitle{International Conference on Learning Representations}
(\byear{2020}).
\burl{https://openreview.net/forum?id=S1e2agrFvS}
\end{bchapter}
\endbibitem

\bibitem[\protect\citeauthoryear{Errica et~al.}{2020}]{errica2019fair}
\begin{bchapter}
\bauthor{\bsnm{Errica}, \binits{F.}},
\bauthor{\bsnm{Podda}, \binits{M.}},
\bauthor{\bsnm{Bacciu}, \binits{D.}},
\bauthor{\bsnm{Micheli}, \binits{A.}}:
\bctitle{A fair comparison of graph neural networks for graph classification}.
In: \bbtitle{International Conference on Learning Representations}
(\byear{2020}).
\burl{https://openreview.net/forum?id=HygDF6NFPB}
\end{bchapter}
\endbibitem

\bibitem[\protect\citeauthoryear{Shaham et~al.}{2018}]{shaham2018spectralnet}
\begin{bchapter}
\bauthor{\bsnm{Shaham}, \binits{U.}},
\bauthor{\bsnm{Stanton}, \binits{K.}},
\bauthor{\bsnm{Li}, \binits{H.}},
\bauthor{\bsnm{Basri}, \binits{R.}},
\bauthor{\bsnm{Nadler}, \binits{B.}},
\bauthor{\bsnm{Kluger}, \binits{Y.}}:
\bctitle{Spectralnet: Spectral clustering using deep neural networks}.
In: \bbtitle{International Conference on Learning Representations}
(\byear{2018}).
\burl{https://openreview.net/forum?id=HJ\_aoCyRZ}
\end{bchapter}
\endbibitem

\bibitem[\protect\citeauthoryear{Botvinick and Cohen}{2014}]{botvinick2014computational}
\begin{barticle}
\bauthor{\bsnm{Botvinick}, \binits{M.M.}},
\bauthor{\bsnm{Cohen}, \binits{J.D.}}:
\batitle{The computational and neural basis of cognitive control: charted territory and new frontiers}.
\bjtitle{Cognitive Science}
\bvolume{38}(\bissue{6}),
\bfpage{1249}--\blpage{1285}
(\byear{2014})
\end{barticle}
\endbibitem

\bibitem[\protect\citeauthoryear{Gu et~al.}{2015}]{gu2015controllability}
\begin{barticle}
\bauthor{\bsnm{Gu}, \binits{S.}},
\bauthor{\bsnm{Pasqualetti}, \binits{F.}},
\bauthor{\bsnm{Cieslak}, \binits{M.}},
\bauthor{\bsnm{Telesford}, \binits{Q.K.}},
\bauthor{\bsnm{Yu}, \binits{A.B.}},
\bauthor{\bsnm{Kahn}, \binits{A.E.}},
\bauthor{\bsnm{Medaglia}, \binits{J.D.}},
\bauthor{\bsnm{Vettel}, \binits{J.M.}},
\bauthor{\bsnm{Miller}, \binits{M.B.}},
\bauthor{\bsnm{Grafton}, \binits{S.T.}}, \betal:
\batitle{Controllability of structural brain networks}.
\bjtitle{Nature Communications}
\bvolume{6}(\bissue{1}),
\bfpage{8414}
(\byear{2015})
\end{barticle}
\endbibitem

\bibitem[\protect\citeauthoryear{McGowan et~al.}{2022}]{mcgowan2022controllability}
\begin{barticle}
\bauthor{\bsnm{McGowan}, \binits{A.L.}},
\bauthor{\bsnm{Parkes}, \binits{L.}},
\bauthor{\bsnm{He}, \binits{X.}},
\bauthor{\bsnm{Stanoi}, \binits{O.}},
\bauthor{\bsnm{Kang}, \binits{Y.}},
\bauthor{\bsnm{Lomax}, \binits{S.}},
\bauthor{\bsnm{Jovanova}, \binits{M.}},
\bauthor{\bsnm{Mucha}, \binits{P.J.}},
\bauthor{\bsnm{Ochsner}, \binits{K.N.}},
\bauthor{\bsnm{Falk}, \binits{E.B.}}, \betal:
\batitle{Controllability of structural brain networks and the waxing and waning of negative affect in daily life}.
\bjtitle{Biological Psychiatry Global Open Science}
\bvolume{2}(\bissue{4}),
\bfpage{432}--\blpage{439}
(\byear{2022})
\end{barticle}
\endbibitem

\bibitem[\protect\citeauthoryear{Medaglia et~al.}{2017}]{medaglia2017brain}
\begin{barticle}
\bauthor{\bsnm{Medaglia}, \binits{J.D.}},
\bauthor{\bsnm{Pasqualetti}, \binits{F.}},
\bauthor{\bsnm{Hamilton}, \binits{R.H.}},
\bauthor{\bsnm{Thompson-Schill}, \binits{S.L.}},
\bauthor{\bsnm{Bassett}, \binits{D.S.}}:
\batitle{Brain and cognitive reserve: Translation via network control theory}.
\bjtitle{Neuroscience \& Biobehavioral Reviews}
\bvolume{75},
\bfpage{53}--\blpage{64}
(\byear{2017})
\end{barticle}
\endbibitem

\bibitem[\protect\citeauthoryear{Chen et~al.}{2018}]{chen2018neuralode}
\begin{botherref}
\oauthor{\bsnm{Chen}, \binits{R.T.}},
\oauthor{\bsnm{Rubanova}, \binits{Y.}},
\oauthor{\bsnm{Bettencourt}, \binits{J.}},
\oauthor{\bsnm{Duvenaud}, \binits{D.K.}}:
Neural ordinary differential equations.
Advances in Neural Information Processing Systems
\textbf{31}
(2018)
\end{botherref}
\endbibitem

\bibitem[\protect\citeauthoryear{Hasani et~al.}{2021}]{hasani2021liquid}
\begin{bchapter}
\bauthor{\bsnm{Hasani}, \binits{R.}},
\bauthor{\bsnm{Lechner}, \binits{M.}},
\bauthor{\bsnm{Amini}, \binits{A.}},
\bauthor{\bsnm{Rus}, \binits{D.}},
\bauthor{\bsnm{Grosu}, \binits{R.}}:
\bctitle{Liquid time-constant networks}.
In: \bbtitle{Proceedings of the AAAI Conference on Artificial Intelligence},
vol. \bseriesno{35},
pp. \bfpage{7657}--\blpage{7666}
(\byear{2021})
\end{bchapter}
\endbibitem

\bibitem[\protect\citeauthoryear{Dan et~al.}{2022}]{dan2022neuro}
\begin{bchapter}
\bauthor{\bsnm{Dan}, \binits{T.}},
\bauthor{\bsnm{Cai}, \binits{H.}},
\bauthor{\bsnm{Huang}, \binits{Z.}},
\bauthor{\bsnm{Laurienti}, \binits{P.}},
\bauthor{\bsnm{Kim}, \binits{W.H.}},
\bauthor{\bsnm{Wu}, \binits{G.}}:
\bctitle{Neuro-RDM: an explainable neural network landscape of reaction-diffusion model for cognitive task recognition}.
In: \bbtitle{International Conference on Medical Image Computing and Computer-Assisted Intervention},
pp. \bfpage{365}--\blpage{374}
(\byear{2022})
\end{bchapter}
\endbibitem

\bibitem[\protect\citeauthoryear{Capouskova et~al.}{2022}]{capouskova2022modes}
\begin{barticle}
\bauthor{\bsnm{Capouskova}, \binits{K.}},
\bauthor{\bsnm{Kringelbach}, \binits{M.L.}},
\bauthor{\bsnm{Deco}, \binits{G.}}:
\batitle{Modes of cognition: Evidence from metastable brain dynamics}.
\bjtitle{NeuroImage}
\bvolume{260},
\bfpage{119489}
(\byear{2022})
\end{barticle}
\endbibitem

\bibitem[\protect\citeauthoryear{Min et~al.}{2020}]{min2020scattering}
\begin{barticle}
\bauthor{\bsnm{Min}, \binits{Y.}},
\bauthor{\bsnm{Wenkel}, \binits{F.}},
\bauthor{\bsnm{Wolf}, \binits{G.}}:
\batitle{Scattering gcn: Overcoming oversmoothness in graph convolutional networks}.
\bjtitle{Advances in Neural Information Processing Systems}
\bvolume{33},
\bfpage{14498}--\blpage{14508}
(\byear{2020})
\end{barticle}
\endbibitem

\bibitem[\protect\citeauthoryear{Aoyagi}{1995}]{aoyagi1995network}
\begin{barticle}
\bauthor{\bsnm{Aoyagi}, \binits{T.}}:
\batitle{Network of neural oscillators for retrieving phase information}.
\bjtitle{Physical Review Letters}
\bvolume{74}(\bissue{20}),
\bfpage{4075}
(\byear{1995})
\end{barticle}
\endbibitem

\bibitem[\protect\citeauthoryear{Liu et~al.}{2011}]{liu2011controllability}
\begin{barticle}
\bauthor{\bsnm{Liu}, \binits{Y.-Y.}},
\bauthor{\bsnm{Slotine}, \binits{J.-J.}},
\bauthor{\bsnm{Barab{\'a}si}, \binits{A.-L.}}:
\batitle{Controllability of complex networks}.
\bjtitle{Nature}
\bvolume{473}(\bissue{7346}),
\bfpage{167}--\blpage{173}
(\byear{2011})
\end{barticle}
\endbibitem

\bibitem[\protect\citeauthoryear{Chandra et~al.}{2019}]{chandra2019continuous}
\begin{barticle}
\bauthor{\bsnm{Chandra}, \binits{S.}},
\bauthor{\bsnm{Girvan}, \binits{M.}},
\bauthor{\bsnm{Ott}, \binits{E.}}:
\batitle{Continuous versus discontinuous transitions in the d-dimensional generalized Kuramoto model: Odd d is different}.
\bjtitle{Physical Review X}
\bvolume{9}(\bissue{1}),
\bfpage{011002}
(\byear{2019})
\end{barticle}
\endbibitem

\bibitem[\protect\citeauthoryear{Strogatz}{2000}]{strogatz2000kuramoto}
\begin{barticle}
\bauthor{\bsnm{Strogatz}, \binits{S.H.}}:
\batitle{From Kuramoto to Crawford: exploring the onset of synchronization in populations of coupled oscillators}.
\bjtitle{Physica D: Nonlinear Phenomena}
\bvolume{143}(\bissue{1-4}),
\bfpage{1}--\blpage{20}
(\byear{2000})
\end{barticle}
\endbibitem

\bibitem[\protect\citeauthoryear{Acebr{\'o}n et~al.}{2005}]{acebron2005kuramoto}
\begin{barticle}
\bauthor{\bsnm{Acebr{\'o}n}, \binits{J.A.}},
\bauthor{\bsnm{Bonilla}, \binits{L.L.}},
\bauthor{\bsnm{P{\'e}rez~Vicente}, \binits{C.J.}},
\bauthor{\bsnm{Ritort}, \binits{F.}},
\bauthor{\bsnm{Spigler}, \binits{R.}}:
\batitle{The Kuramoto model: A simple paradigm for synchronization phenomena}.
\bjtitle{Reviews of Modern Physics}
\bvolume{77}(\bissue{1}),
\bfpage{137}--\blpage{185}
(\byear{2005})
\end{barticle}
\endbibitem


\bibitem[\protect\citeauthoryear{Rorden}{2025}]{rorden2025surfice}
\begin{barticle}
\bauthor{\bsnm{Rorden}, \binits{C.}}:
\batitle{Surfice: visualizing neuroimaging meshes, tractography streamlines and connectomes}.
\bjtitle{Nature Methods}
\bvolume{22},
\bfpage{1615}--\blpage{1616}
(\byear{2025})
\end{barticle}
\endbibitem

\bibitem[\protect\citeauthoryear{Ahrens, Geveci and Law}{2005}]{ahrens2005paraview}
\begin{barticle}
\bauthor{\bsnm{Ahrens}, \binits{J.}},
\bauthor{\bsnm{Geveci}, \binits{B.}},
\bauthor{\bsnm{Law}, \binits{C.}}:
\batitle{Paraview: An end-user tool for large data visualization}.
\bjtitle{The Visualization Handbook}
\bvolume{717}(\bissue{8}),
\bfpage{717}--\blpage{731}
(\byear{2005})
\end{barticle}
\endbibitem


\end{thebibliography}
\end{document}